\documentstyle[draft]{mn}

\def\ltsima{$\; \buildrel < \over \sim \;$}
\def\simlt{\lower.5ex\hbox{\ltsima}}
\def\gtsima{$\; \buildrel > \over \sim \;$}
\def\simgt{\lower.5ex\hbox{\gtsima}}
\def\arcsec{^{\prime\prime}}

\part*{}

\title[ISOPHOT imaging of pre-stellar cores]{The 
initial conditions of isolated star formation. V: \\
ISOPHOT imaging and the
temperatures and energy balance of pre-stellar cores}

\author[Ward-Thompson, Andr\'e \& Kirk]{D. Ward-Thompson$^{1}$,
P. Andr\'e$^{2}$ \& J. M. Kirk$^{1}$ \\
$^1$Department of Physics \& Astronomy, Cardiff University,
PO Box 913, 5 The Parade, Cardiff CF2 3YB \\
$^2$CEA, DSM, DAPNIA, Service d'Astrophysique, 
C.E. Saclay, F-91191 Gif-sur-Yvette Cedex, France
}

\date{Accepted 2001 September 7; received 2001 August 7; in original form
2000 December 21.} 
\pagerange{000--000} 
\def\LaTeX{L\kern-.36em\raise.3ex\hbox{a}\kern-.15em    
T\kern-.1667em\lower.7ex\hbox{E}\kern-.125emX} 

\begin{document} 
\label{firstpage} 

\maketitle

   \begin{abstract}
ISO data taken with the long-wavelength imaging photo-polarimeter ISOPHOT
are presented of 18 pre-stellar cores at three far-infrared
wavelengths -- 90, 170 \& 200 $\mu$m. Most of the cores are detected clearly 
at 170 and 200 $\mu$m, but only one is
detected strongly at 90 $\mu$m, indicating that mostly
they are very cold, with typical temperatures of only $\sim$10--20~K.
Colour temperature images are constructed for each of the cores.
Most of the cores are seen to be either isothermal, or to
have associated temperature gradients from the core centres to their edges,
with all except one being cooler at the centre.
We compare the data with previous ISOCAM absorption data and calculate
the energy balance for those cores in common between the two samples.
We find that the energy radiated by each core
in the far-IR is similar to that absorbed at shorter wavelengths.
Hence there is no evidence for a central 
heating source in any of the cores -- even those for which previous
evidence for core contraction exists.
This is all consistent with external heating of the cores by the local 
interstellar radiation field, confirming their pre-stellar nature.
   \end{abstract}


\section{Introduction}

The process of star formation is currently only partially understood. One
of the main problems is that we do not know the initial conditions which
pertain in the clouds from which stars form. The initial form of the radial
variation of density, temperature, velocity and magnetic field are
crucial to the protostellar collapse phase of star formation. This gap in
our knowledge is currently one the major limiting factors in our ability to 
understand the star formation process for relatively low-mass stars (0.2--2
M$_\odot$).

The main protostellar collapse phase was identified observationally by
Andr\'e, Ward-Thompson \& Barsony (1993 -- hereafter AWB93), and labelled
the Class 0 stage. The subsequent Class I stage 
(Lada \& Wilking 1984; Lada 1987) represents the late
accretion phase during which the remnant circumstellar envelope accretes
onto the central protostar and disk (Andr\'e 1994; 1997; Ward-Thompson 1996).
The final pre-main-sequence stages of Classes II \& III 
(Lada \& Wilking 1984; Lada 1987) correspond to the
Classical T Tauri (CTT) and Weak-line T Tauri (WTT) stages respectively
(Andr\'e \& Montmerle 1994).

These protostellar and pre-main-sequence stages are understood at least in
outline (for a review, see: Andr\'e, Ward-Thompson \& Barsony 2000 --
hereafter AWB00).
Infall has been reported in Class 0 sources by a number of authors
(e.g. Zhou et al. 1993; Ward-Thompson et al. 1996). However, the manner of
the collapse remains a matter for debate. The old ideas of inside-out
collapse (e.g. Shu et al. 1987) have been disputed by many authors (e.g.
Foster \& Chevalier 1993;
Whitworth et al. 1996), and recently appear to have been finally
disproved observationally (Tafalla et al. 1998). Broad general agreement
now appears to be emerging that the collapse occurs at a non-constant
rate, which decreases with time throughout the accretion phase (e.g.
Kenyon \& Hartmann 1995; Henriksen, Andr\'e \& Bontemps 1997;
Safier, McKee \& Stahler 1997).

\begin{table*}
\caption{Source positions, observing details and raw flux densities. 
In this table, and in all subsequent figures and tables, the sources
are listed in increasing order of Right Ascension. The positions quoted are 
the 200-$\mu$m peak positions. In all cases the 170-$\mu$m peak positions 
agree to within the telescope resolution. The target dedicated time (TDT)
designation of each observation is listed (see text). The 170- \& 200-$\mu$m
observations have the same TDT as they were observed together. The
90-$\mu$m observations were taken at a different time.
The peak flux densities are
quoted in MJy/sr, and the total flux densities are
quoted in Jy, to an accuracy of 3 significant figures. The absolute
calibration errors are $\sim\pm$30\% (see text for discussion).
Note that these flux densities have not had any background removal,
so they include flux from the core, the extended cloud and other backgrounds
(including Zodiacal). The
90-$\mu$m peak upper limit quoted in each case is simply the 
value at the brightest point, which may not be
exactly coincident with the source,
and also includes extended cloud emission. This is shown as an upper limit
in all cases except L1689A (which has a detection) and L1517C (which 
has no data).}
\begin{tabular}{lcccccccccc}
\hline
Source & R.A. & Dec. &  TDT & TDT &
\multicolumn{3}{c}{Peak flux} & \multicolumn{3}{c}{Total flux densities} \\
Name  & (2000) & (2000) & 90$\mu$m & 170/200$\mu$m &
\multicolumn{3}{c}{densities (MJy/sr)} &
\multicolumn{3}{c}{in mapped areas (Jy)} \\
 & & & & & 90$\mu$m & 170$\mu$m & 200$\mu$m &
90$\mu$m & 170$\mu$m & 200$\mu$m \\ \hline
L1498  & 04$^{\rm h}$ 10$^{\rm m}$ 52.2$^{\rm s}$ &
 $+$25$^{\circ}$ 09$^{\prime}$ 12$^{\prime\prime}$ & 
 65400701 & 67900714 &
$\leq$21.0 & 73.9 & 81.7 &
146 & 707 & 714 \\
L1517C & 04$^{\rm h}$ 54$^{\rm m}$ 45.3$^{\rm s}$ &
 $+$30$^{\circ}$ 35$^{\prime}$ 46$^{\prime\prime}$ & 
 --    & 68600815   &
 -- & 84.6 & 78.3 &
 --  & 840 & 859 \\
L1517A & 04$^{\rm h}$ 55$^{\rm m}$ 11.7$^{\rm s}$ & 
 $+$30$^{\circ}$ 33$^{\prime}$ 14$^{\prime\prime}$ &  
 68001003   & 68600716   &
 $\leq$28.9 & 88.6 & 91.1 &
195  & 874 & 862 \\
L1517B & 04$^{\rm h}$ 55$^{\rm m}$ 16.7$^{\rm s}$ &
 $+$30$^{\circ}$ 37$^{\prime}$ 08$^{\prime\prime}$ &  
 68601413   & 68200926   &
 $\leq$26.7 & 77.4 & 90.5 &
177  & 886 & 956 \\
L1512  & 05$^{\rm h}$ 04$^{\rm m}$ 07.6$^{\rm s}$ &
 $+$32$^{\circ}$ 44$^{\prime}$ 24$^{\prime\prime}$ &  
 65200804   & 65200717   &
 $\leq$29.6 & 81.9 & 93.2 &
181  & 798 & 815 \\
L1544  & 05$^{\rm h}$ 04$^{\rm m}$ 17.4$^{\rm s}$ &
 $+$25$^{\circ}$ 11$^{\prime}$ 29$^{\prime\prime}$ & 
 68601205   & 68600618   & 
 $\leq$32.7 & 104 & 116 &
219  & 986 & 975 \\
L1582A & 05$^{\rm h}$ 32$^{\rm m}$ 00.0$^{\rm s}$ &
 $+$12$^{\circ}$ 31$^{\prime}$ 47$^{\prime\prime}$ &  
 69201506   & 69201419   &
 $\leq$102 & 315 & 374 &
463  & 2280 & 2550 \\
L183   & 15$^{\rm h}$ 54$^{\rm m}$ 12.5$^{\rm s}$ &
 $-$02$^{\circ}$ 52$^{\prime}$ 04$^{\prime\prime}$ &  
 62701701   & 62701813   &
 $\leq$18.3 & 76.2 & 77.2 &
107  & 501 & 475 \\
L1696A & 16$^{\rm h}$ 28$^{\rm m}$ 32.7$^{\rm s}$ &
 $-$24$^{\circ}$ 18$^{\prime}$ 14$^{\prime\prime}$ &  
 12002302   & 12002214   &
 $\leq$191 & 308 & 368 &
1050  & 2120 & 2610 \\
L1709A & 16$^{\rm h}$ 30$^{\rm m}$ 48.6$^{\rm s}$ &
 $-$23$^{\circ}$ 41$^{\prime}$ 34$^{\prime\prime}$ &  
 12001407   & 12001619   &
 $\leq$122 & 185 & 229 &
605  & 1340 & 1580 \\
L1689A & 16$^{\rm h}$ 32$^{\rm m}$ 14.3$^{\rm s}$ &
 $-$25$^{\circ}$ 03$^{\prime}$ 29$^{\prime\prime}$ &  
 11800703   & 11800415   &
 252 & 649 & 701 &
1240  & 3380 & 4060 \\
L1709C & 16$^{\rm h}$ 33$^{\rm m}$ 57.8$^{\rm s}$ &
 $-$23$^{\circ}$ 40$^{\prime}$ 33$^{\prime\prime}$ &  
 12002408   & 12001520   &
 $\leq$89.5 & 179 & 224 &
445  & 1090 & 1440 \\
L1689B & 16$^{\rm h}$ 34$^{\rm m}$ 46.2$^{\rm s}$ &
 $-$24$^{\circ}$ 37$^{\prime}$ 46$^{\prime\prime}$ & 
 12001904   & 11800516   & 
 $\leq$138 & 260 & 235 &
697  & 1630 & 1450 \\
L204B  & 16$^{\rm h}$ 47$^{\rm m}$ 39.1$^{\rm s}$ &
 $-$11$^{\circ}$ 58$^{\prime}$ 02$^{\prime\prime}$ & 
 11400309   & 11400421   & 
 $\leq$60.0 & 145 & 191 &
329  & 1020 & 1220 \\
L63    & 16$^{\rm h}$ 50$^{\rm m}$ 16.0$^{\rm s}$ &
 $-$18$^{\circ}$ 04$^{\prime}$ 40$^{\prime\prime}$ & 
 11800605   & 11800317   & 
 $\leq$54.5 & 120 & 131 &
290  & 709 & 656 \\
B68    & 17$^{\rm h}$ 22$^{\rm m}$ 37.9$^{\rm s}$ &
 $-$23$^{\circ}$ 50$^{\prime}$ 24$^{\prime\prime}$ & 
 11800910   & 11800822   & 
 $\leq$65.4 & 109 & 148 &
316  & 657 & 918 \\
B133   & 19$^{\rm h}$ 06$^{\rm m}$ 11.6$^{\rm s}$ &
 $-$06$^{\circ}$ 53$^{\prime}$ 03$^{\prime\prime}$ & 
 11402306   & 12001318   &
 $\leq$58.5 & 134 & 184 &
285  & 747 & 980 \\
L1155C & 20$^{\rm h}$ 43$^{\rm m}$ 13.0$^{\rm s}$ &
 $+$67$^{\circ}$ 52$^{\prime}$ 00$^{\prime\prime}$ &  
 12000212   & 12000324   &
 $\leq$22.8 & 71.2 & 75.6 &
116  & 442 & 419 \\
\hline
\end{tabular}
\end{table*}

However, the exact form of the collapse depends almost entirely on the initial
conditions (Whitworth \& Summers 1985; Foster \& Chevalier 1993;
Whitworth et al. 1996). 
A decreasing accretion rate is obtained when the initial radial density 
is relatively flat in the centre, and steepens towards the edge
(Foster \& Chevalier 1993; Henriksen et al 1997; Whitworth \& 
Ward-Thompson 2001). 

In this series of papers we are exploring observationally the initial 
conditions for star formation in order to determine empirically the physical 
parameters in molecular cloud cores which are on the verge of forming stars.
We selected for study the regions which were identified by Myers and 
co-workers (e.g. Myers \& Benson 1983; Myers, Linke \& Benson 1983; Benson
\& Myers 1989) from the catalogues of Lynds and others (e.g. Lynds 1962),
and which were observed in various transitions of NH$_3$, CO and other 
molecules (Benson \& Myers 1989 and references therein). But
we selected from these a sub-sample of cores which do not contain 
IRAS (Infra-Red Astronomical Satellite)
sources (Beichman et al. 1986; 1988), 
on the grounds that these should be at
an earlier evolutionary stage, with the cores with IRAS sources having 
already formed protostars at their centres.

In Paper I (Ward-Thompson et al. 1994)
we coined the term pre-protosellar
cores to refer to those cores without embedded stars (the `starless cores' 
of Myers and co-workers) which appeared to be sufficiently centrally
condensed to be about to form stars. We have subsequently shortened this
term to pre-stellar cores (c.f. AWB93). Paper I discussed the results of a 
submillimetre study of some pre-stellar cores, and found 
that the cores all appeared to follow a similar form of density profile 
to that predicted by theory to produce a decreasing accretion rate with
time (Foster \& Chevalier 1993).

In Papers II \& III (Andr\'e, 
Ward-Thompson \& Motte 1996; Ward-Thompson, 
Motte \& Andr\'e 1999) we presented studies at 
1.3-millimetre wavelength 
of a number of pre-stellar cores, and compared the findings in detail with
models which predict evolution of molecular cloud cores by ambipolar
diffusion (e.g. Basu \& Mouschovias 1995). It was found that, 
although the radial
density profiles of the cores appear similar to those predicted by ambipolar 
diffusion models, the details of the time-scales required by the models at
different stages did not appear to match the life-times we calculated from
the numbers we detected at each evolutionary stage. However, these results
were based on relatively small number statistics, so we intend to improve
the accuracy of the results by enlarging our statistical sample. 

In Paper IV
(Jessop \& Ward-Thompson 2001) we carried out a detailed spectroscopic
study of one of the cores, and showed that a temperature gradient from
the outside-in was required to explain the data. If this is true for all
of the cores, then this will affect any density gradient measurements
inferred from submillimetre and millimetre data. Therefore, 
it appears necessary to explore such temperature gradient effects in
a wide sample of cores using infra-red wavelengths.

\begin{table*}
\caption{Source background level estimates (including Zodiacal background)
in MJy/sr quoted to 3 significant figures. 
B1 represents the lowest background
emission level in each region. B2 is the level of emission
in the extended cloud, not associated with the pre-stellar
core. For L1517A, B \& C a common B2 area was taken.
The quoted errors are the 1$\sigma$ variations in each background region.
Absolute calibration errors are $\pm$30\% (see text for details).}
\begin{tabular}{lcccccc}
\hline
Source & \multicolumn{3}{c}{Background level -- B1 (MJy/sr)} &
\multicolumn{3}{c}{Background level -- B2 (MJy/sr)} \\
Name & 90$\mu$m & 170$\mu$m & 200$\mu$m & 90$\mu$m & 170$\mu$m & 200$\mu$m \\
\hline
L1498    & 17.2$\pm$0.9 & 39.5$\pm$2.5 & 43.4$\pm$1.4 & 
 20.3$\pm$0.2 & 67.2$\pm$0.4 & 67.8$\pm$0.7 \\
L1517C   &     --       & 60.0$\pm$0.7 & 55.7$\pm$0.9 &
           -- & 69.0$\pm$2.1 & 71.4$\pm$2.3 \\
L1517A   & 22.2$\pm$7.3 & 59.4$\pm$1.4 & 53.7$\pm$1.4 &
 25.0$\pm$5.3 & 69.0$\pm$2.1 & 71.4$\pm$2.3 \\
L1517B   & 22.2$\pm$7.3 & 56.7$\pm$2.1 & 61.7$\pm$2.1 &
 25.0$\pm$5.3 & 69.0$\pm$2.1 & 71.4$\pm$2.3 \\
L1512    & 24.2$\pm$0.7 & 47.7$\pm$2.1 & 44.5$\pm$1.3 &
 26.0$\pm$0.9 & 75.1$\pm$3.2 & 83.1$\pm$3.8 \\
L1544    & 26.7$\pm$0.7 & 65.3$\pm$1.3 & 56.0$\pm$2.3 &
 29.9$\pm$0.9 & 71.8$\pm$2.3 & 60.9$\pm$4.6 \\
L1582A   & 45.0$\pm$2.3 &  72.4$\pm$14 &  81.5$\pm$15 &
 74.5$\pm$3.4 &  200$\pm$5.7 &  206$\pm$6.7 \\
L183     & 14.9$\pm$0.2 & 36.7$\pm$3.4 & 28.3$\pm$4.4 &
 16.4$\pm$0.7 & 63.4$\pm$2.5 & 56.4$\pm$2.5 \\
L1696A   &  135$\pm$4.8 &  216$\pm$5.5 &   252$\pm$11 &
 178$\pm$2.5  &  270$\pm$5.0 &  336$\pm$5.5 \\
L1709A   & 94.3$\pm$1.3  & 160$\pm$9.1  & 176$\pm$5.9 &
 110$\pm$2.3  & 175$\pm$1.3  &  202$\pm$4.4 \\
L1689A   & 154$\pm$8.0 & 280$\pm$15    & 373$\pm$14   &
 212$\pm$10   & 376$\pm$12   & 465$\pm$17   \\
L1709C   & 60.9$\pm$0.9 &  109$\pm$2.3 &  141$\pm$2.5 &
 85.2$\pm$2.7 &  163$\pm$1.9 &  217$\pm$3.2 \\
L1689B   &  108$\pm$2.5 &  173$\pm$3.2 &  140$\pm$4.4 &
 117$\pm$1.3  &  213$\pm$5.5 &  192$\pm$2.7 \\
L204B    & 50.6$\pm$0.9 &  100$\pm$3.9 &  107$\pm$6.1 &
 55.0$\pm$6.5 & 127$\pm$3.8 &  138$\pm$11   \\
L63      & 40.1$\pm$1.9 & 67.4$\pm$3.4 & 55.5$\pm$2.7 &
 49.3$\pm$1.3 & 75.1$\pm$1.4 & 56.0$\pm$1.9 \\
B68      & 47.0$\pm$1.9 & 67.2$\pm$1.1 & 98.9$\pm$1.1 &
 58.7$\pm$1.1 & 85.4$\pm$1.1 &  114$\pm$2.1 \\
B133     & 45.7$\pm$1.1 & 69.0$\pm$1.3 & 82.7$\pm$5.3 &
 55.7$\pm$0.7 &  100$\pm$1.4 &  130$\pm$5.0 \\
L1155C   & 16.7$\pm$0.7 & 40.5$\pm$1.1 & 32.2$\pm$0.9 &
 19.0$\pm$0.7 & 53.7$\pm$5.7 & 46.6$\pm$6.5 \\
\hline
\end{tabular}
\end{table*}

One of the difficulties associated with studying pre-stellar cores is that 
they have not previously been detected in the infra-red, either from the
ground, or by the IRAS satellite out to 100~$\mu$m -- this is one of the 
selection criteria of these sources. 
However, Papers I--III have shown us that the spectral
energy distributions increase with decreasing wavelength from 1.3~mm to
350~$\mu$m. Consequently the spectra must peak between
100 and 350~$\mu$m. So we decided to observe the pre-stellar cores with the
Infrared Space Observatory (ISO) at far-infrared wavelengths from 90 to
200~$\mu$m using the long wavelength imaging
photo-polarimeter instrument ISOPHOT (see Ward-Thompson et al. 
1998 for preliminary results). In this paper we present the complete results 
of our ISOPHOT study of pre-stellar cores.

\section{Observations}

The Infrared Space Observatory (ISO) was launched by the European Space Agency
(ESA) on 1995 November 17 and began routine observations for the community on 
1996 February 4. It finally ceased astronomical observations on 1998 April 8,
almost a year longer than originally envisaged (Kessler et al., 1996).
Our observations were carried out with the imaging photo-polarimeter ISOPHOT
(Lemke et al., 1996)
during a number of different orbits in the course of two different observing
programmes. 

Observations for the first programme, which was allocated time
in the first call for observing proposals, took place predominantly
during the period 1996 March 10--16 in orbits 114--120, and covered the
sources in the R.A. range 16--20$^{\rm h}$. The source L183, at
an R.A. of 15$^{\rm h}$,
was observed during orbit 627 on 1997 August 4. The second programme,
which was allocated time in the second call for observing proposals, 
took place during the period from 1997 August 29 to October 8
in orbits 652--692, and observed all of our sources in the R.A. range
04--05$^{\rm h}$.

Table 1 lists our target sources.
A total of 18 sources in
16 dark clouds were observed at each of the three wavelengths, 90,
170 \& 200~$\mu$m (except for the L1517C part of the L1517 cloud, 
which was not observed at 90~$\mu$m
due to time constraints). The total time taken for all of the observations
was about 52,000 seconds.
The target dedicated time (TDT) of every observation is listed in Table 1.
The TDT is a unique 8-digit
reference number for every observation that is stored in
the ISO data archive (Arviset \& Prusti 1999;
Kessler et al., 2000). The first 3 digits of the TDT
specify the orbit number in which the observations were taken and the
remaining digits specify the observation number within that orbit.
In each case the 170- \& 200-$\mu$m
observations have the same TDT as they were observed together. Usually the
90-$\mu$m observations were taken at a different time.

All of the observations were carried out using
ISOPHOT in its over-sampled mapping mode -- astronomical observation 
template (AOT) reference code PHT32. In this mode 
the source is mapped by using the chopper to position the source
on the detector array at a series of positions, separated by less than the
detector resolution, intermediate between successive spacecraft pointings.
We used a chopper throw of 30 arcsec at the two longer wavelengths 
(C-200 and C-160 -- see below) and 15 arcsec at the shorter wavelength 
(C-90 -- see below) such that we were guaranteed a factor of at least
2 over-sampling at all wavelengths.
A map was built up of a series of such scans in a raster fashion (for
details, see: Klaas et al., 1994).

The filters used were the C-200 filter, which has a reference wavelength
of 200~$\mu$m and bandwidth full width at half maximum (FWHM) 
$\sim$30~$\mu$m, the C-160 filter, which
actually has a peak wavelength of 170~$\mu$m and bandwidth FWHM of 
$\sim$50~$\mu$m, and the C-90 filter, which has a peak wavelength of 90~$\mu$m
and FWHM $\sim$40~$\mu$m (Klaas et al. 1994). The C-200 and C-160
filters are associated with the PHT-C200 camera, which has four pixels in a 
2$\times$2 array with each pixel 90$\times$90 arcsec square. The C-90
filter is associated with the PHT-C100 camera, which has nine pixels in a 
3$\times$3 array with each pixel 45$\times$45 arcsec square (Klaas et al.
1994). Since all of the data were mapped in over-sampled
mapping mode, it was decided to re-bin all data uniformly onto a grid
of 45$\times$45 arcsec square pixels in Right Ascension and Declination.
These data are reproduced in Figures 1--8 as isophotal contour maps
superposed on grey-scale images.

The data were reduced using version 7 (V7) of the PHOT Interactive Analysis
(PIA) software (Gabriel et al. 1997;
Laureijs et al. 1998; 2001) in the standard way.
This procedure takes into account instrumental effects such as detector
responsivity and linearity, transient behaviour, dark current, detector
saturation, bias, stray light and chopper vignetting (Laureijs 
et al. 2001). 

\begin{table*}
\caption{Source extended flux densities and full width at half maximum
(FWHM) sizes. The FWHM are estimated from the 200-$\mu$m data using the 
`Best est.' (see below) background subtraction 
(the differences at other wavelengths are insignificant).
The columns headed `Best est.' represent our best estimates of the
background-subtracted flux densities of the pre-stellar cores measured
within an aperture size equal to the FWHM sizes.
For the 200- and 170-$\mu$m data: the columns headed 
B1 \& B2 represent the flux densities in the FWHM using
different methods for removing
the background flux densities due to the general far-infrared background, 
and the extended cloud emission respectively (see text for discussion).
The best estimate columns use a first-order polynomial
background across the field,
while B1 \& B2 can be thought of as representing the full range of
possible variation. The error-bars quoted are the residual errors from
the background removal process, and these are generally dominated by 
fluctuations in this background. The absolute calibration errors are 
$\pm$30\%. For the 90-$\mu$m data, upper limits are recorded for all sources,
which are the 3$\sigma$ variations in the background,
except L1517C (which wasn't observed at 90$\mu$m)
and L1689A (which was detected at 90~$\mu$m). Sources marked with a $*$
are either marginal detections at 90~$\mu$m ($\leq$5$\sigma$) or else 
contain a peak
which is not coincident with the source at 170 and
200~$\mu$m. The 90-$\mu$m fluxes of these secondary peaks
are: L1582A -- 33$\pm$3.8Jy;
L1709C -- 15$\pm$3.2Jy; B68 -- 6$\pm$1.5Jy; and B133 -- 15$\pm$5Jy.}
\begin{tabular}{lcccccccc}
\hline
Source &  FWHM & 90-$\mu$m (Jy) &
\multicolumn{3}{c}{170-$\mu$m FWHM flux densities (Jy)} &
\multicolumn{3}{c}{200-$\mu$m FWHM flux densities (Jy)} \\ 
Name  & (arcmin) & Best est. & 
 Best est. & (B1) & (B2) & 
 Best est. & (B1) & (B2) \\ \hline
L1498 & 7$\times$6 & $\leq$6.0
& 68.4$\pm$1  & 132$\pm$1   & 3.04$\pm$0.08 
& 100$\pm$0.8 & 124$\pm$0.7 & 7.06$\pm$0.1 \\
L1517C & 7$\times$5 & --
& 16.6$\pm$0.7 & 21.9$\pm$0.2 & 10.1$\pm$0.6 
& 34.9$\pm$4   & 42.6$\pm$0.2 & 15.2$\pm$0.6 \\
L1517A & 7$\times$5 & $\leq$3.6
& 36.8$\pm$0.9 & 42.9$\pm$0.5 & 19.1$\pm$0.6 
& 55.9$\pm$4   & 74.1$\pm$0.5 & 25.2$\pm$0.6 \\
L1517B & 4$\times$3 & $\leq$0.5
& 13.3$\pm$0.6 & 14.1$\pm$0.4 & 5.83$\pm$0.4 
& 17.9$\pm$3   & 27.8$\pm$0.4 & 9.45$\pm$0.4 \\
L1512 & 7$\times$5 & $\leq$2.2
& 77.7$\pm$0.5 & 145$\pm$1   & 3.23$\pm$0.6 
& 93.6$\pm$0.8 & 192$\pm$0.6 & 4.60$\pm$0.7 \\
L1544 & 8$\times$4 &$\leq$4.6
& 60.3$\pm$0.4 & 131$\pm$0.6 & 17.2$\pm$0.1 
& 72.4$\pm$0.6  & 202$\pm$1   & 19.9$\pm$0.3 \\
L1582A & 7.5$\times$4 & $\leq$11$^*$
& 262$\pm$3 & 756$\pm$6 & 164$\pm$2 
& 290$\pm$3 & 880$\pm$7 & 261$\pm$2 \\
L183 & 7$\times$5 & $\leq$1.3
& 52.9$\pm$0.3 & 191$\pm$2 & 25.7$\pm$0.9 
& 63.2$\pm$0.7 & 231$\pm$2 & 46.6$\pm$1   \\
L1696A & 5$\times$5 & $\leq$8.9
& 381$\pm$2 & 428$\pm$3 & 12.9$\pm$0.8 
& 516$\pm$2 & 606$\pm$6 & 11.1$\pm$0.6 \\
L1709A & 8$\times$5 & $\leq$17 
& 47.1$\pm$6 & 68.0$\pm$4 & 12.8$\pm$0.4 
& 77.2$\pm$2 & 138$\pm$2  & 31.9$\pm$1 \\
L1689A & 4$\times$4 & 70.8$\pm$2 
& 564$\pm$7 & 849$\pm$6 & 470$\pm$4 
& 591$\pm$6 & 852$\pm$5 & 473$\pm$6 \\
L1709C & 8$\times$4 & $\leq$9.5$^*$
& 65.5$\pm$1 & 225$\pm$1 & 3.32$\pm$0.2 
& 66.0$\pm$2 & 306$\pm$1 & 1.59$\pm$0.4 \\
L1689B & 7$\times$4 & $\leq$7.4
& 108$\pm$2 & 194$\pm$1 & 47.8$\pm$1 
& 107$\pm$2 & 226$\pm$2 & 43.2$\pm$0.7 \\
L204B & 10$\times$5 & $\leq$9.0 
& 56.7$\pm$1 & 151$\pm$2 & 34.8$\pm$1 
& 140$\pm$2  & 272$\pm$3 & 126$\pm$4 \\
L63 & 6$\times$6 & $\leq$5.7 
& 79.4$\pm$1 & 115$\pm$1 & 60.0$\pm$0.4 
& 118$\pm$0.8 & 156$\pm$1 & 105$\pm$1.0 \\
B68 & 4$\times$4 & $\leq$4.5$^*$
& 25.8$\pm$0.3 & 65.8$\pm$0.3 & 16.8$\pm$0.3 
& 34.8$\pm$0.6 & 85.3$\pm$0.3 & 29.2$\pm$0.5  \\
B133 & 8$\times$3 & $\leq$15$^*$
& 72.1$\pm$0.9 & 121$\pm$0.4 & 25.6$\pm$0.3 
& 108$\pm$1    & 185$\pm$2   & 41.5$\pm$1 \\
L1155C & 4$\times$3 & $\leq$0.7 
& 15.0$\pm$0.2 & 107$\pm$0.4 & 4.66$\pm$0.2 
& 13.2$\pm$2   & 143$\pm$0.4 & 4.60$\pm$0.4 \\ \hline
\end{tabular}
\end{table*}

We followed the standard pipeline data reduction procedure, starting with 
the edited raw data (ERD) files. The responsivity was calculated using the 
actual fine calibration source (FCS) responsivity calibration by
averaging between adjacent FCS measurements and interpolating. Destructive
readouts were discarded in calibrating the data from each individual 
integration ramp. Saturated readouts and glitches were discarded likewise.
Ramps were fitted using a first-order polynomial. Dark currents were 
subtracted. Drift handling was enabled and chopper vignetting was accounted
for. Detector non-linearities were also taken into account. For full
details of the data reduction procedure, see Laureijs et al. (2001).

Subsequent upgrades to the
calibration routines have not significantly altered the flux estimates of
extended sources such as ours. This was checked by re-reducing one source
at random using the final version 10 (V10) of the data reduction
pipeline. This is the version with which the final ISO data archive will
be calibrated in its entirety. The empirical point spread function
had been updated between V7 and V10, but this only affected the calibration
of point sources for the C-200 camera, so did not affect our data. The only
other significant change was the non-linearity response function for
very bright sources, which also did not affect our data.

The final absolute calibration errors were
estimated to be $\pm$30 per cent. This is similar to that found by others
working with ISOPHOT (e.g. Klaas et al 1997; Lehtinen et al.
1998). We cross-checked our calibration accuracy by
comparison with the Cosmic Background Explorer satellite (COBE) and this
confirmed our error estimate (see below). 

\section{Results}

\subsection{Images and flux densities}

Figures 1--8 present images of our sample of pre-stellar cores at each of
the three wavelengths of 90, 170 and 200 $\mu$m. 
The grey-scale range from white (faint) to black (bright)
for any given source is set to be the same at all three 
wavelengths using the full range of the 
200-$\mu$m data (see Tables 1 \& 2). In this way the absolute brightness
variation from one wavelength to another can be seen clearly. However,
the contour levels are based on the dynamic range at each wavelength
so that the structure can be discerned in the fainter images.

\begin{figure*}
\setlength{\unitlength}{1mm}
\begin{picture}(230,230)
\includegraphics{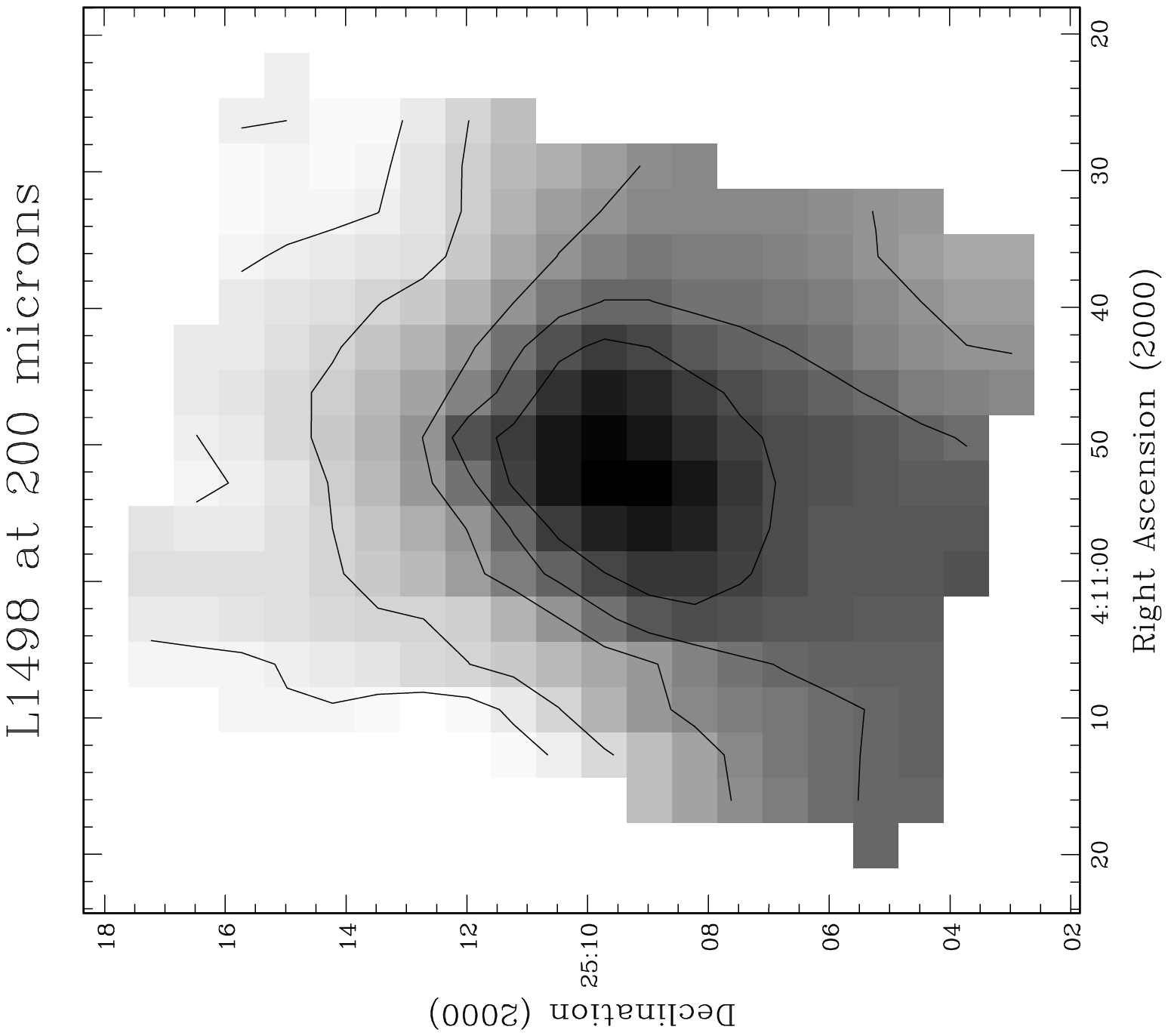}
\includegraphics{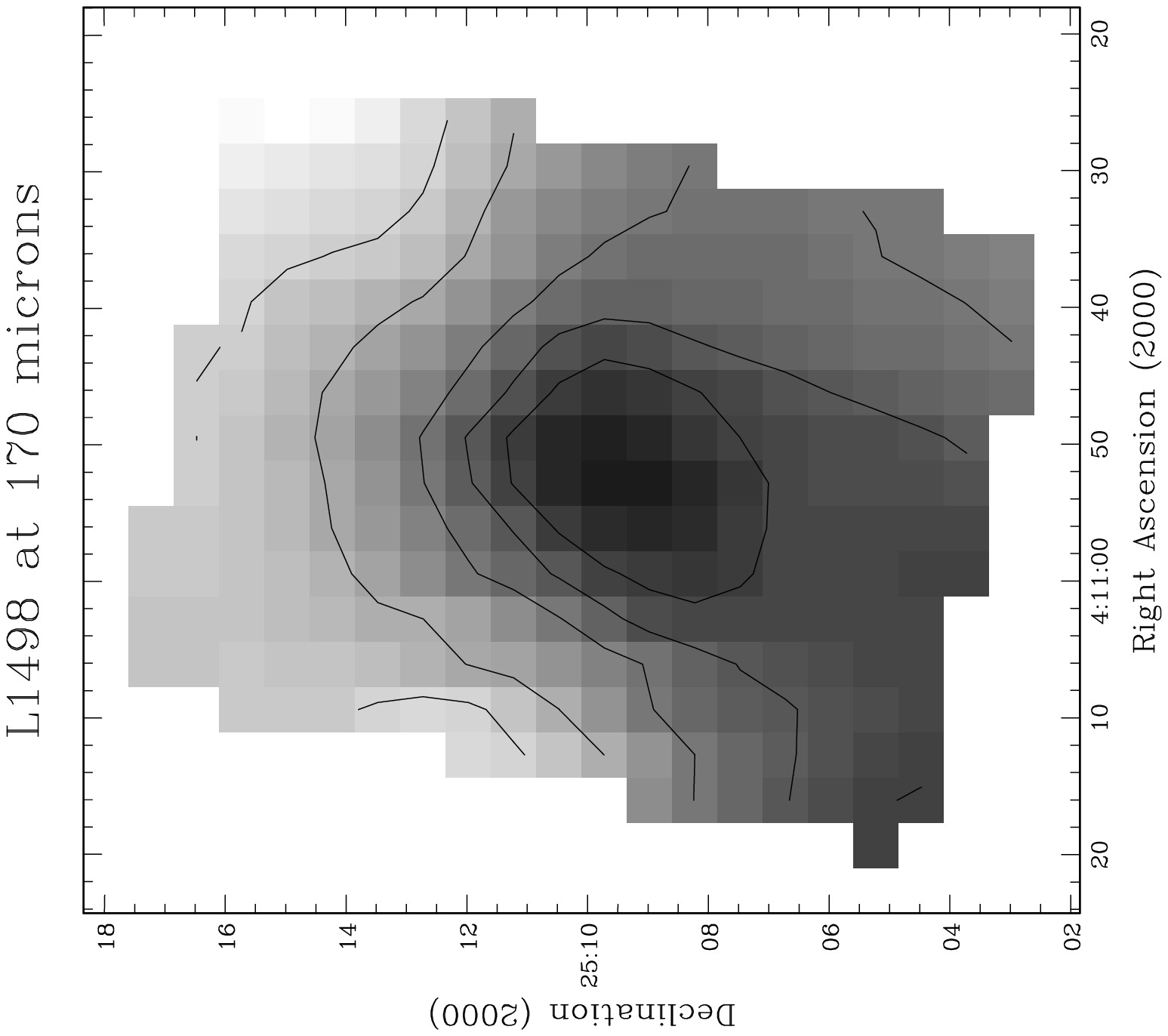}
\includegraphics{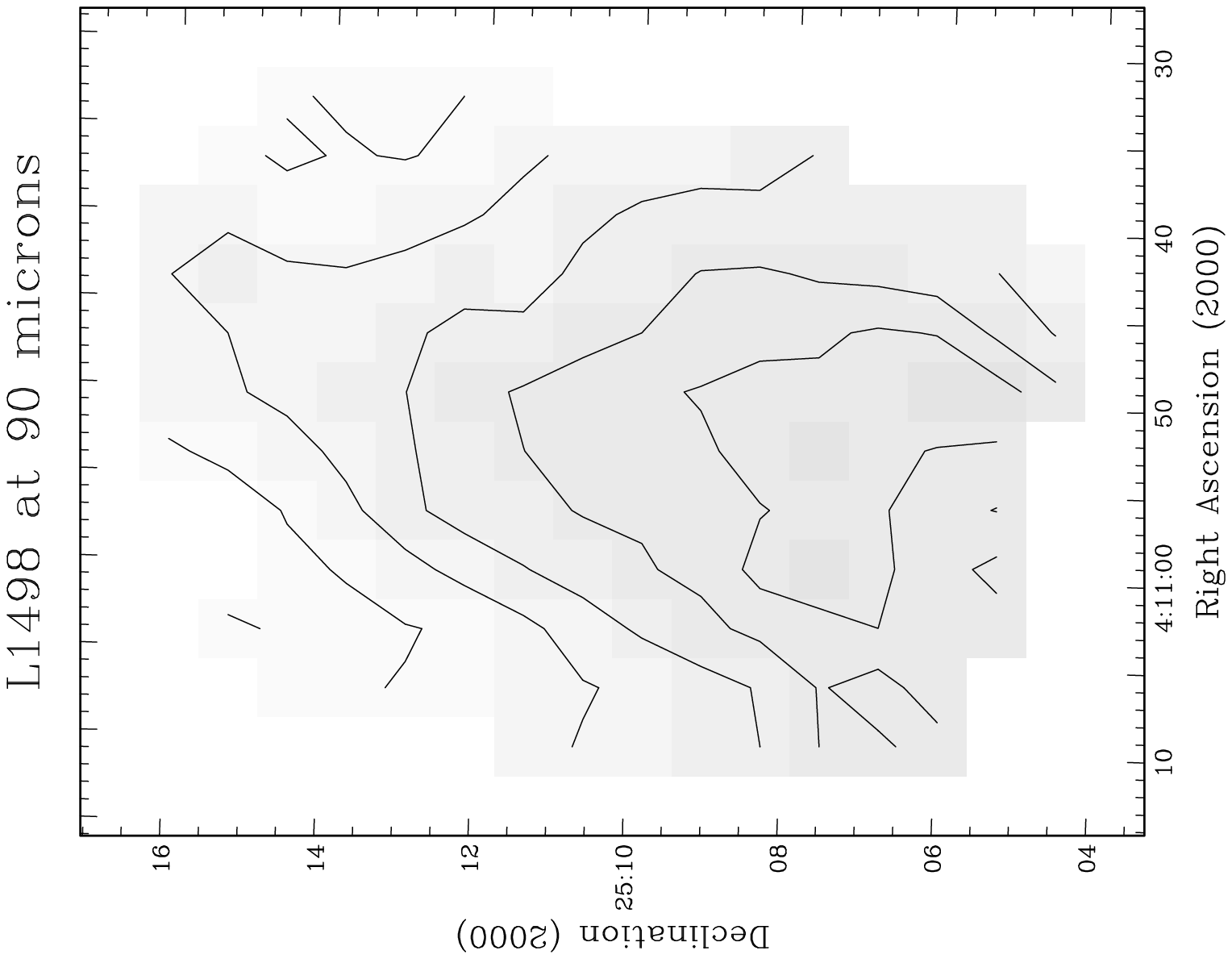}
\includegraphics{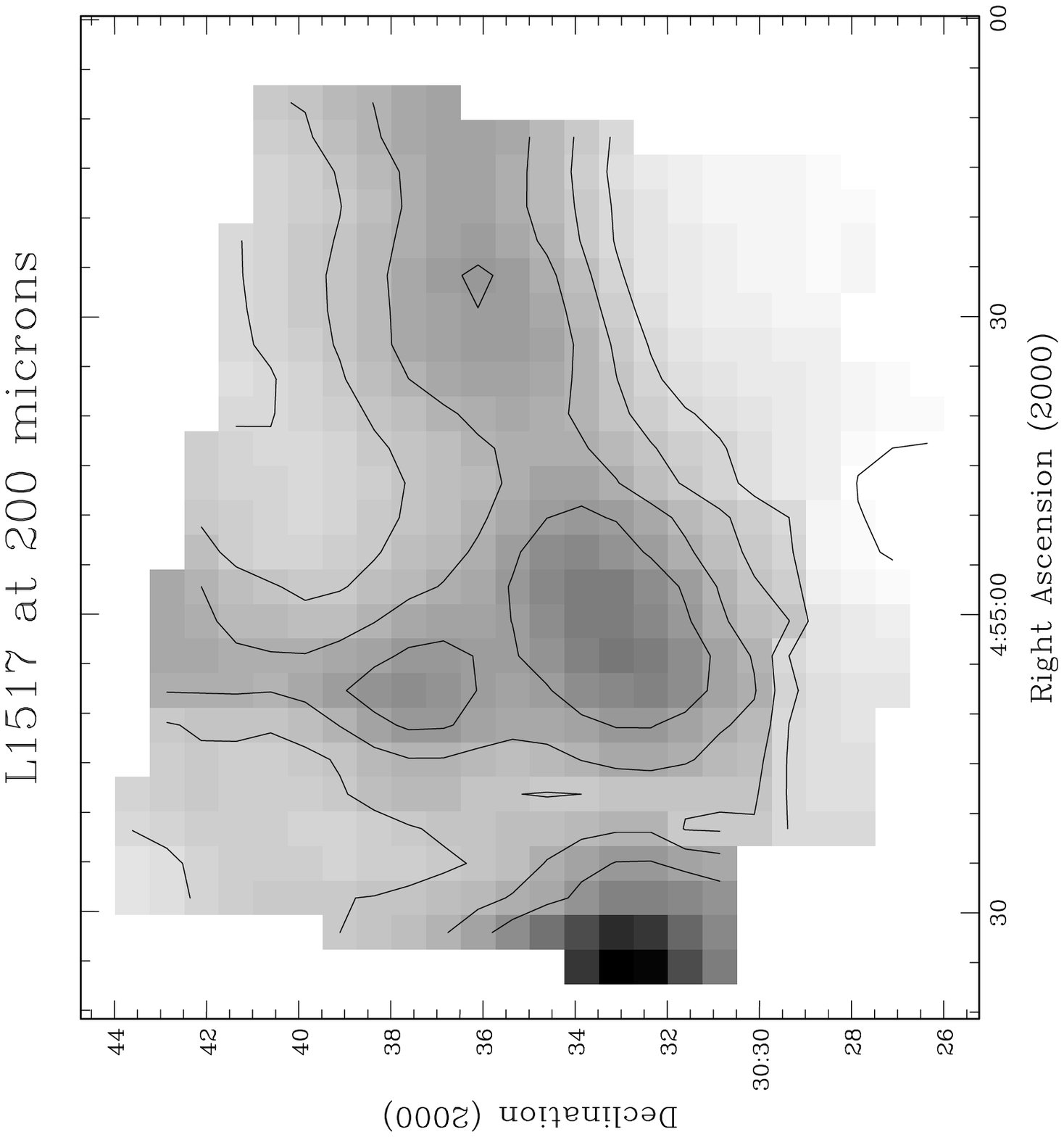}
\includegraphics{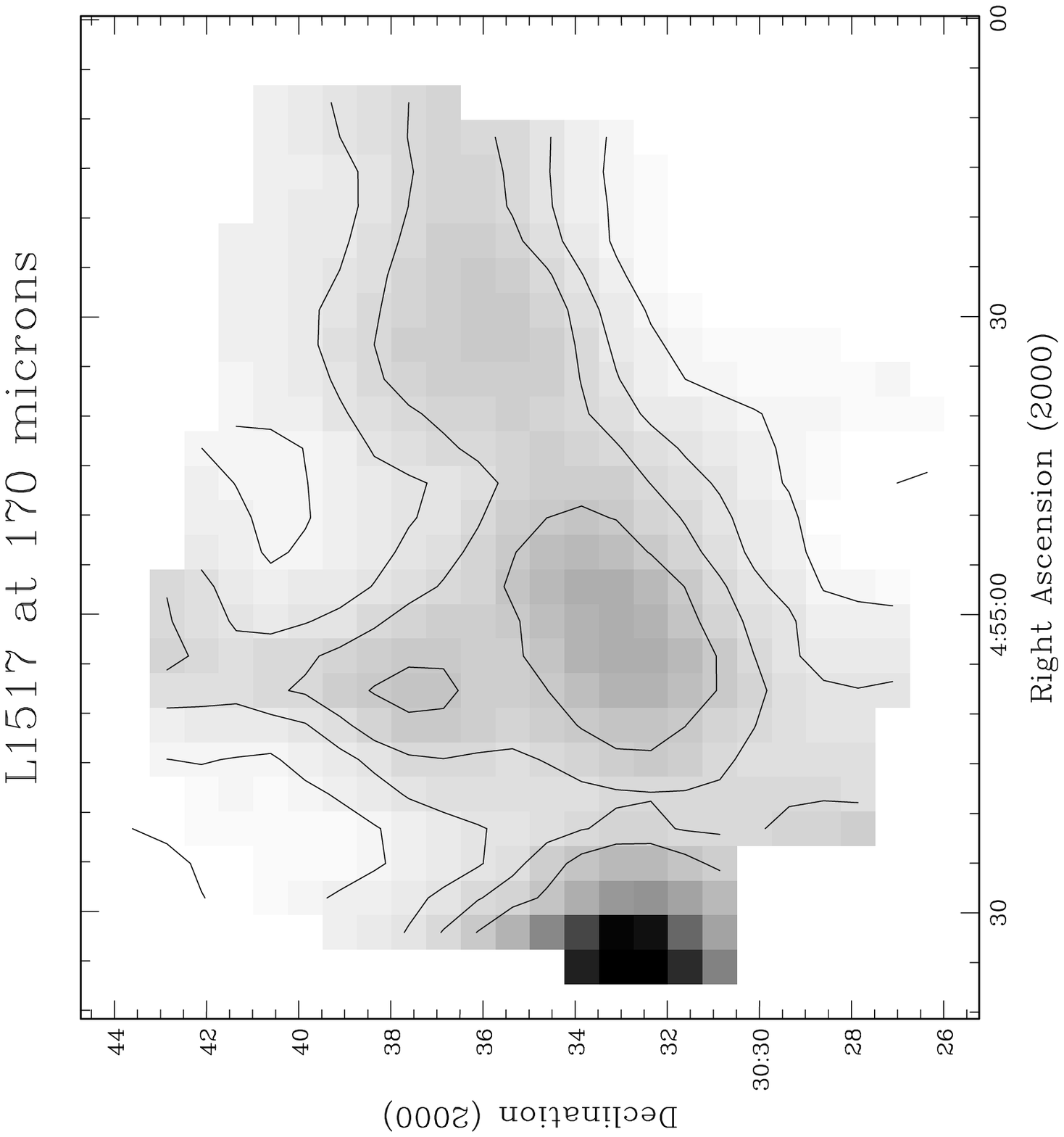}
\includegraphics{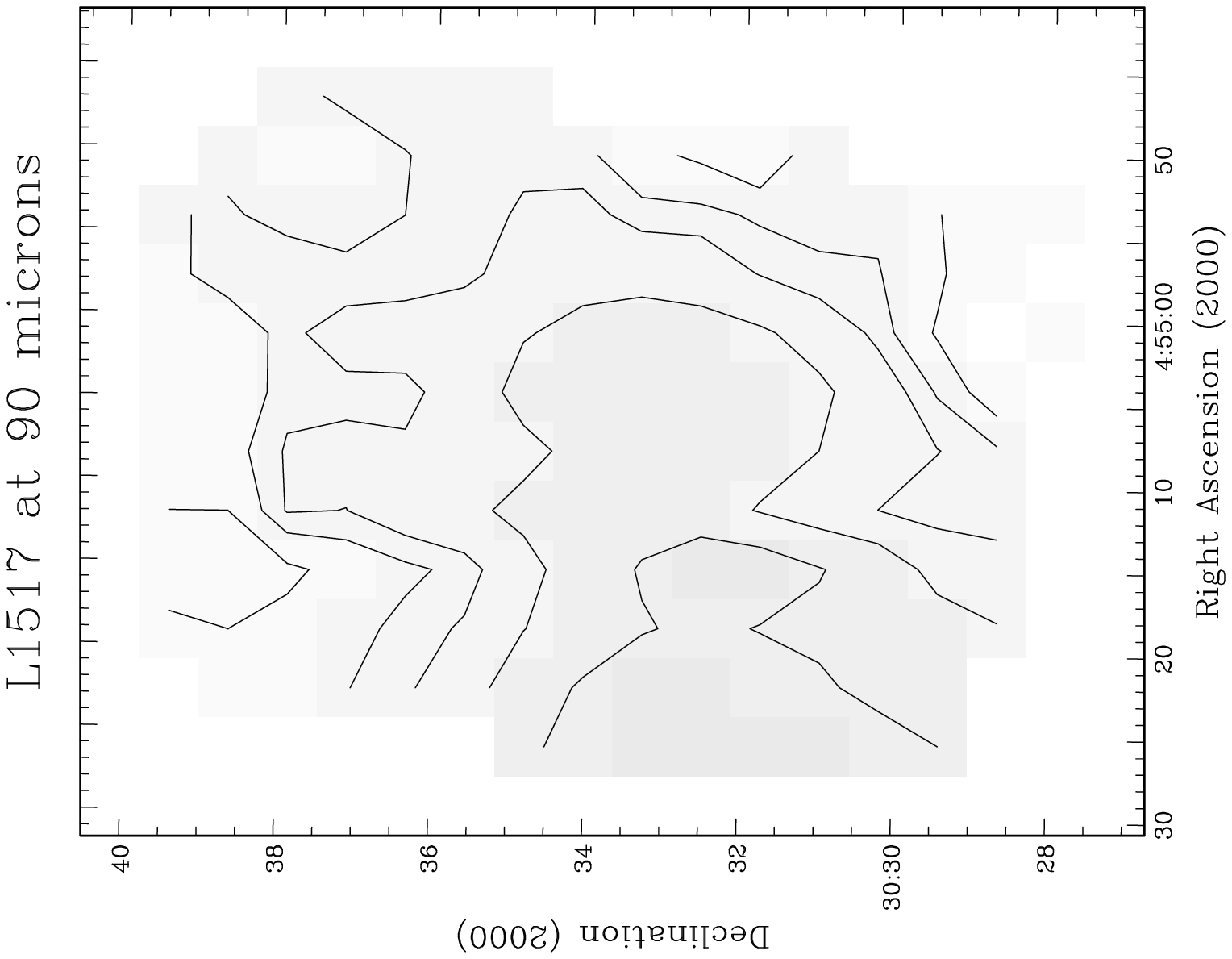}
\end{picture}
\caption{Grey-scale flux density images, with isophotal
contour maps superposed, of L1498 (left) and L1517 (right),
at 200 (top), 170 (middle) and 90 $\mu$m (bottom).
In these images, and in 
Figures 2--8, the grey-scale range from faint
(white) to bright (black) for each source is set
to be the same at all wavelengths using the full range of the 
200-$\mu$m data (see Tables 1 \& 2). The contour levels 
are 10, 30, 50, 70 \& 90\% of the dynamic range at each wavelength.}
\end{figure*}

\begin{figure*}
\setlength{\unitlength}{1mm}
\begin{picture}(230,230)
\includegraphics{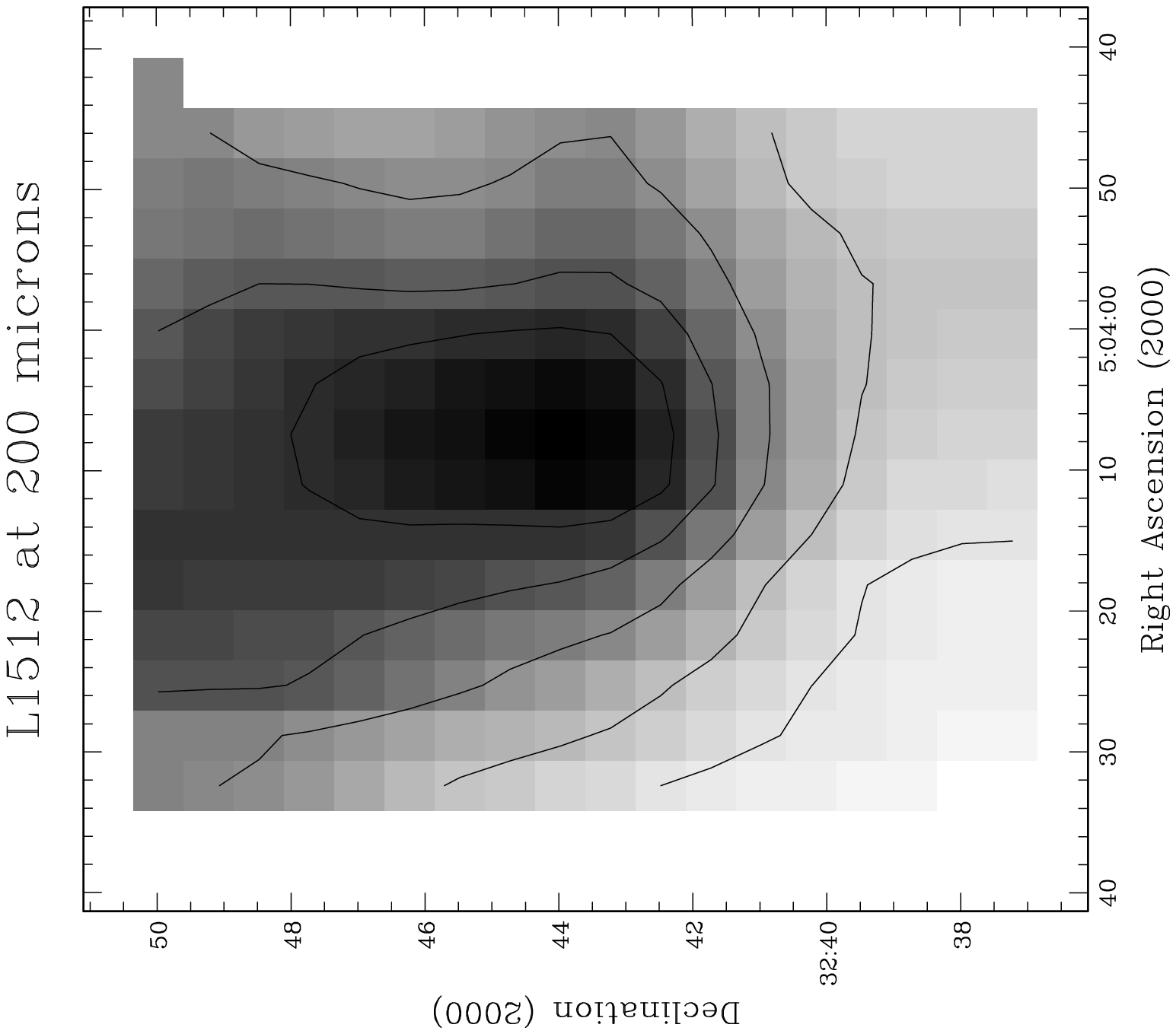}
\includegraphics{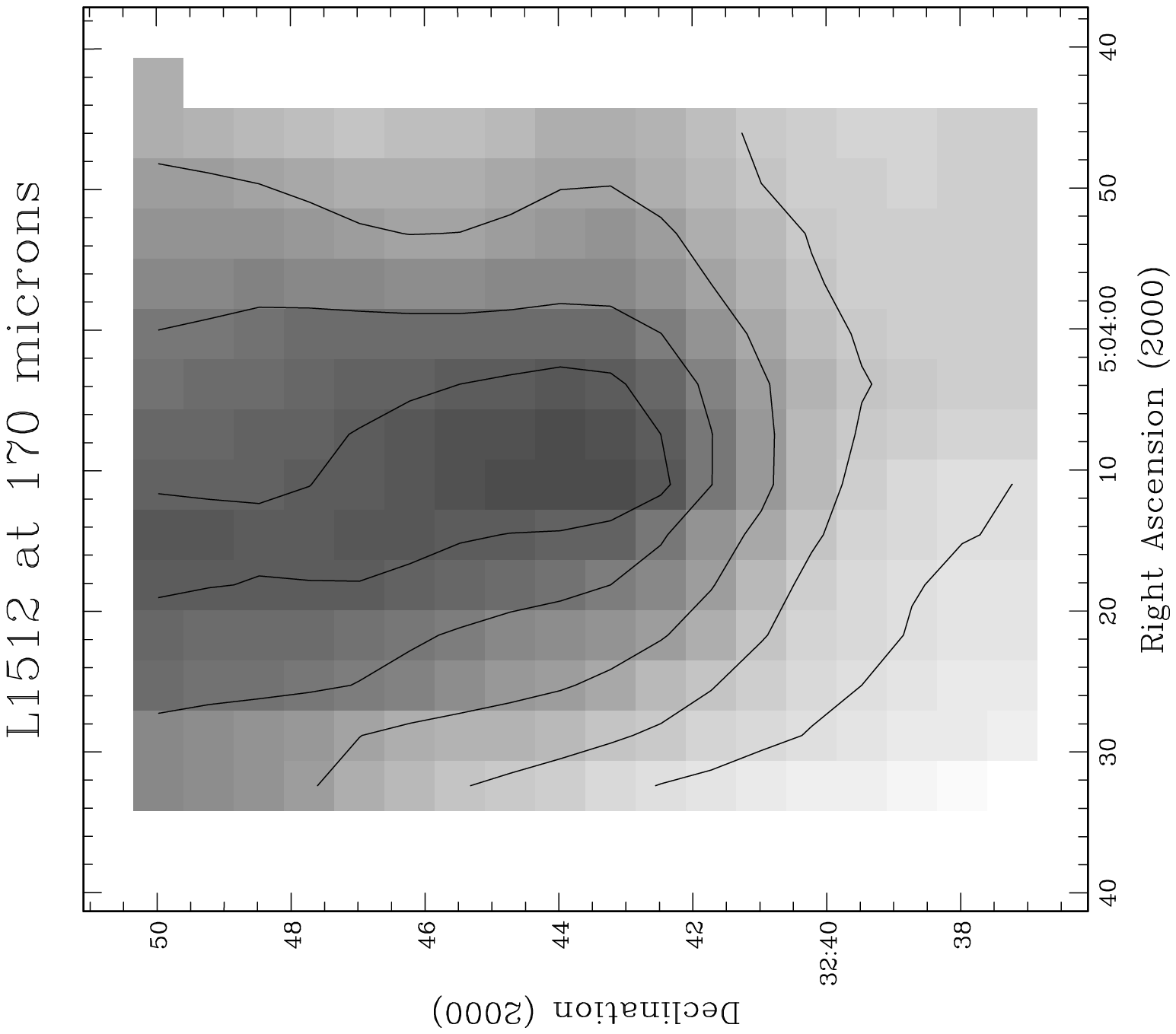}
\includegraphics{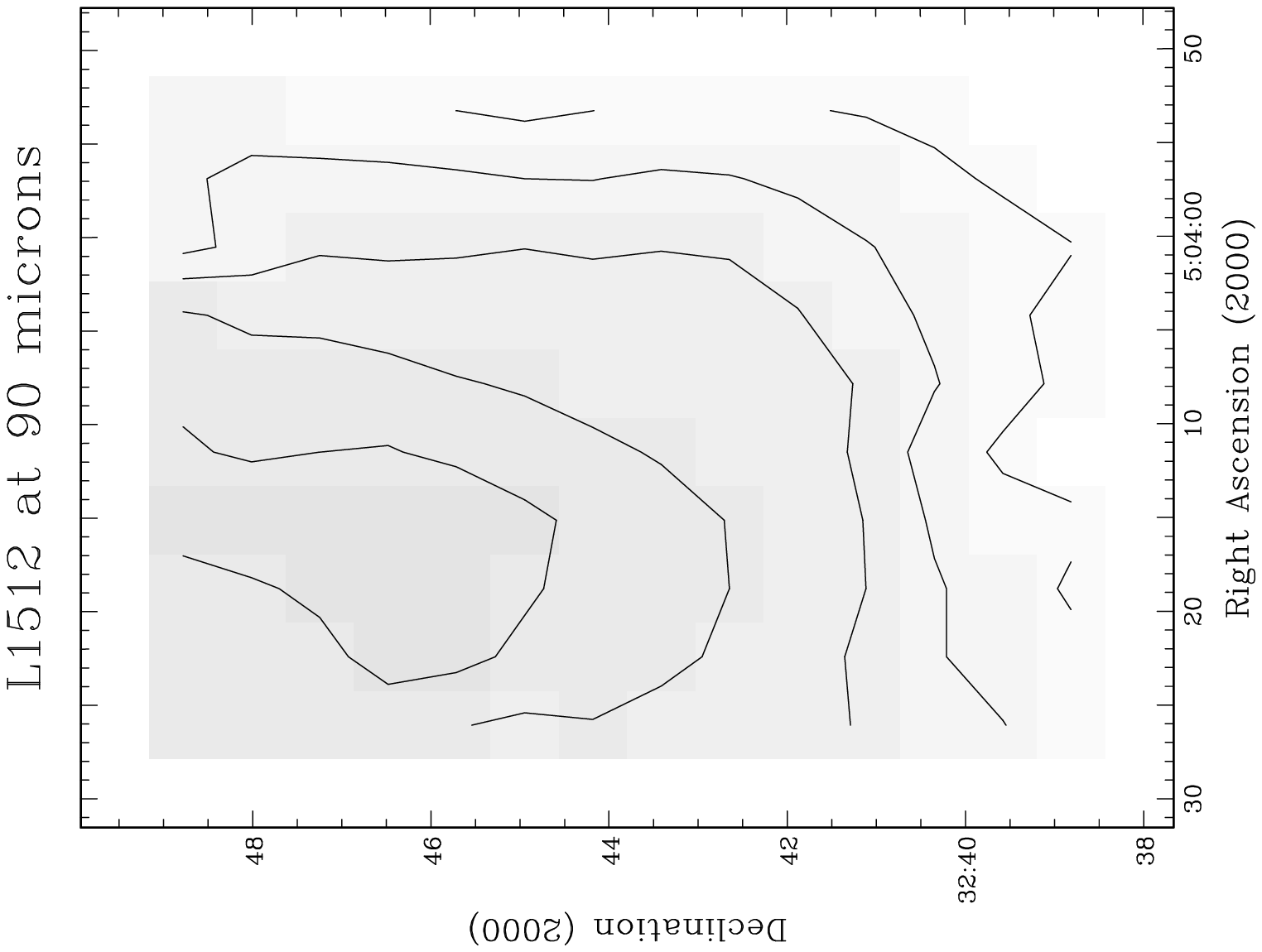}
\includegraphics{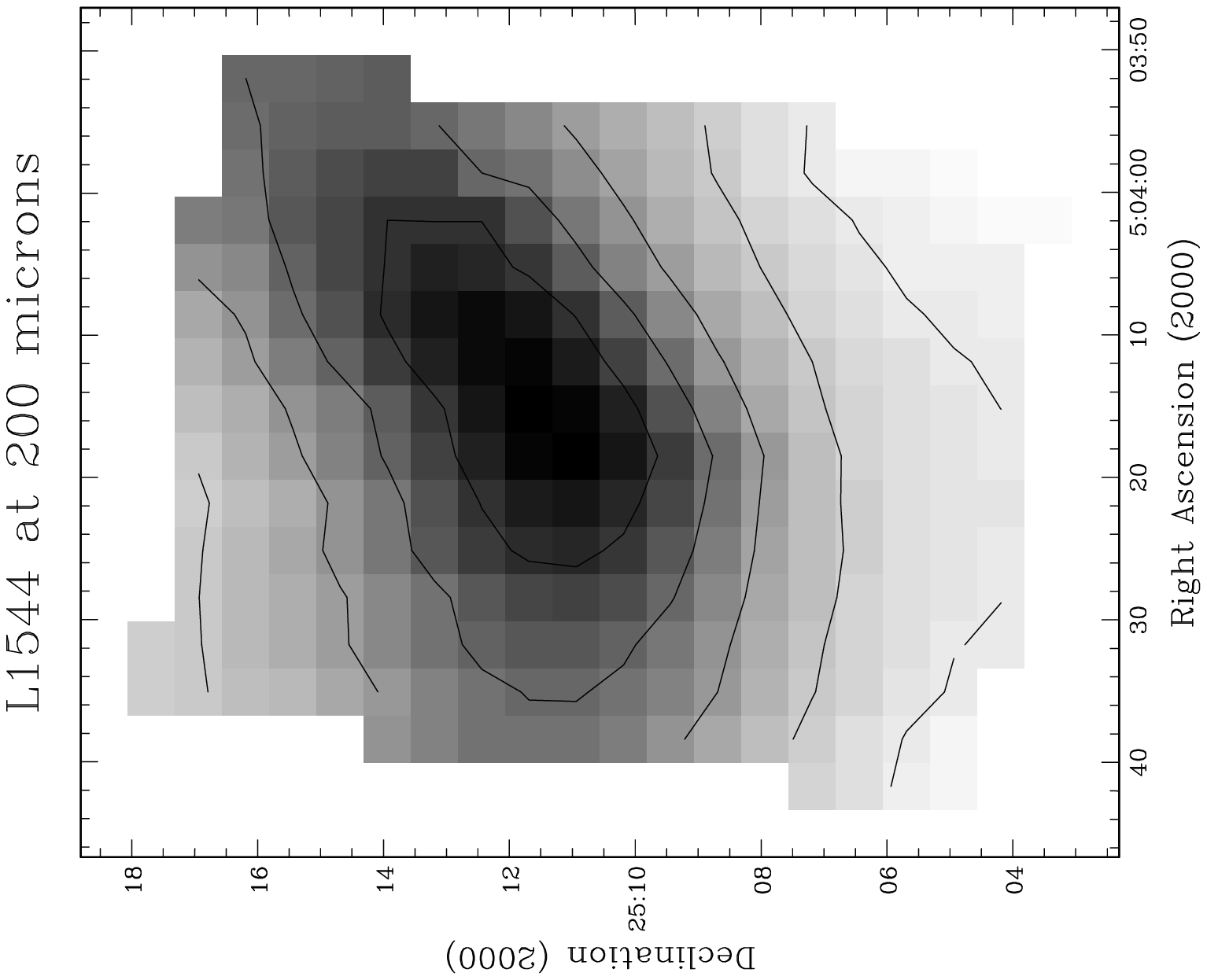}
\includegraphics{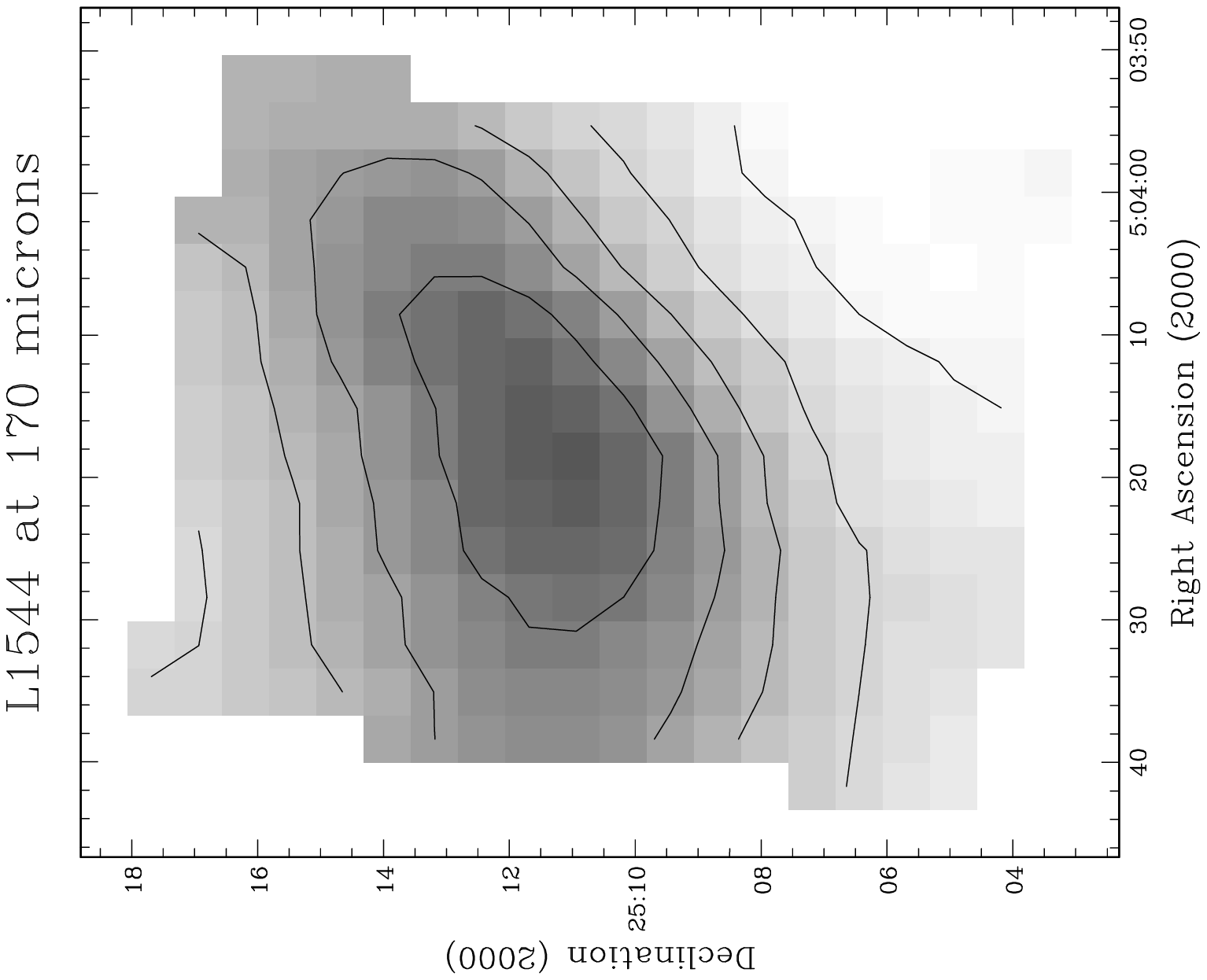}
\includegraphics{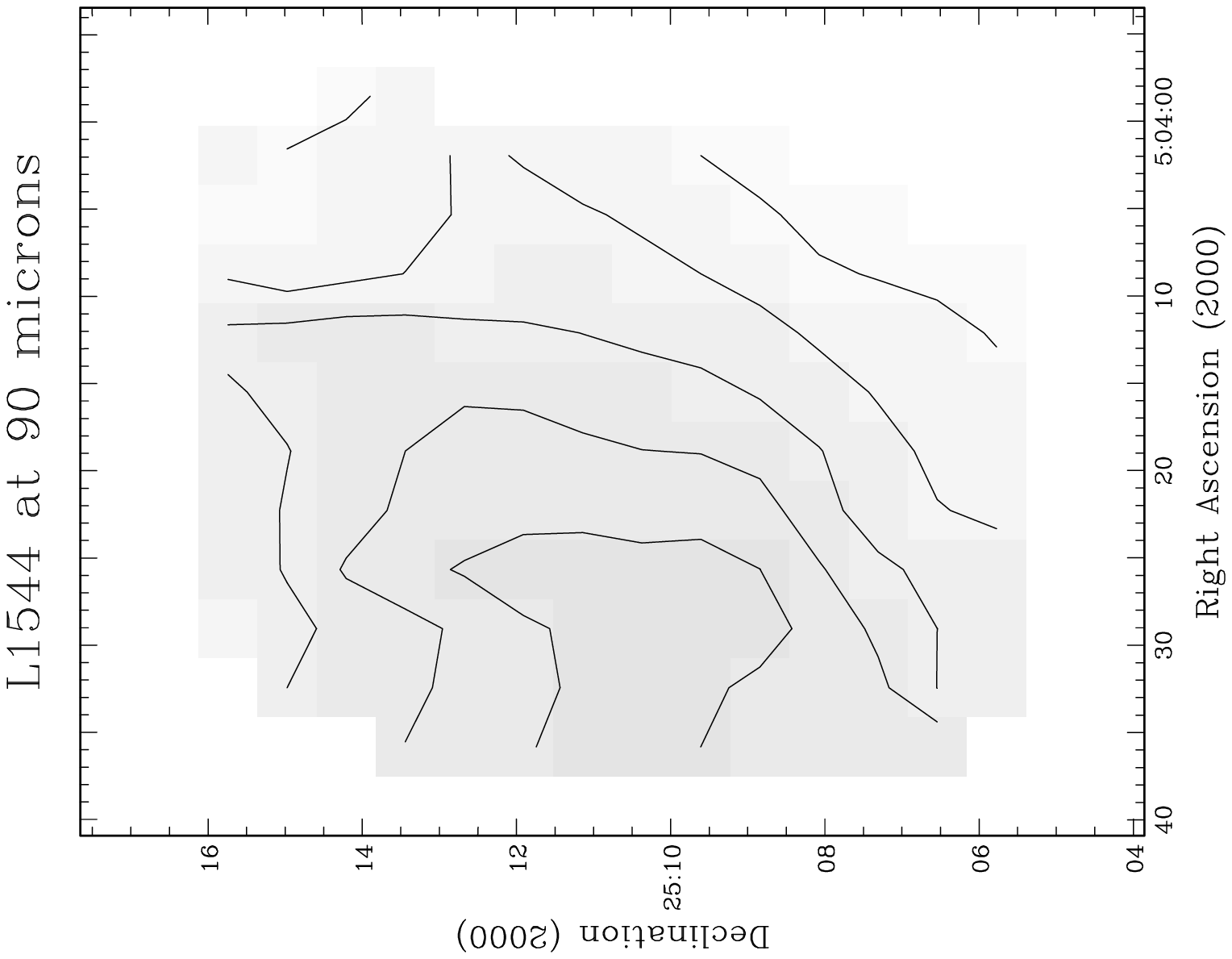}
\end{picture}
\caption{Images of L1512 (left) and L1544 (right). 
Details as in Figure 1.}
\end{figure*}

\begin{figure*}
\setlength{\unitlength}{1mm}
\begin{picture}(230,230)
\includegraphics{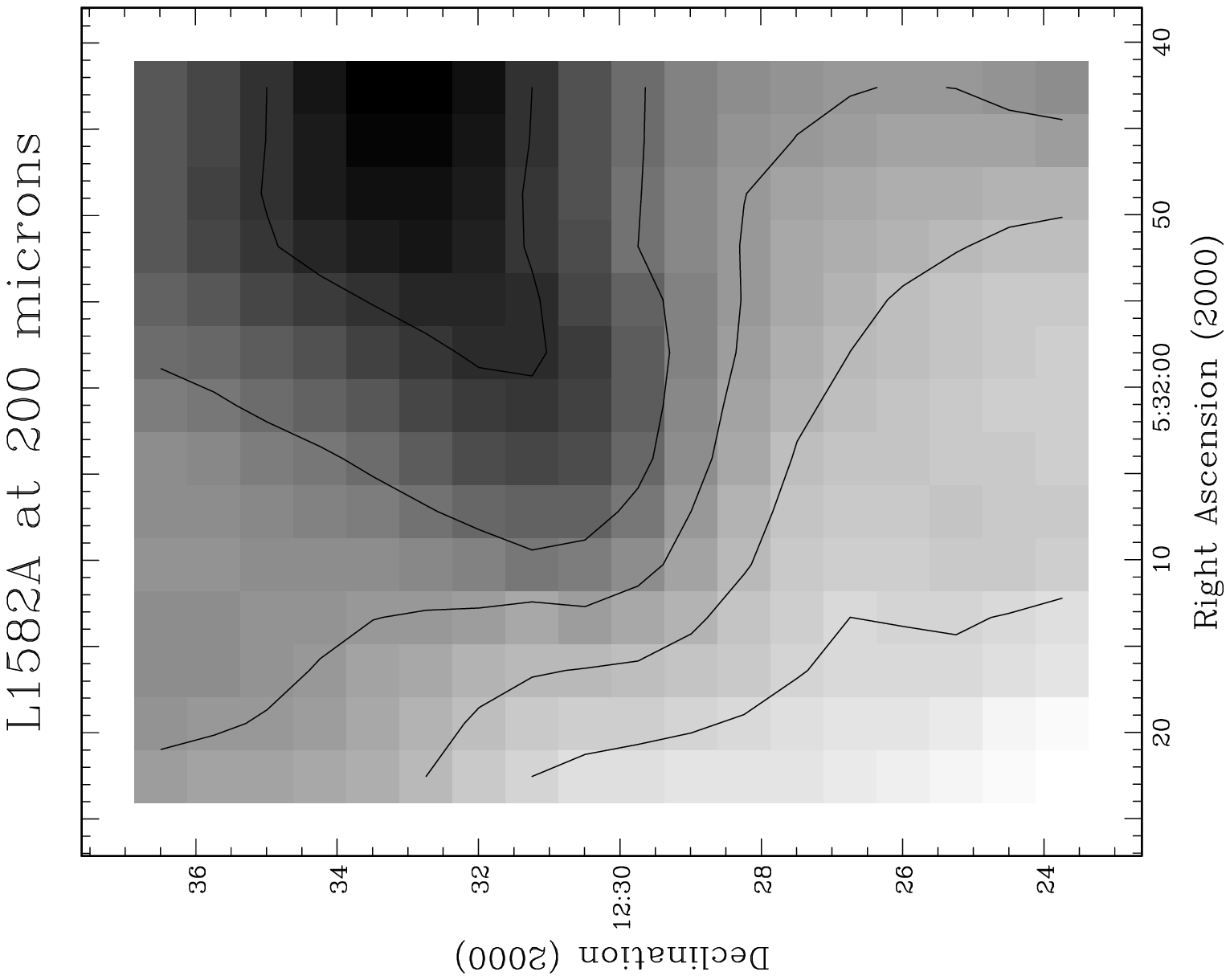}
\includegraphics{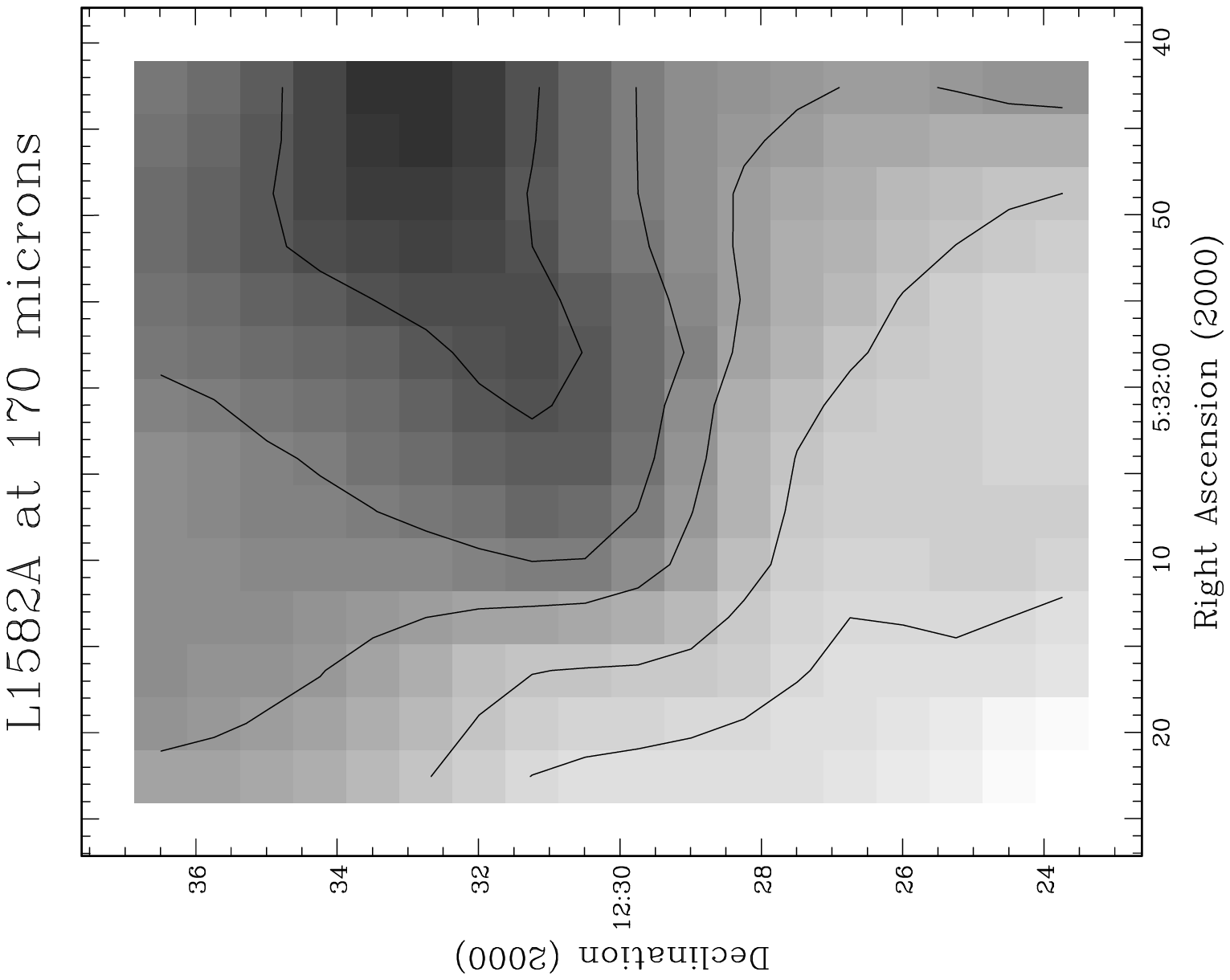}
\includegraphics{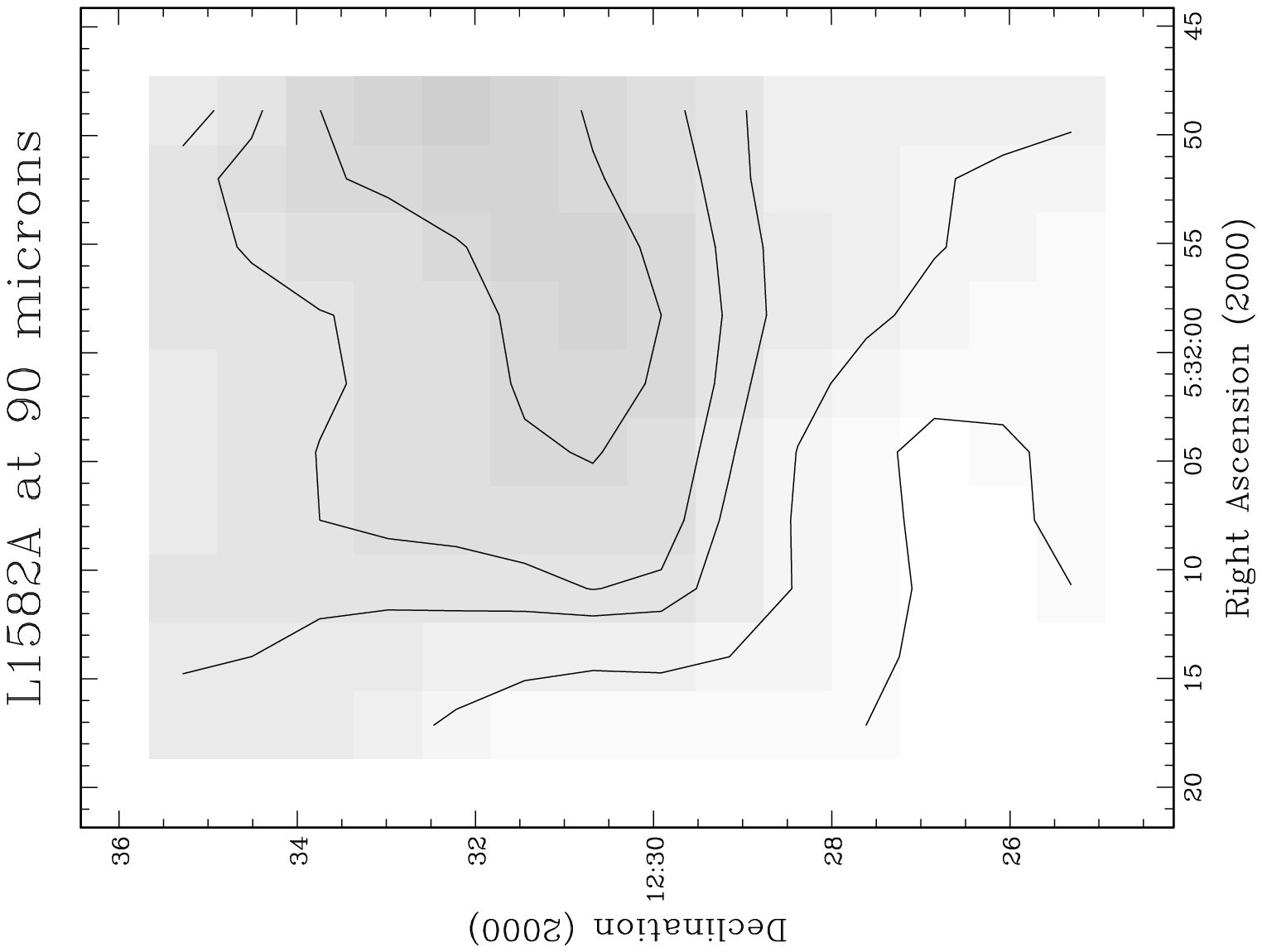}
\includegraphics{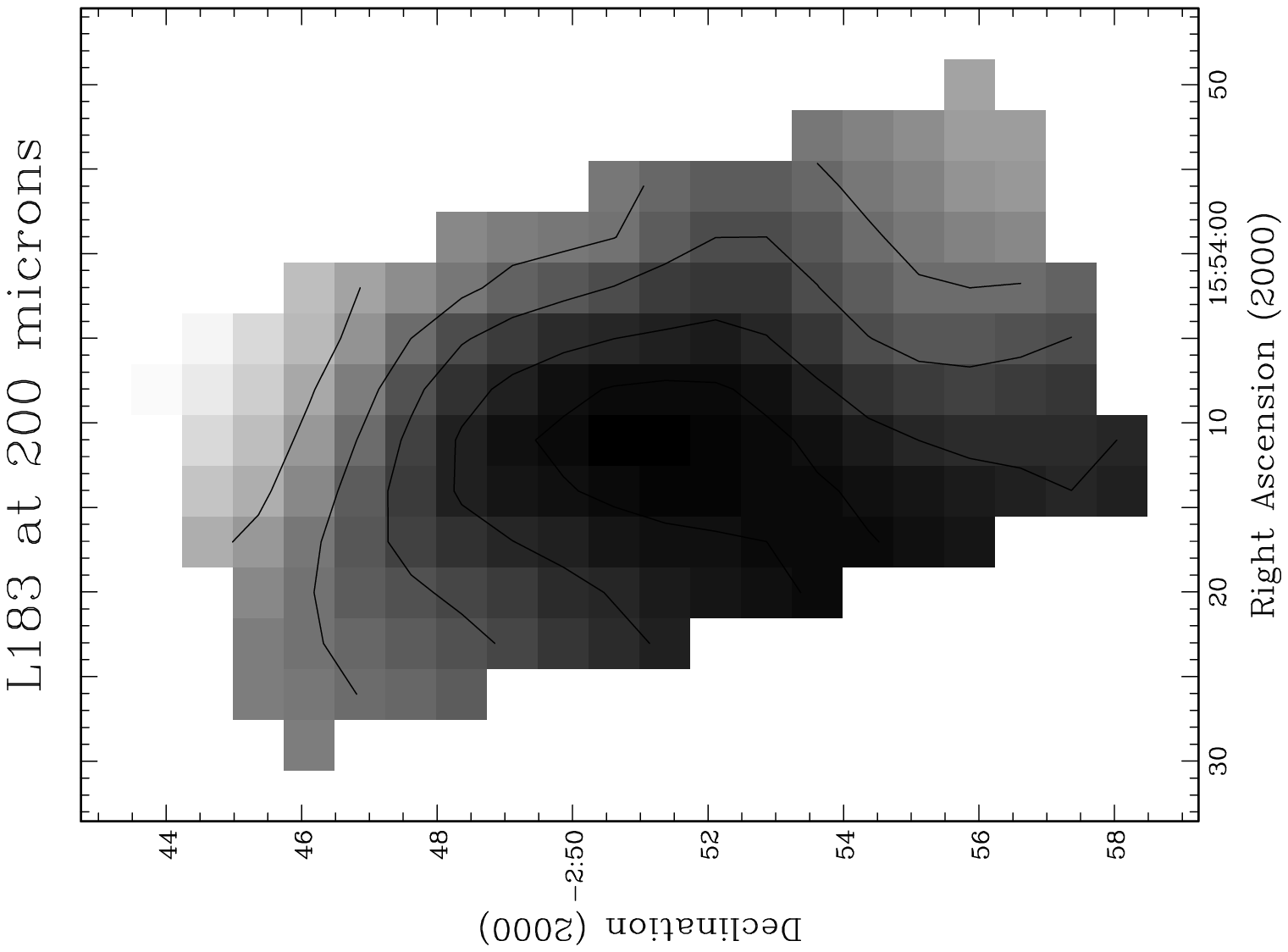}
\includegraphics{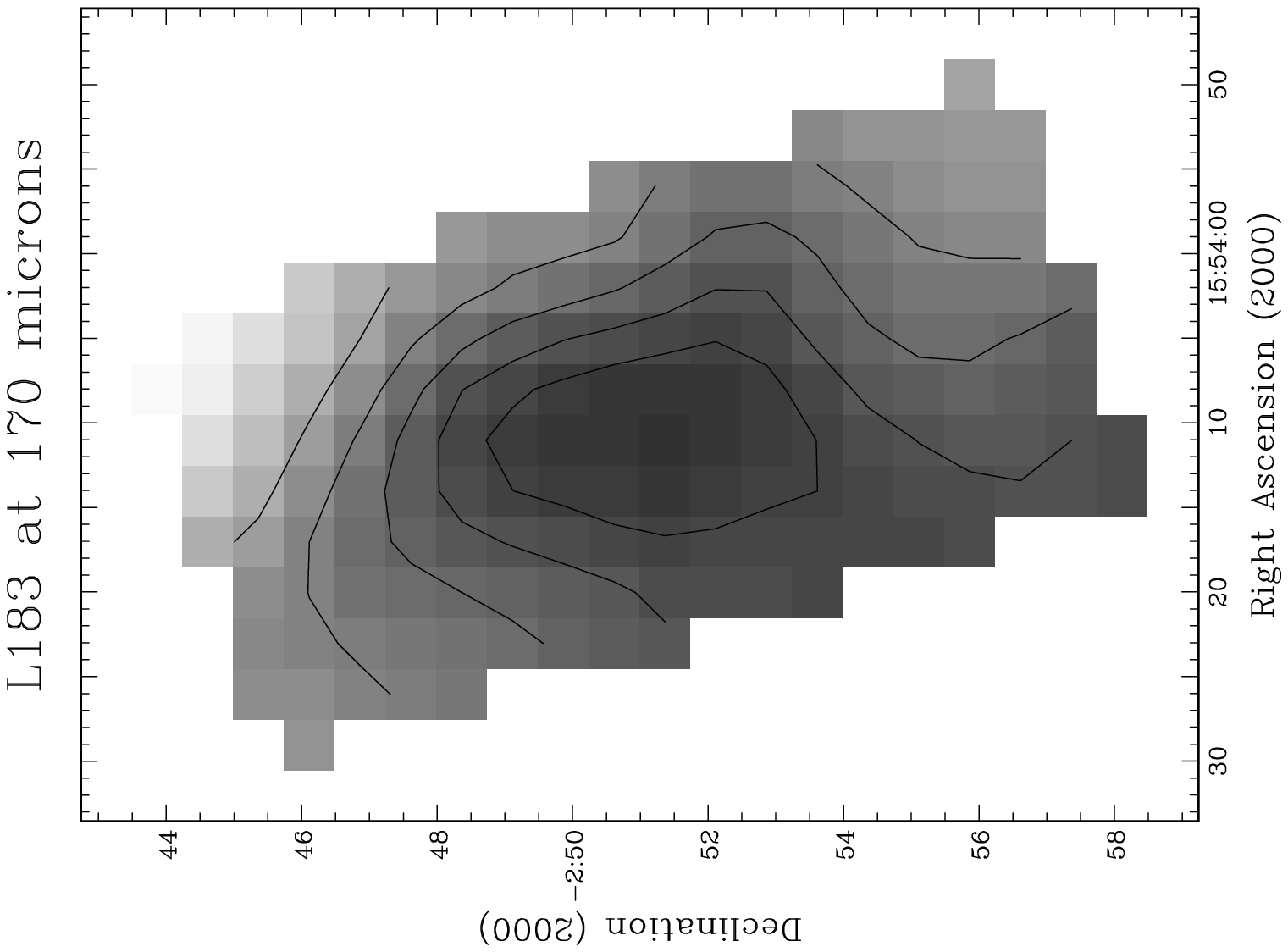}
\includegraphics{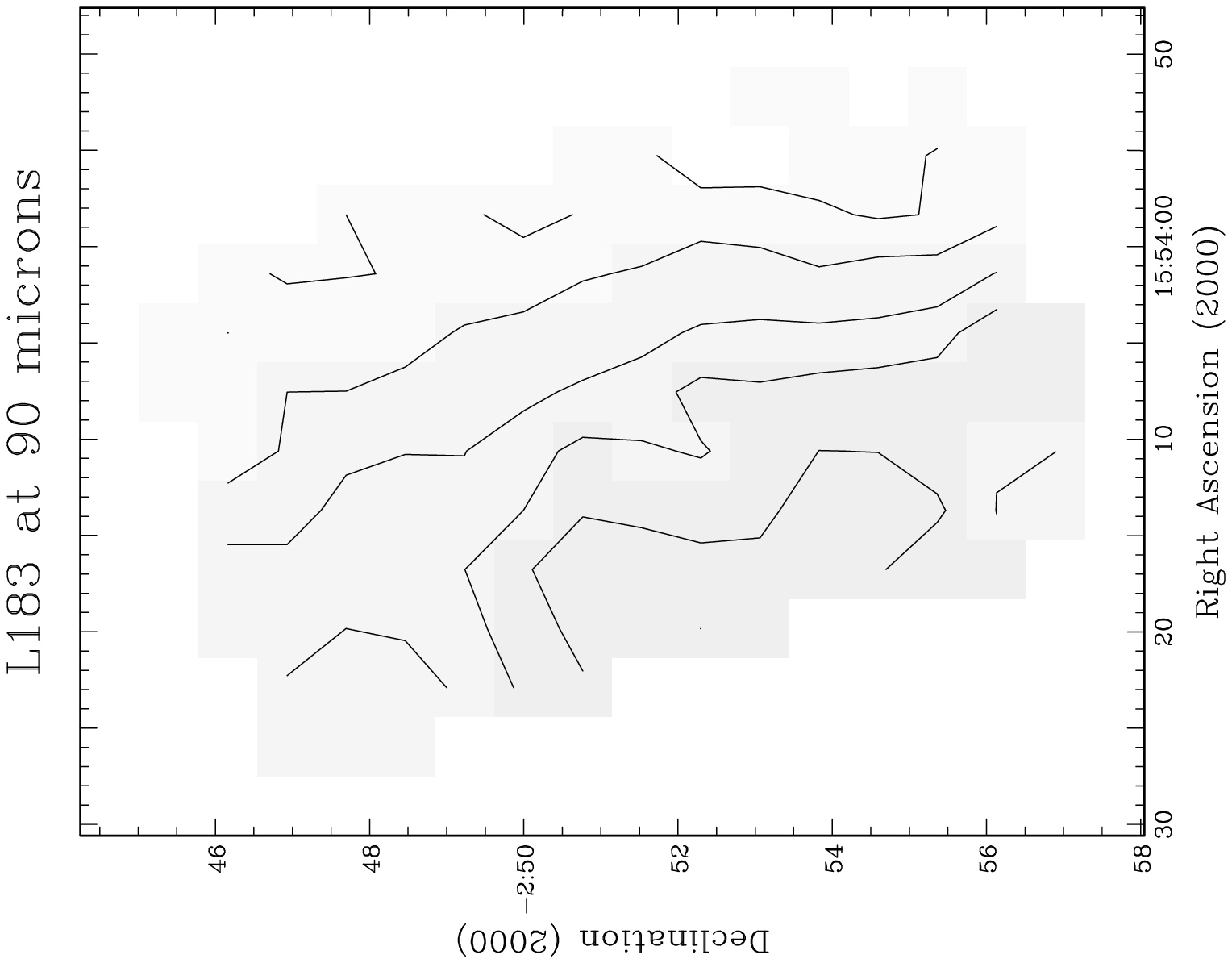}
\end{picture}
\caption{Images of L1582A (left) and L183 (right). Details as in Figure 1.}
\end{figure*}

\begin{figure*}
\setlength{\unitlength}{1mm}
\begin{picture}(230,230)
\includegraphics{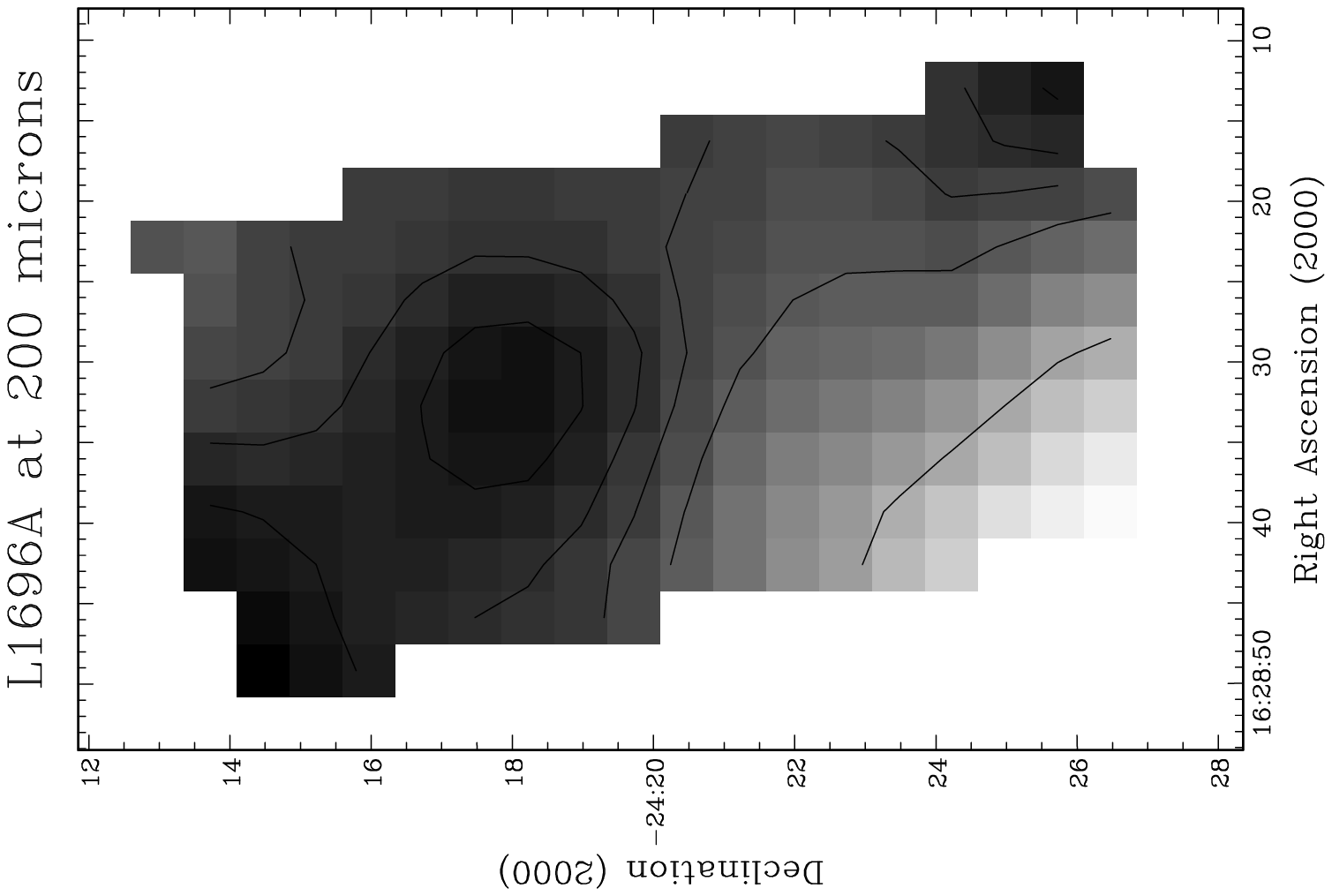}
\includegraphics{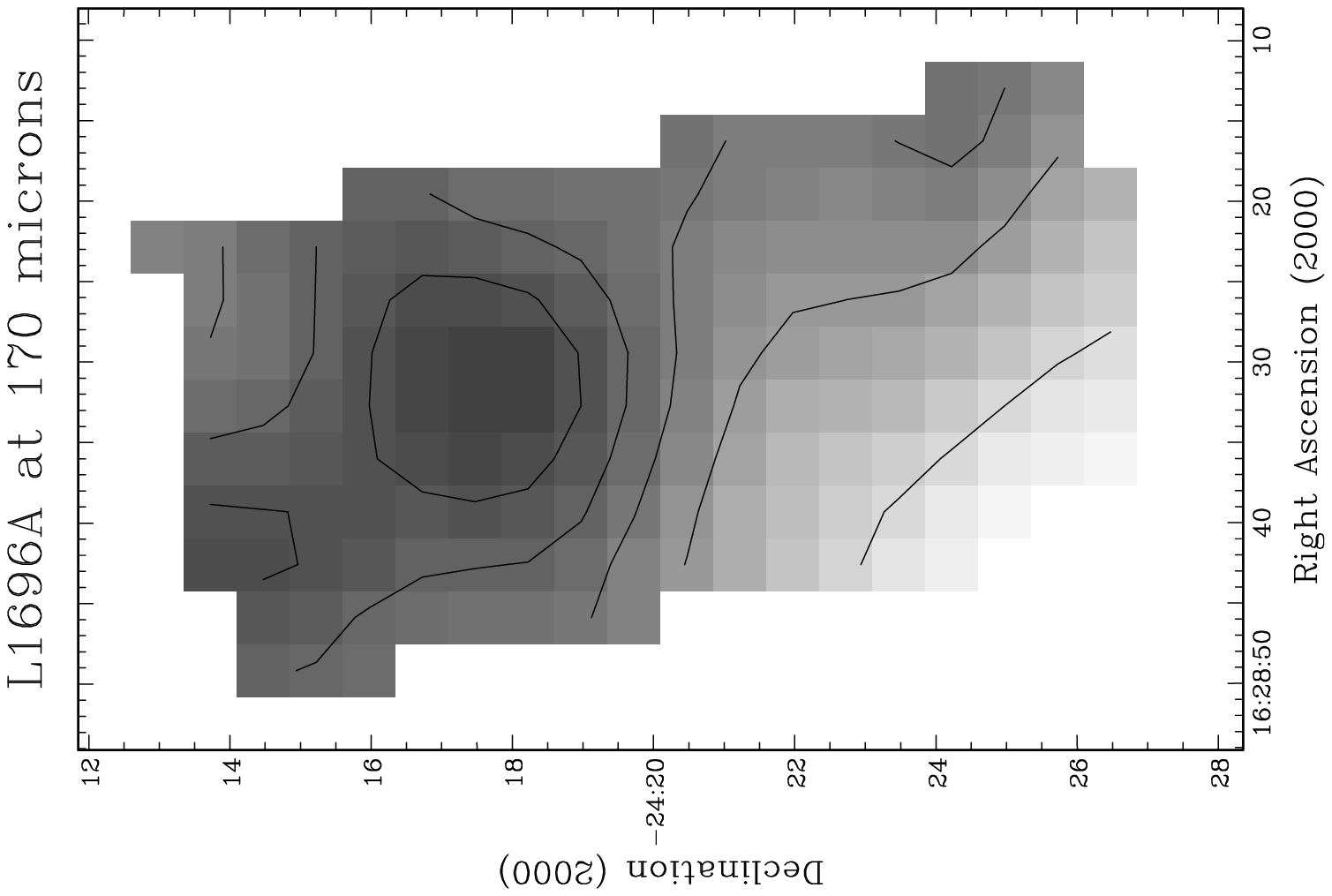}
\includegraphics{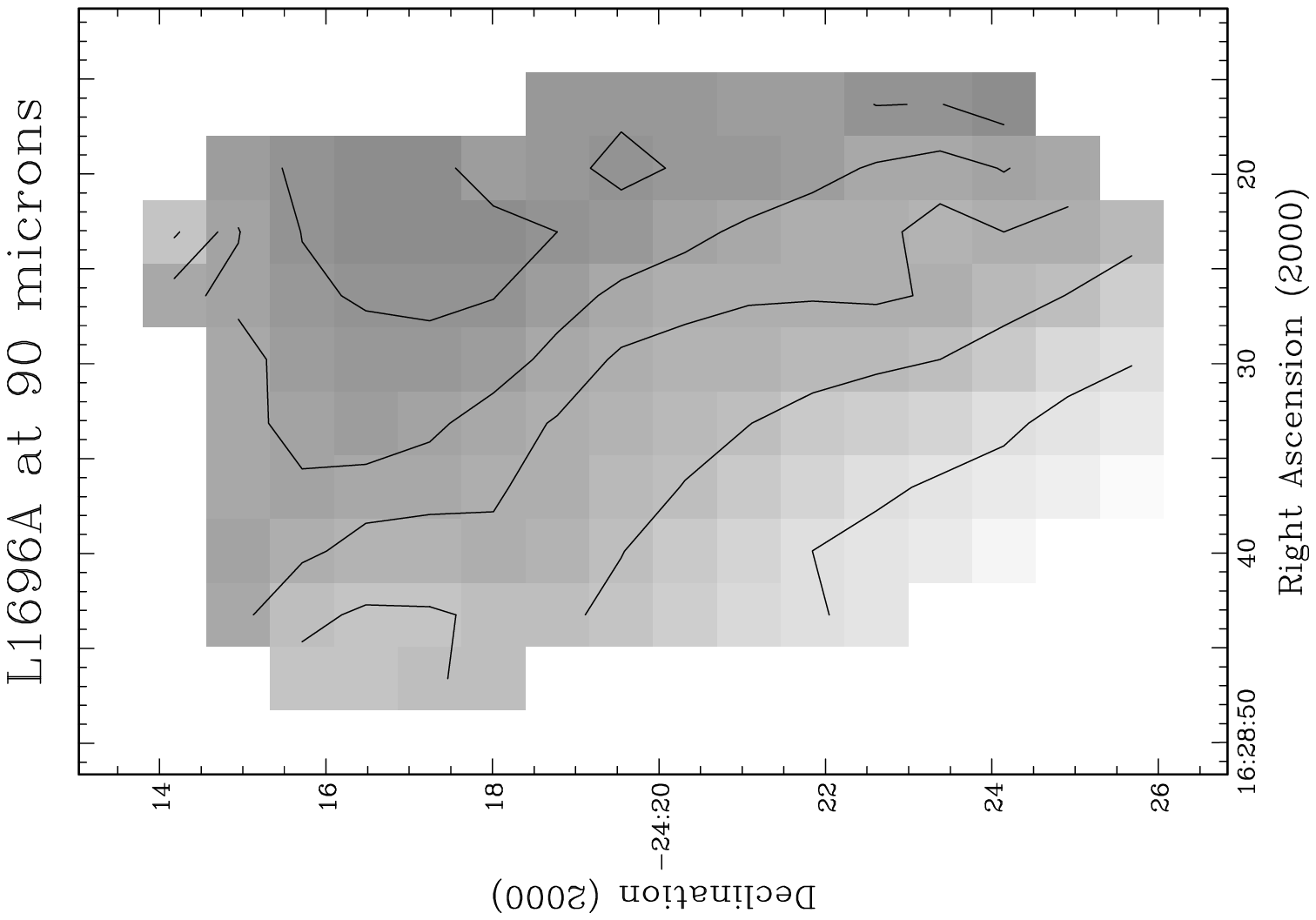}
\includegraphics{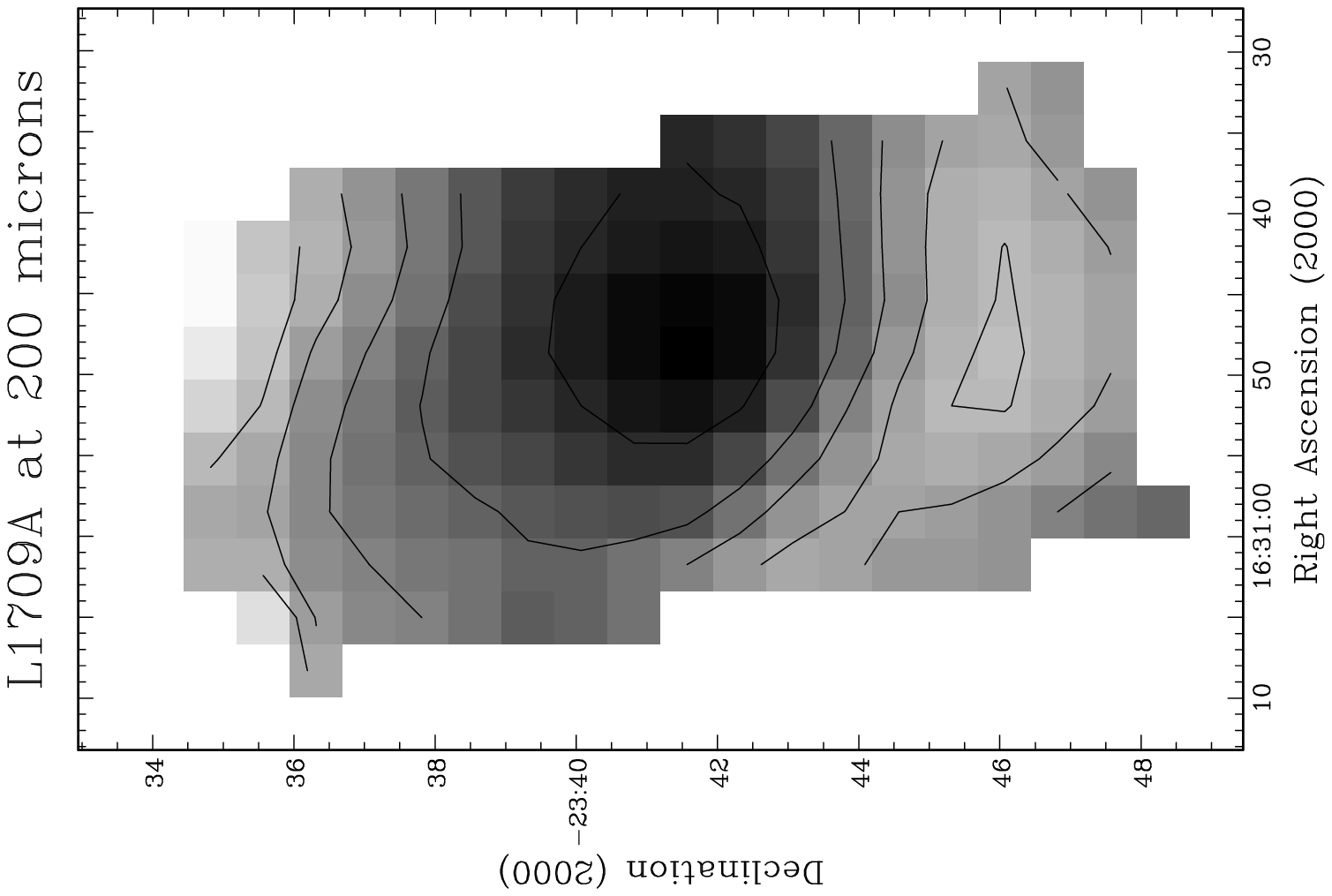}
\includegraphics{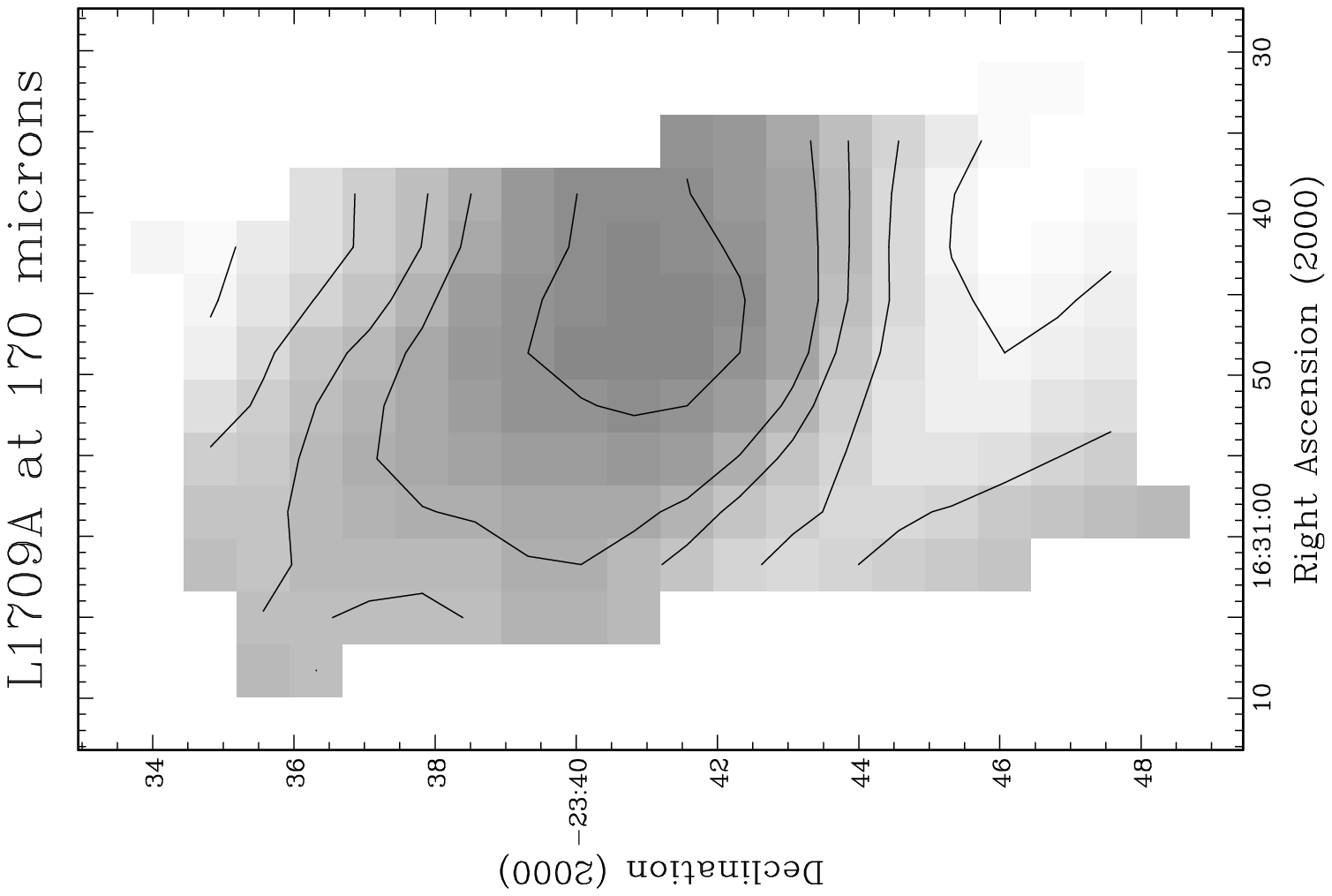}
\includegraphics{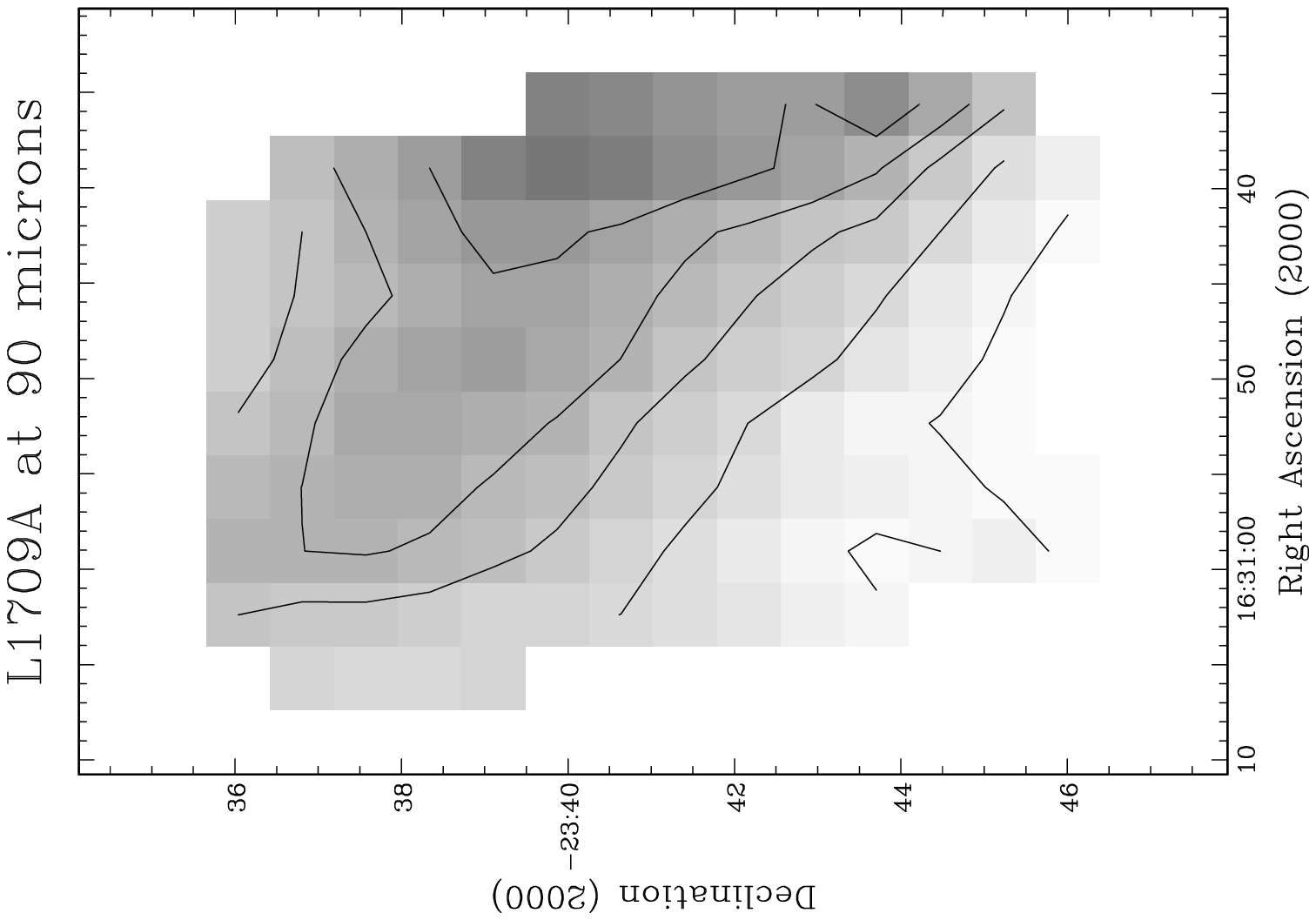}
\end{picture}
\caption{Images of L1696A (left) and L1709A (right). Details as in Figure 1.}
\end{figure*}

\begin{figure*}
\setlength{\unitlength}{1mm}
\begin{picture}(230,230)
\includegraphics{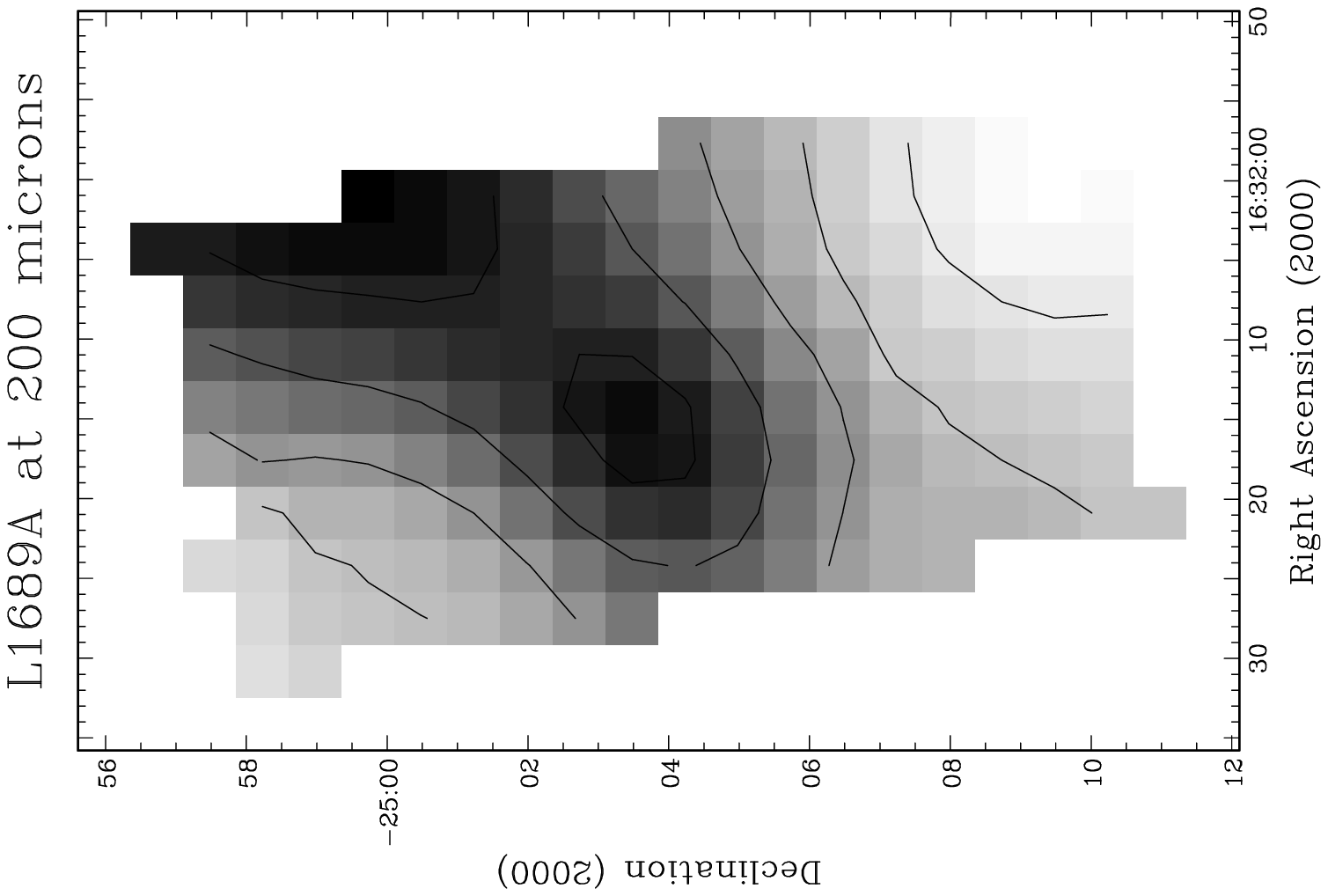}
\includegraphics{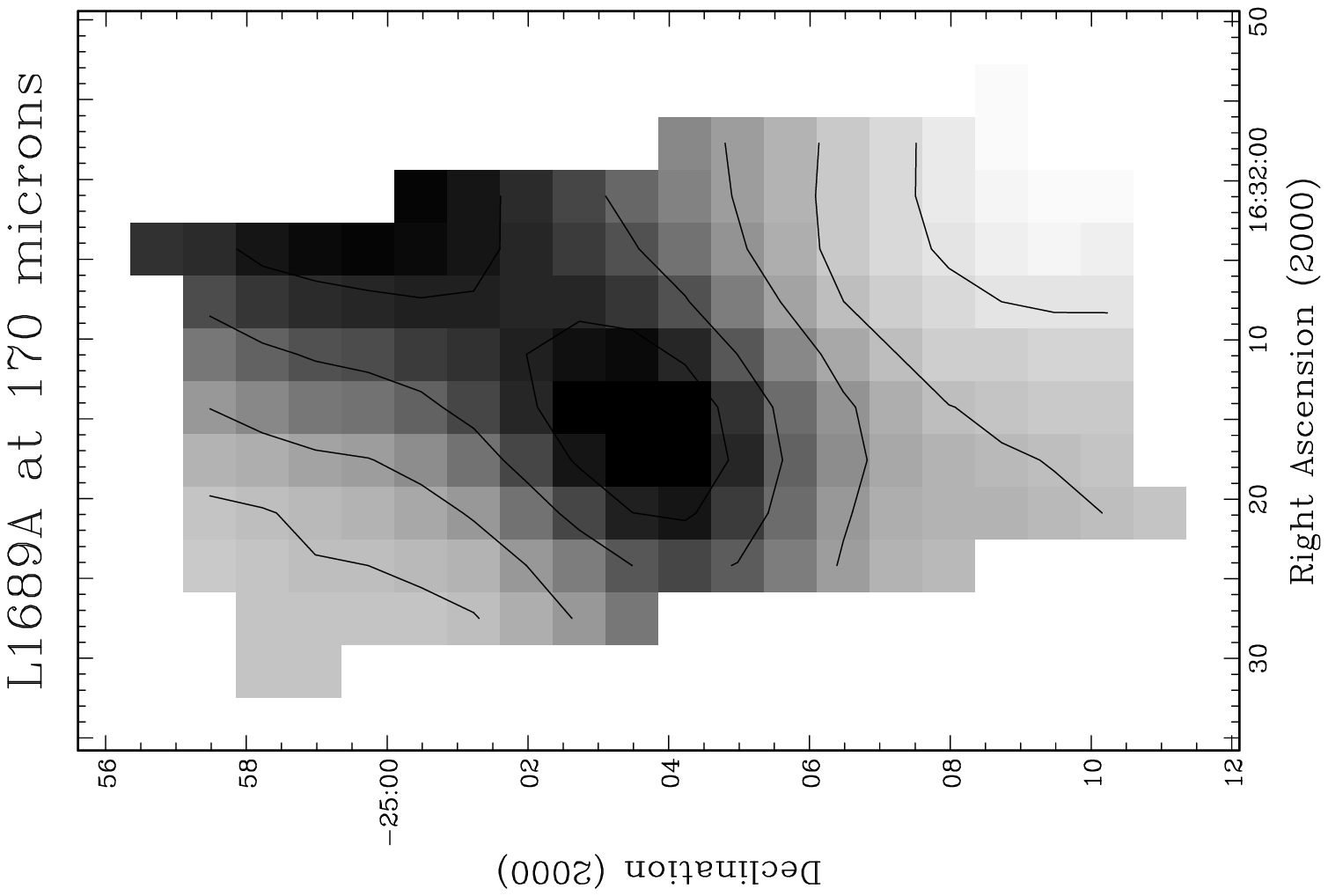}
\includegraphics{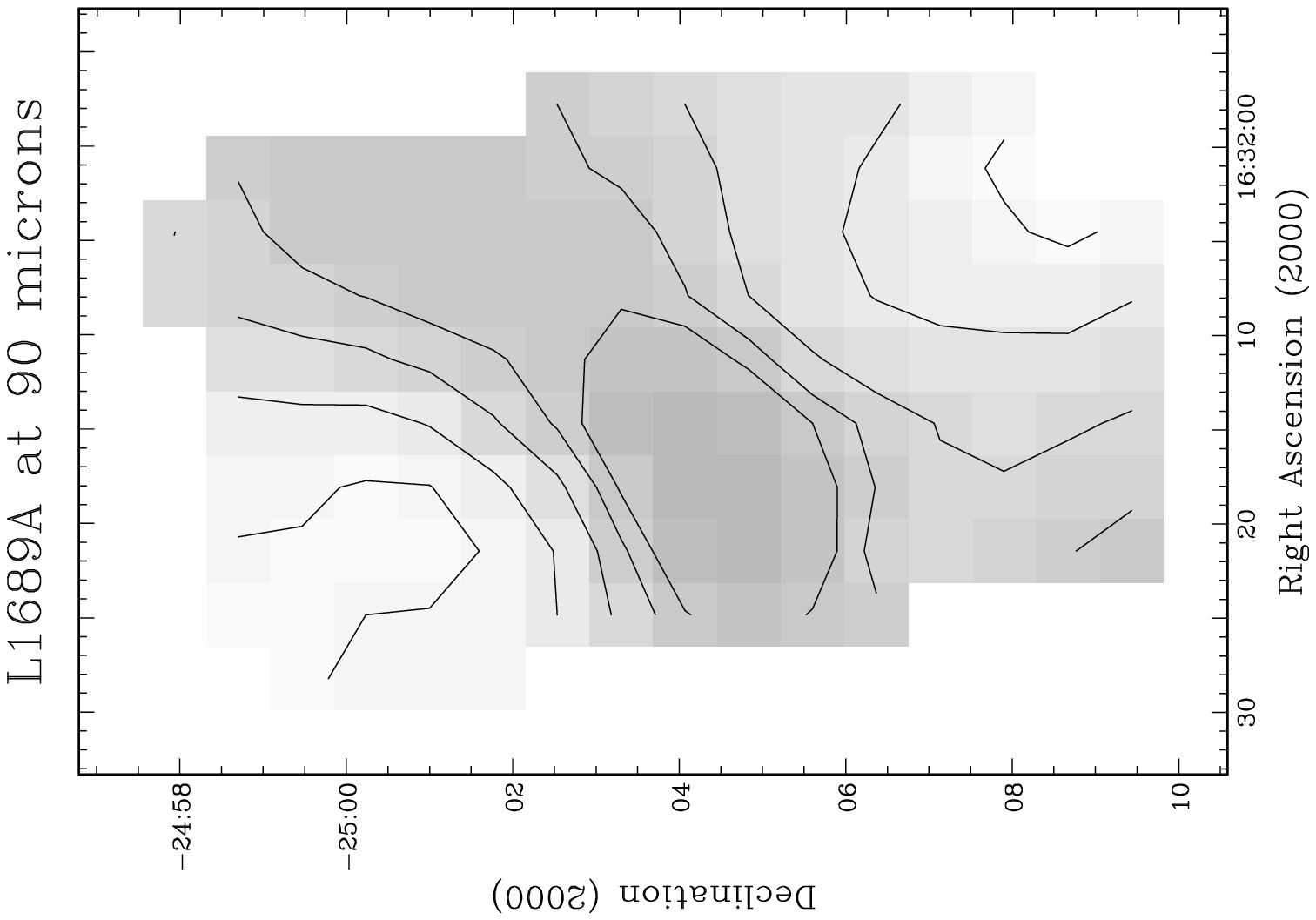}
\includegraphics{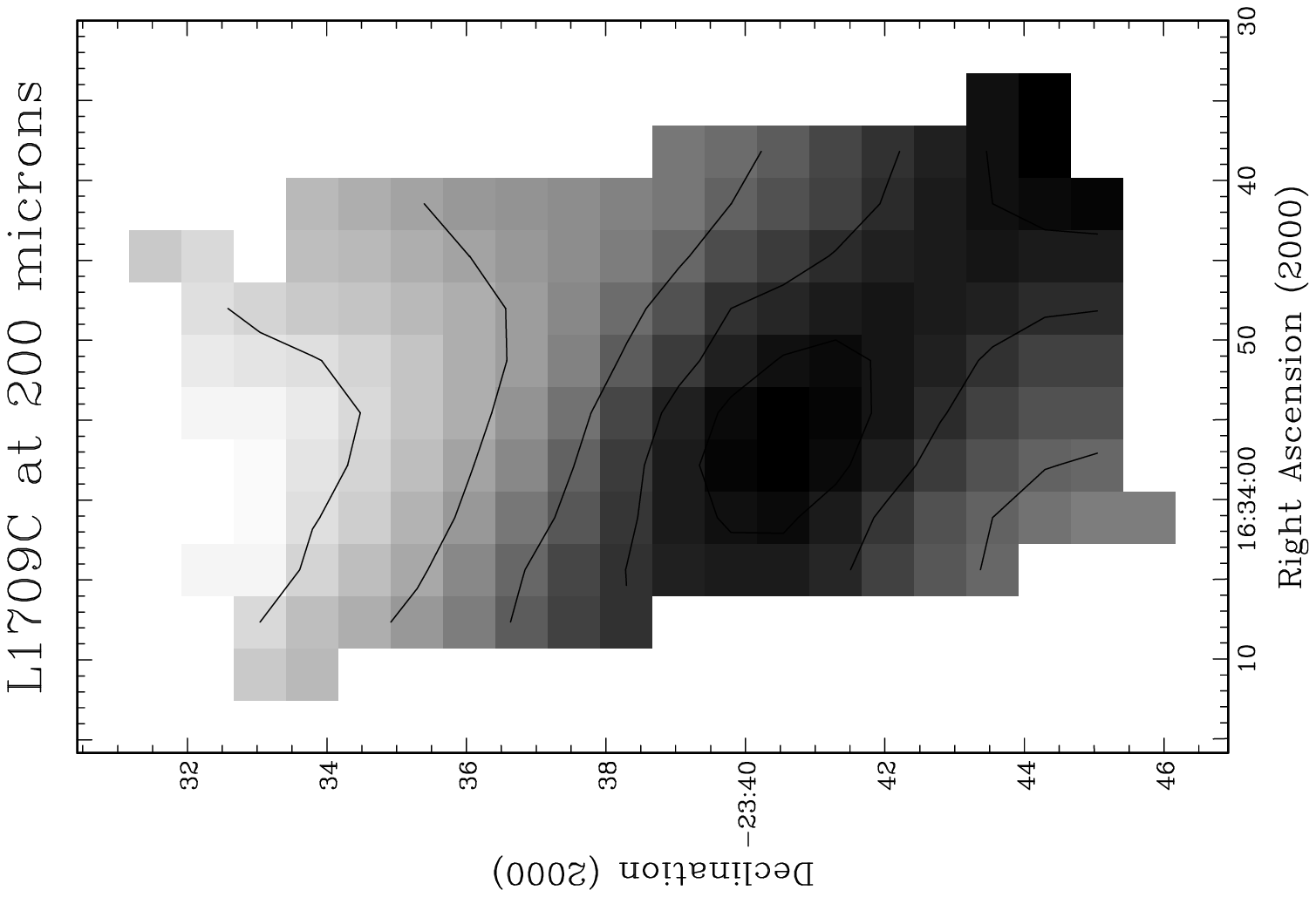}
\includegraphics{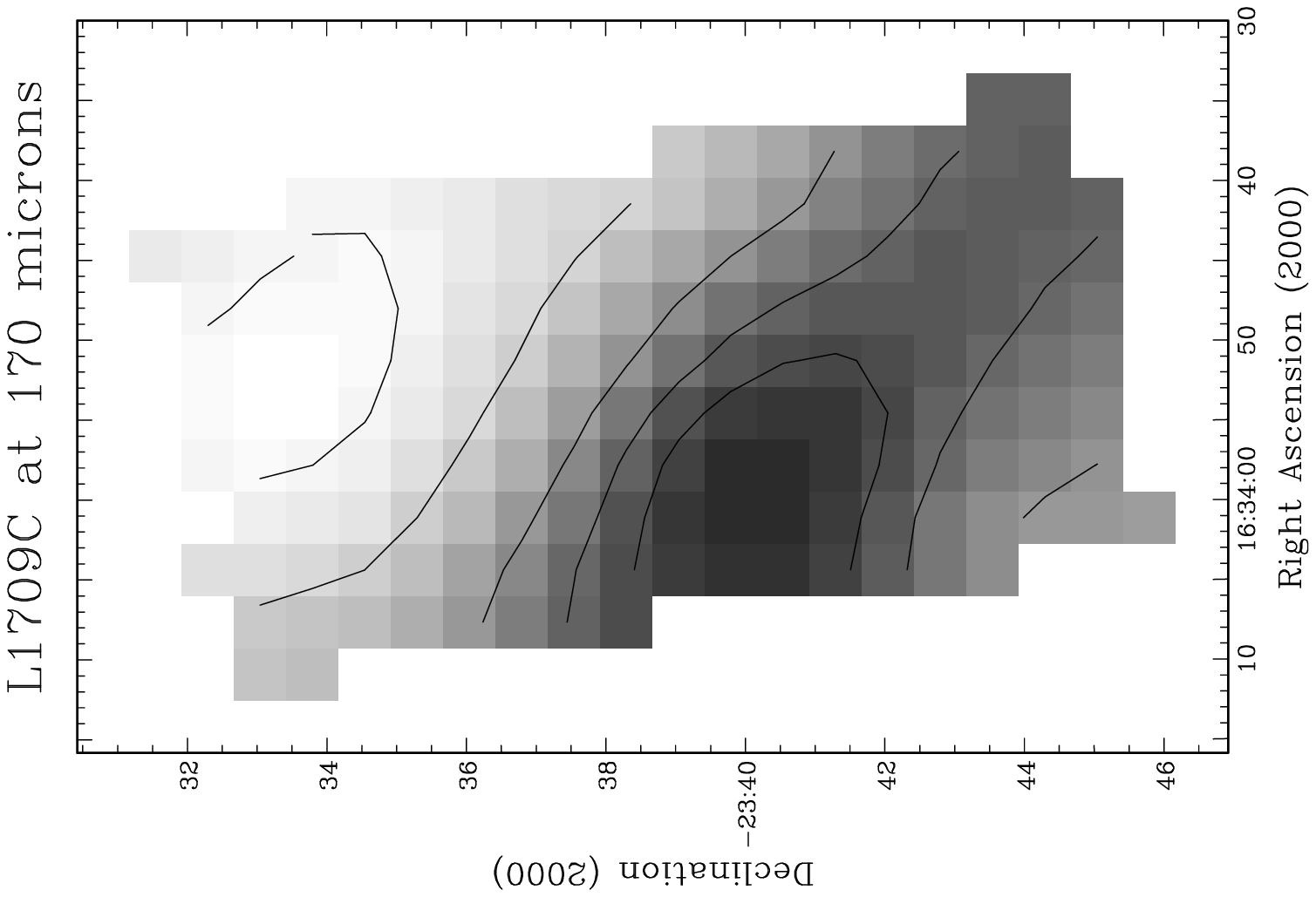}
\includegraphics{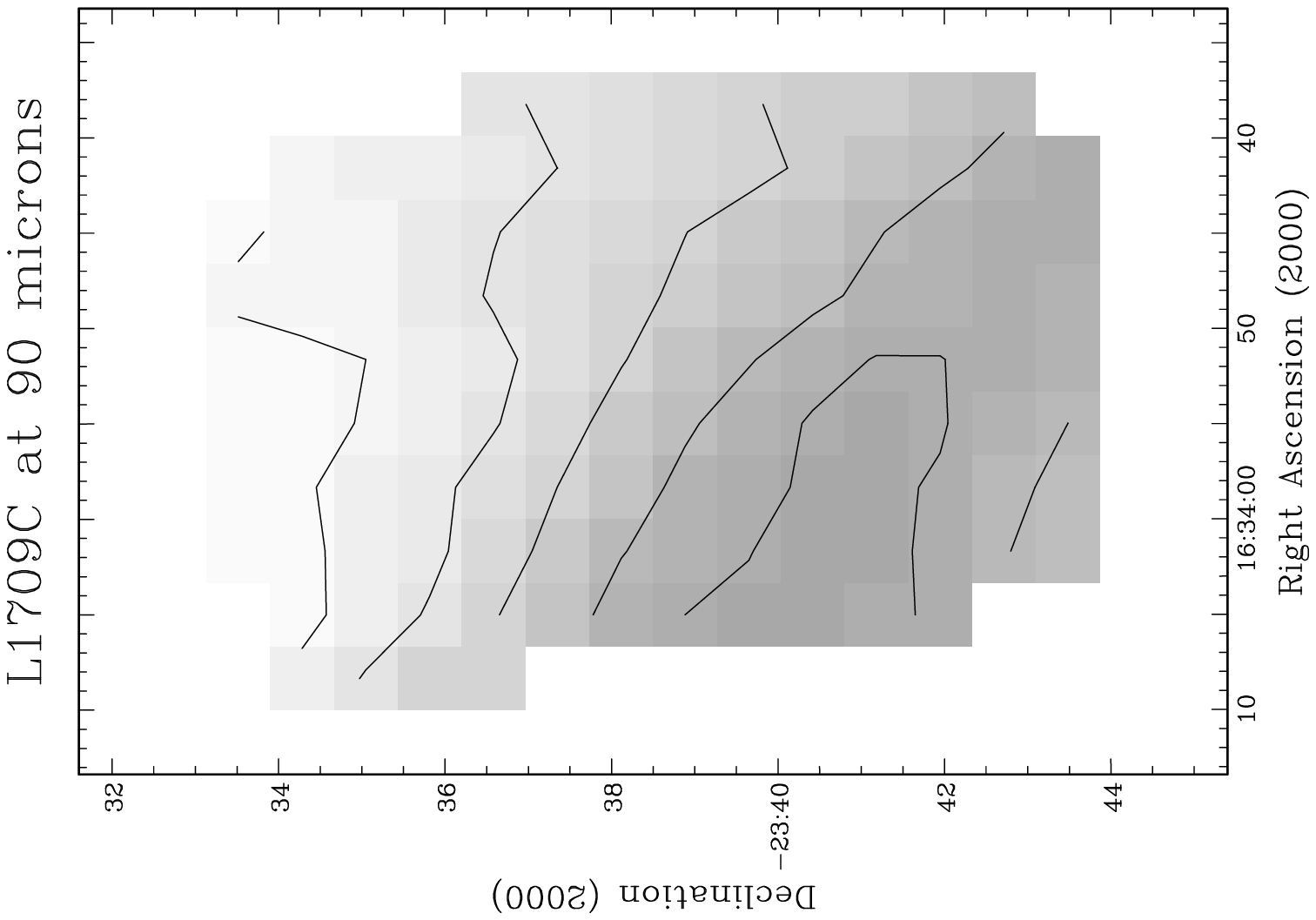}
\end{picture}
\caption{Images of L1689A (left) and L1709C (right). 
Details as in Figure 1.}
\end{figure*}

\begin{figure*}
\setlength{\unitlength}{1mm}
\begin{picture}(230,230)
\includegraphics{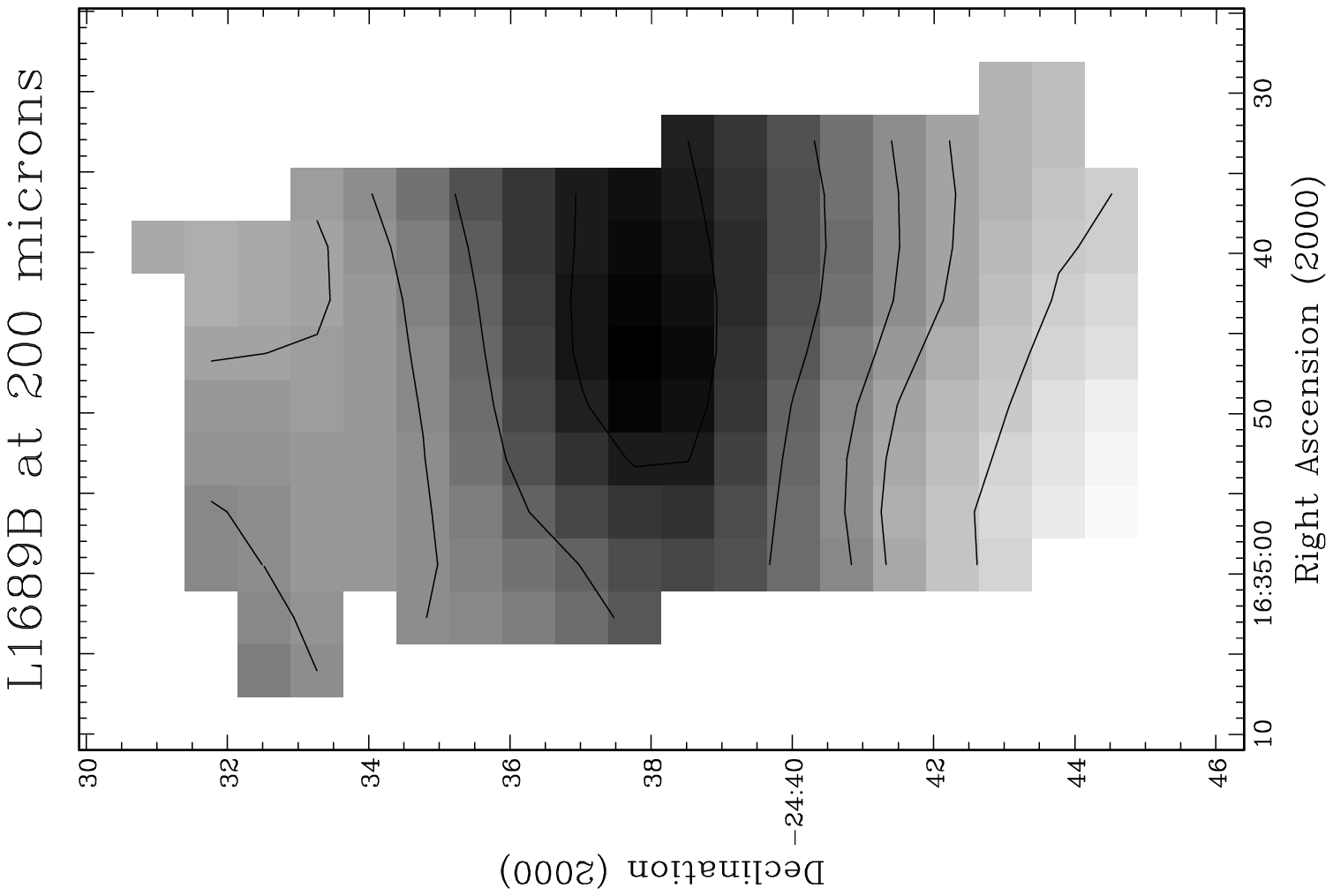}
\includegraphics{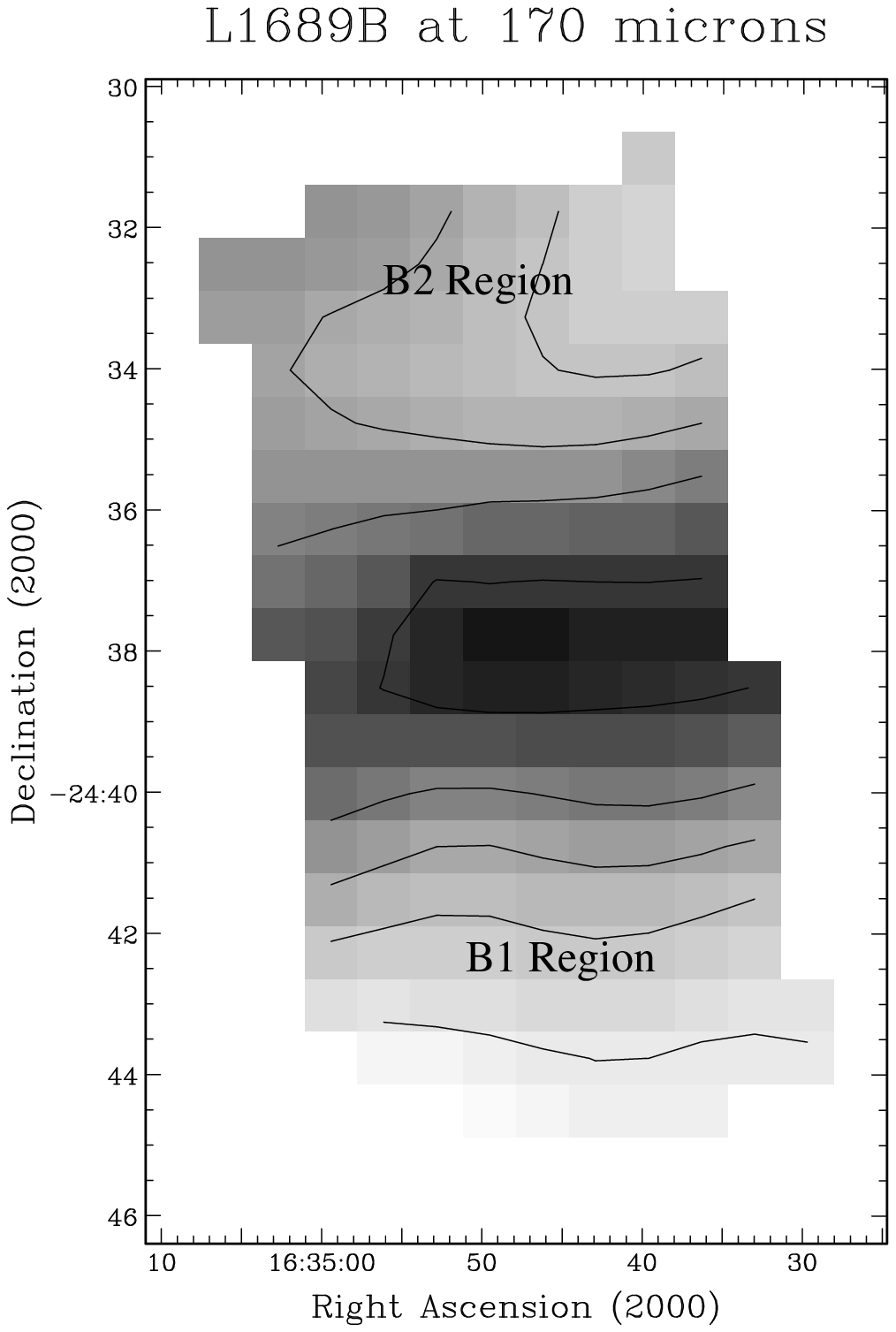}
\includegraphics{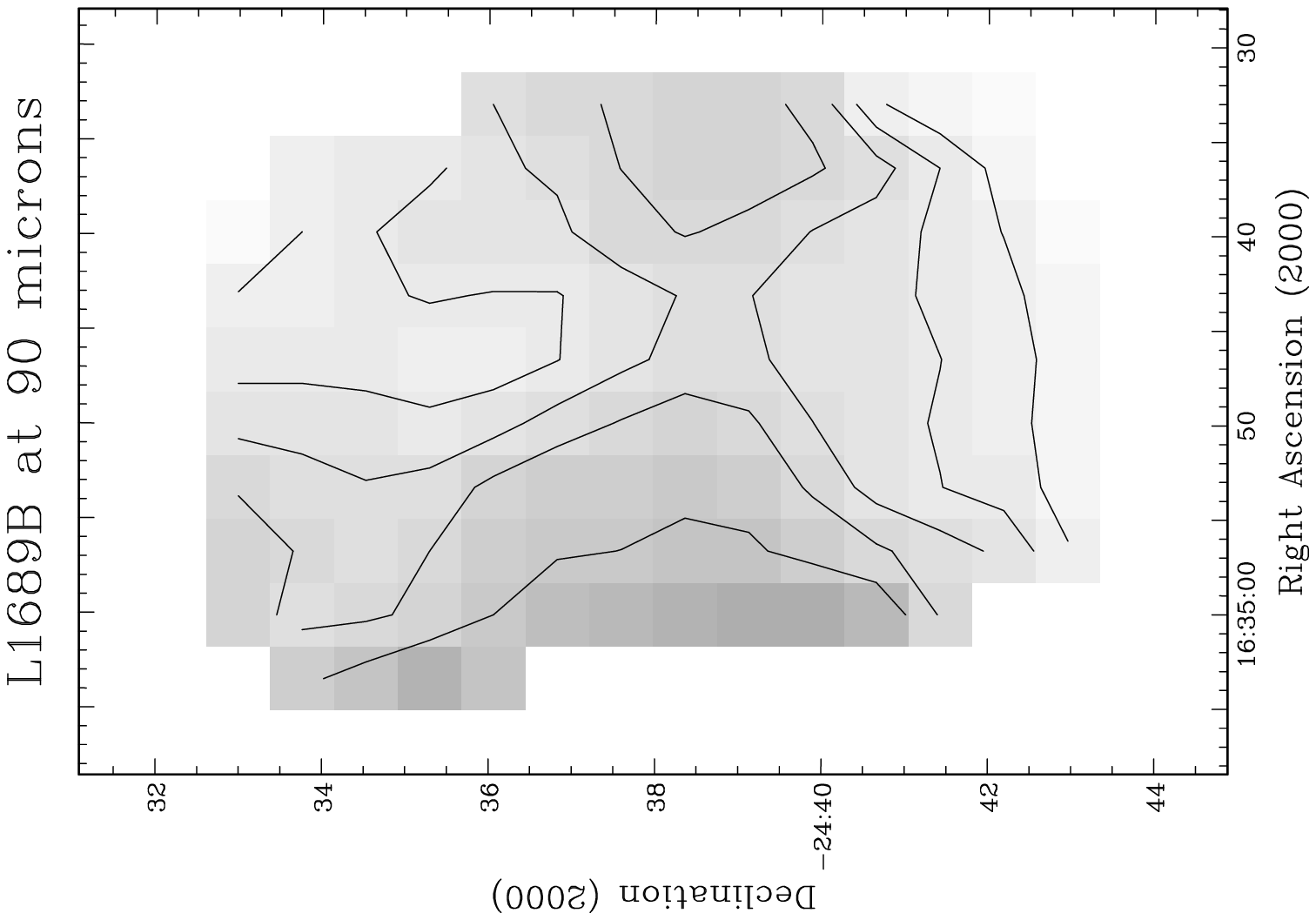}
\includegraphics{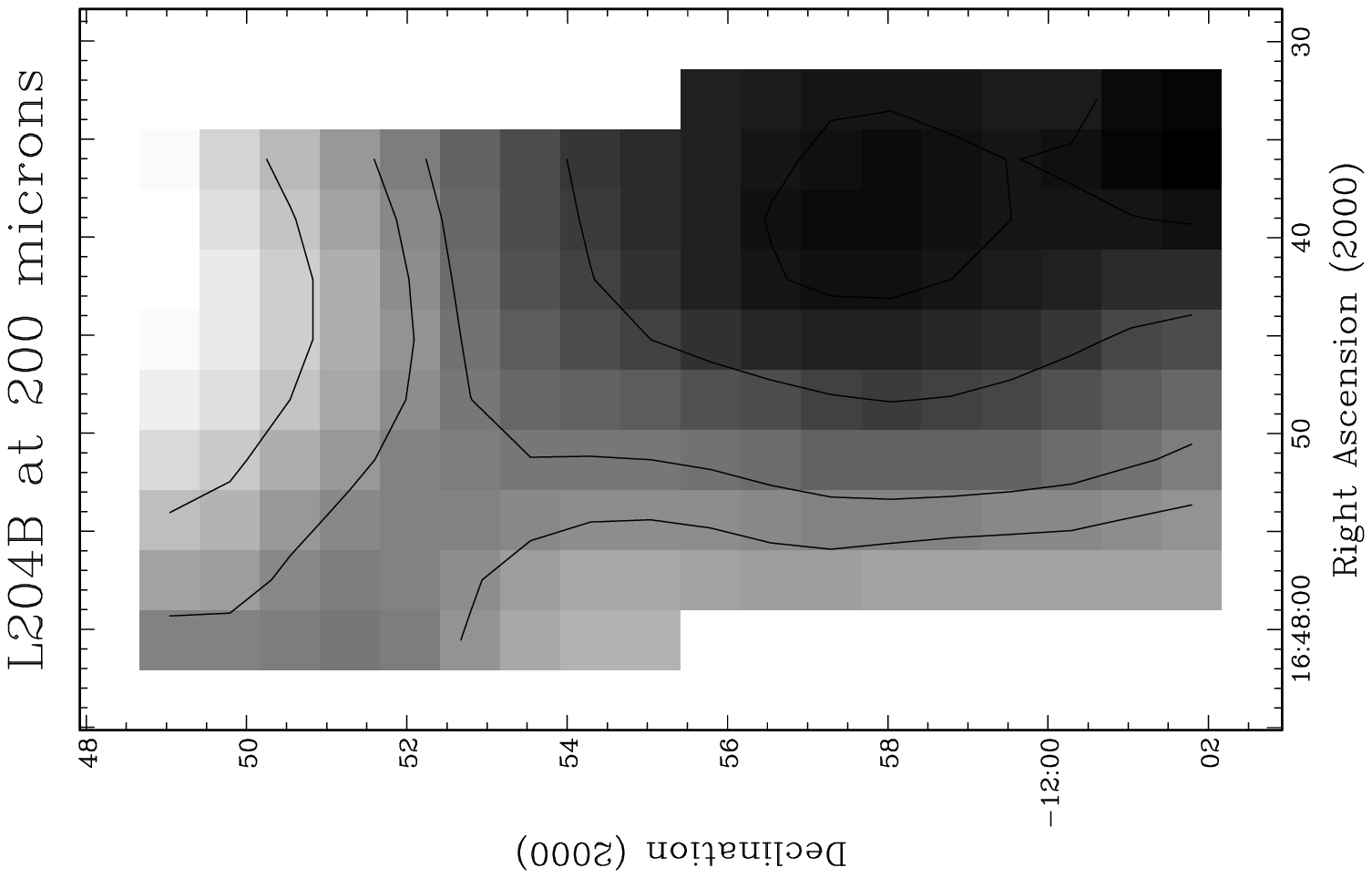}
\includegraphics{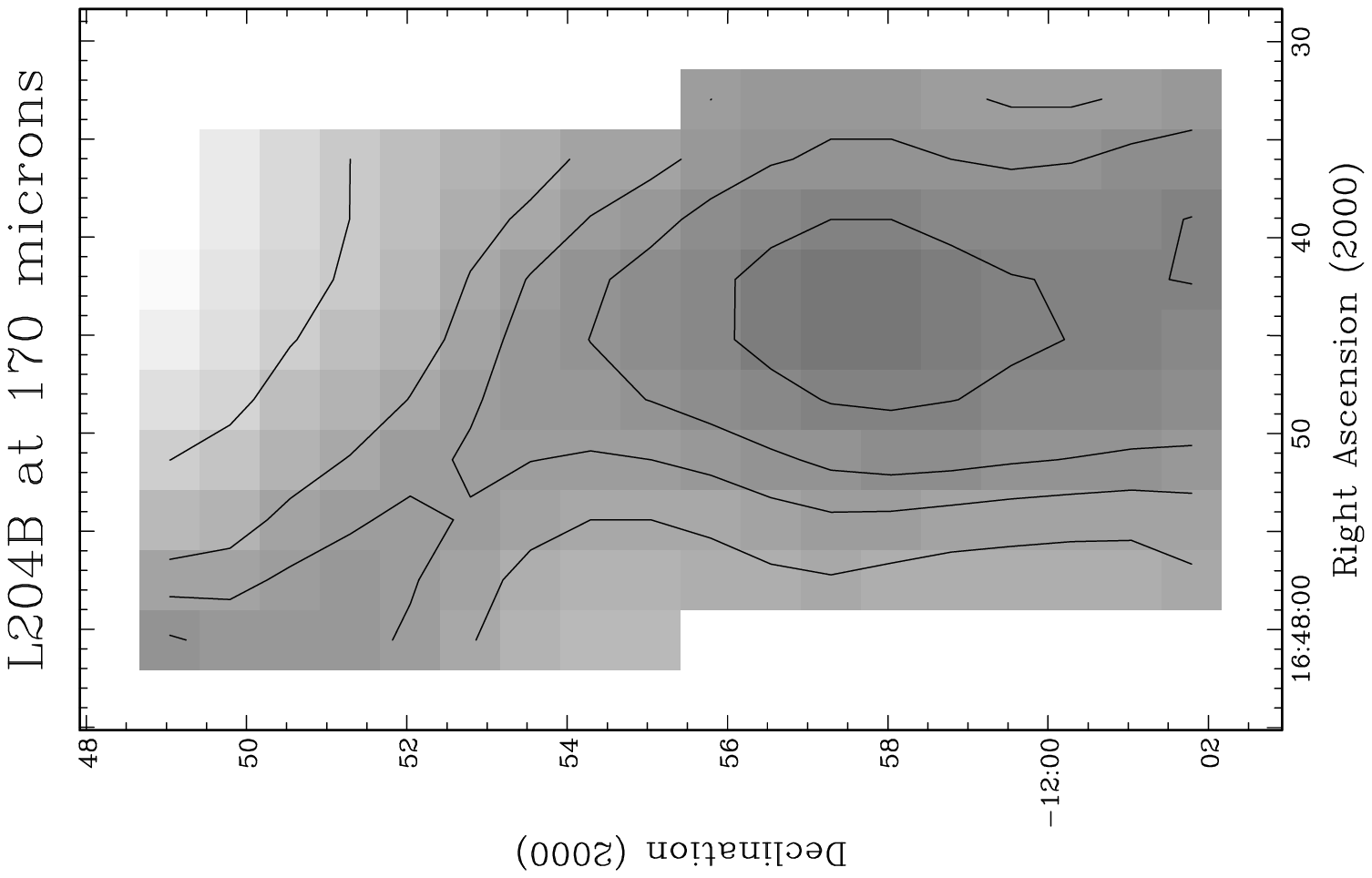}
\includegraphics{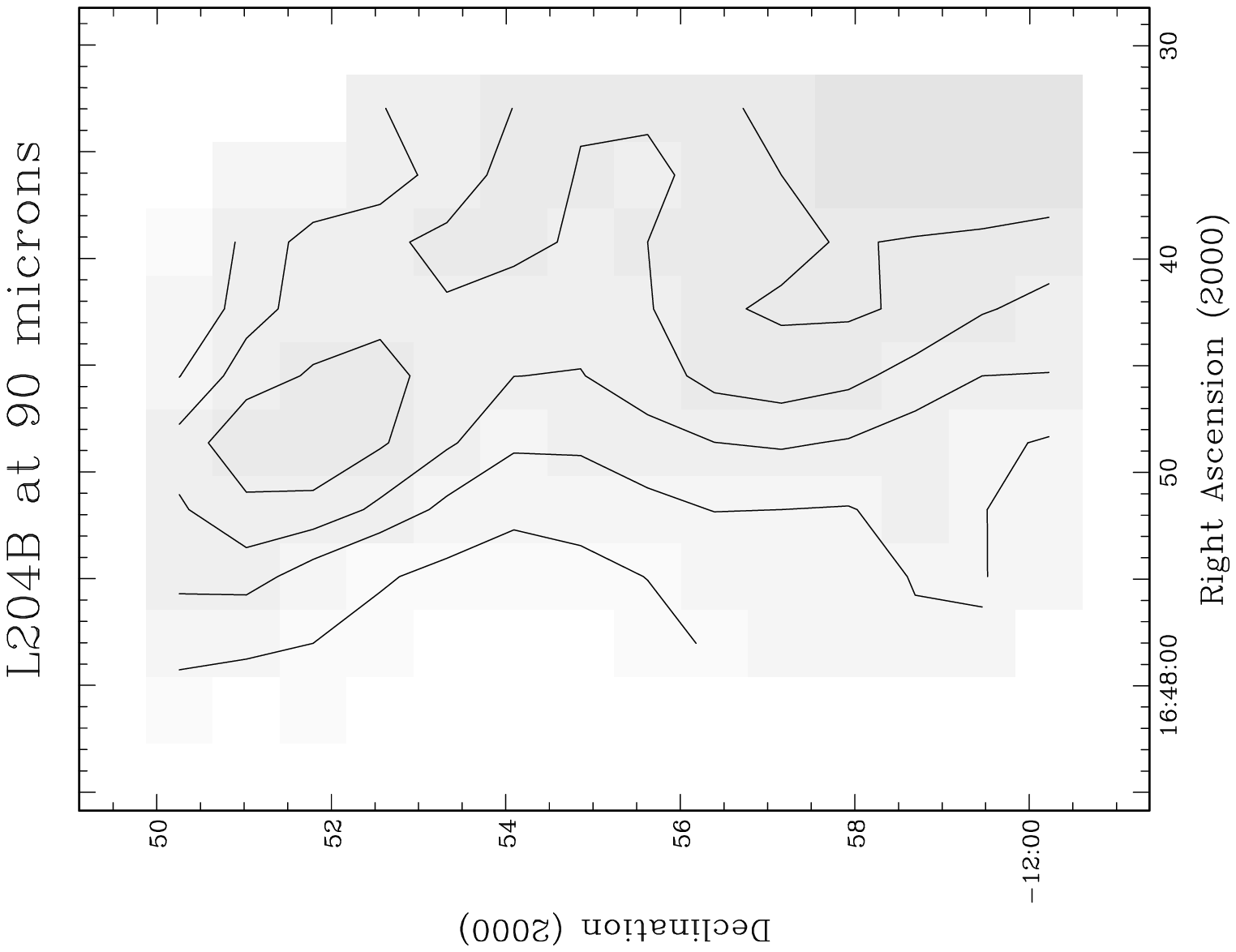}
\end{picture}
\caption{Images of L1689B (left) and L204B (right). Details as in Figure 1.}
\end{figure*}

\begin{figure*}
\setlength{\unitlength}{1mm}
\begin{picture}(230,230)
\includegraphics{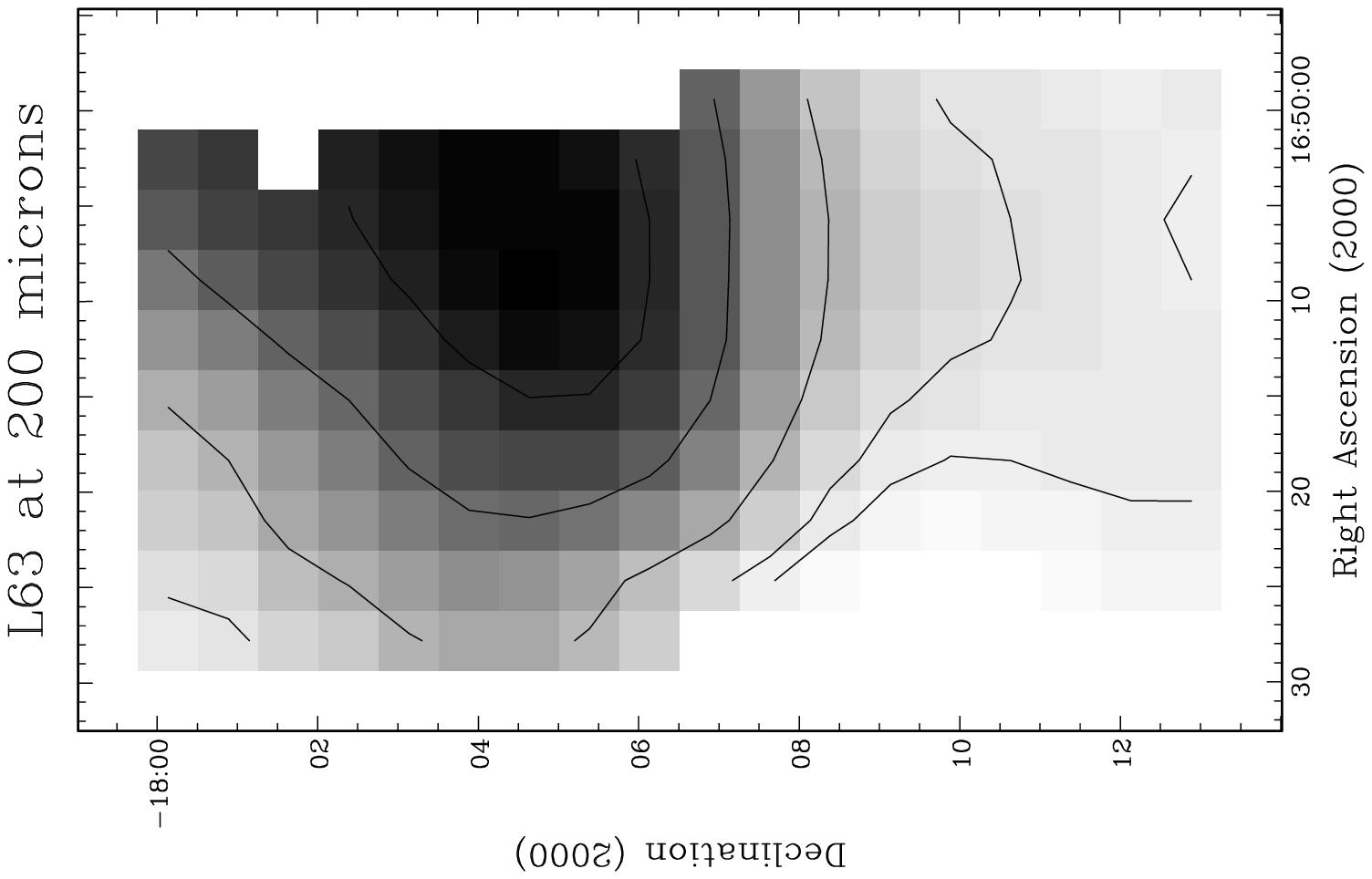}
\includegraphics{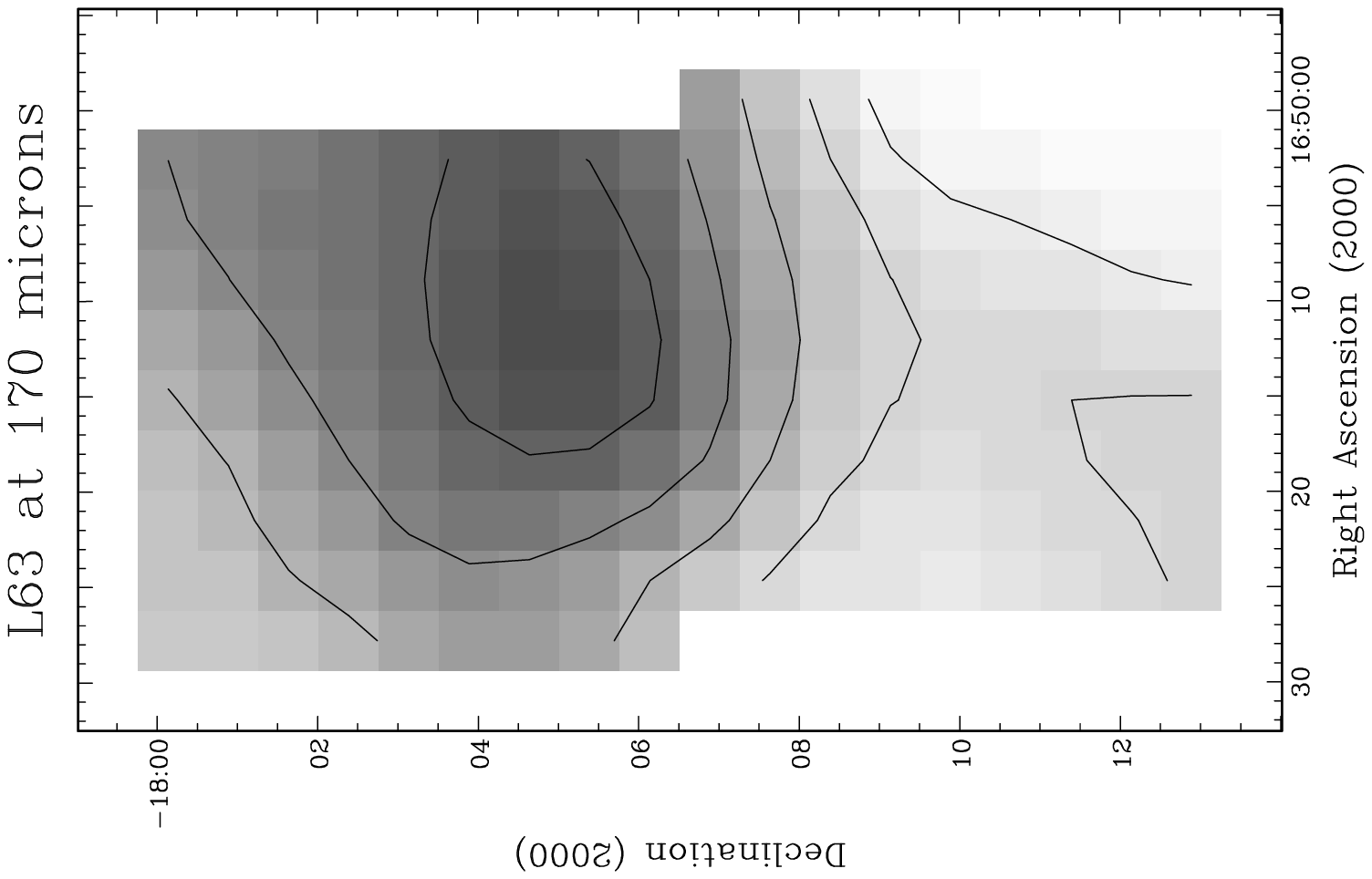}
\includegraphics{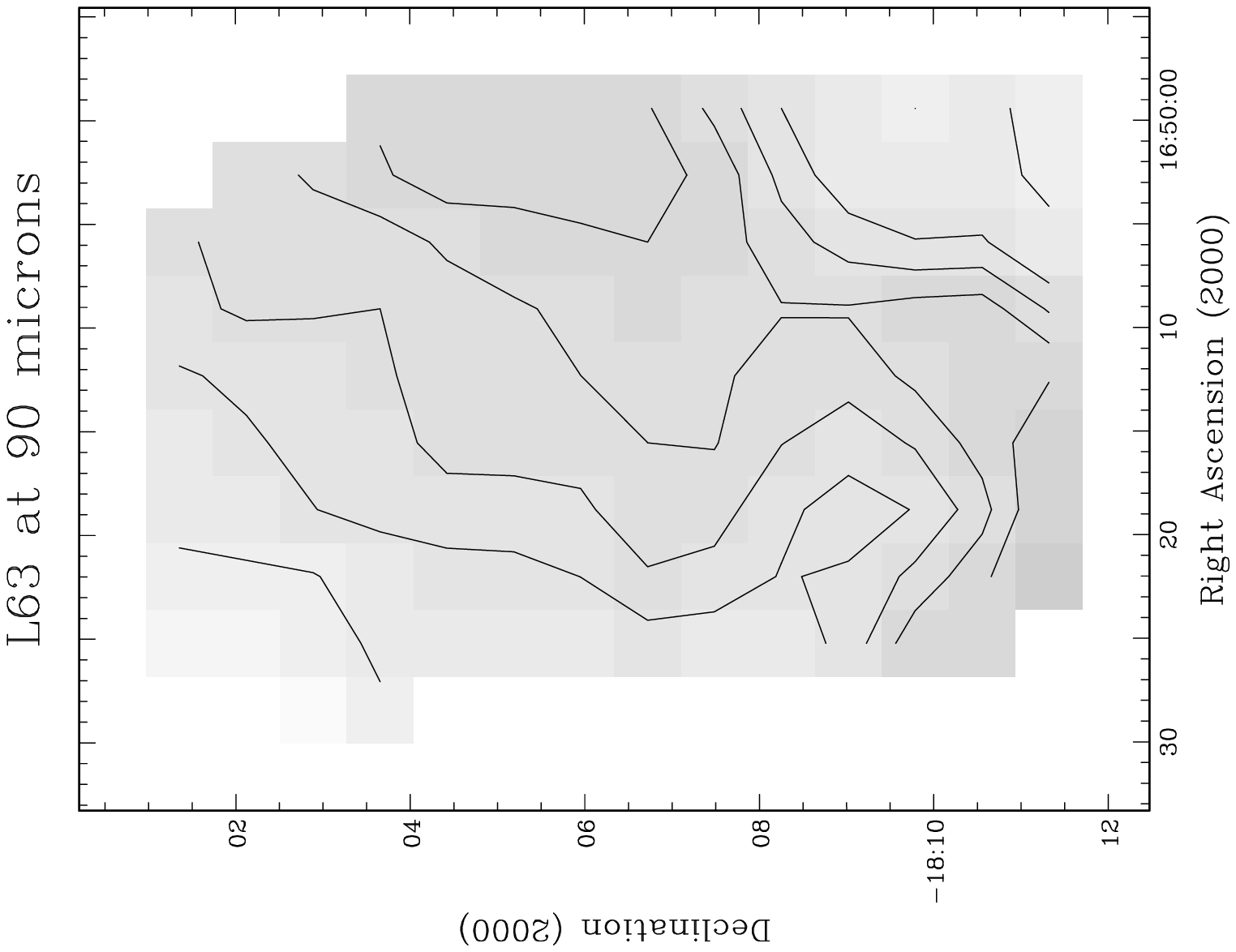}
\includegraphics{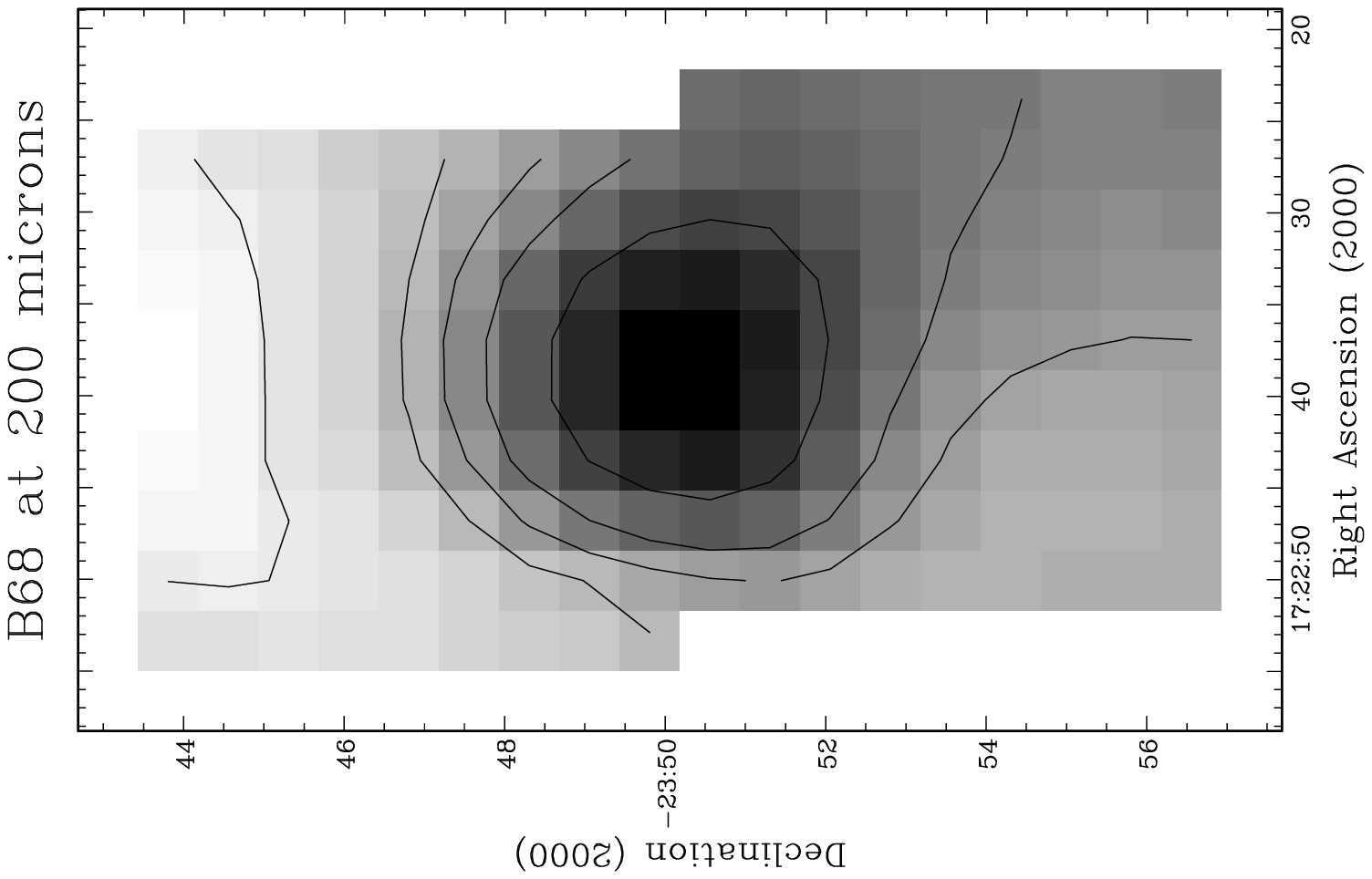}
\includegraphics{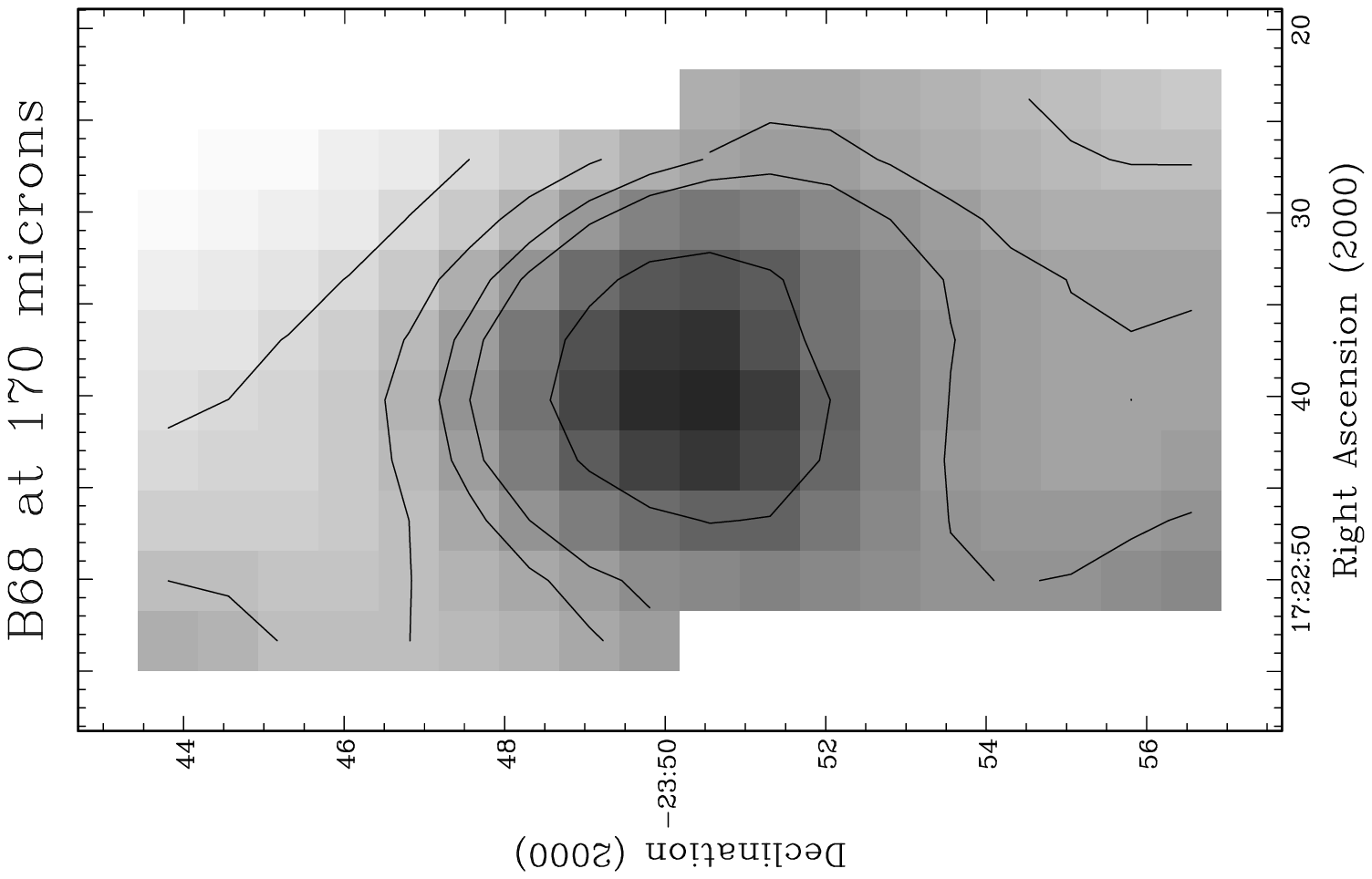}
\includegraphics{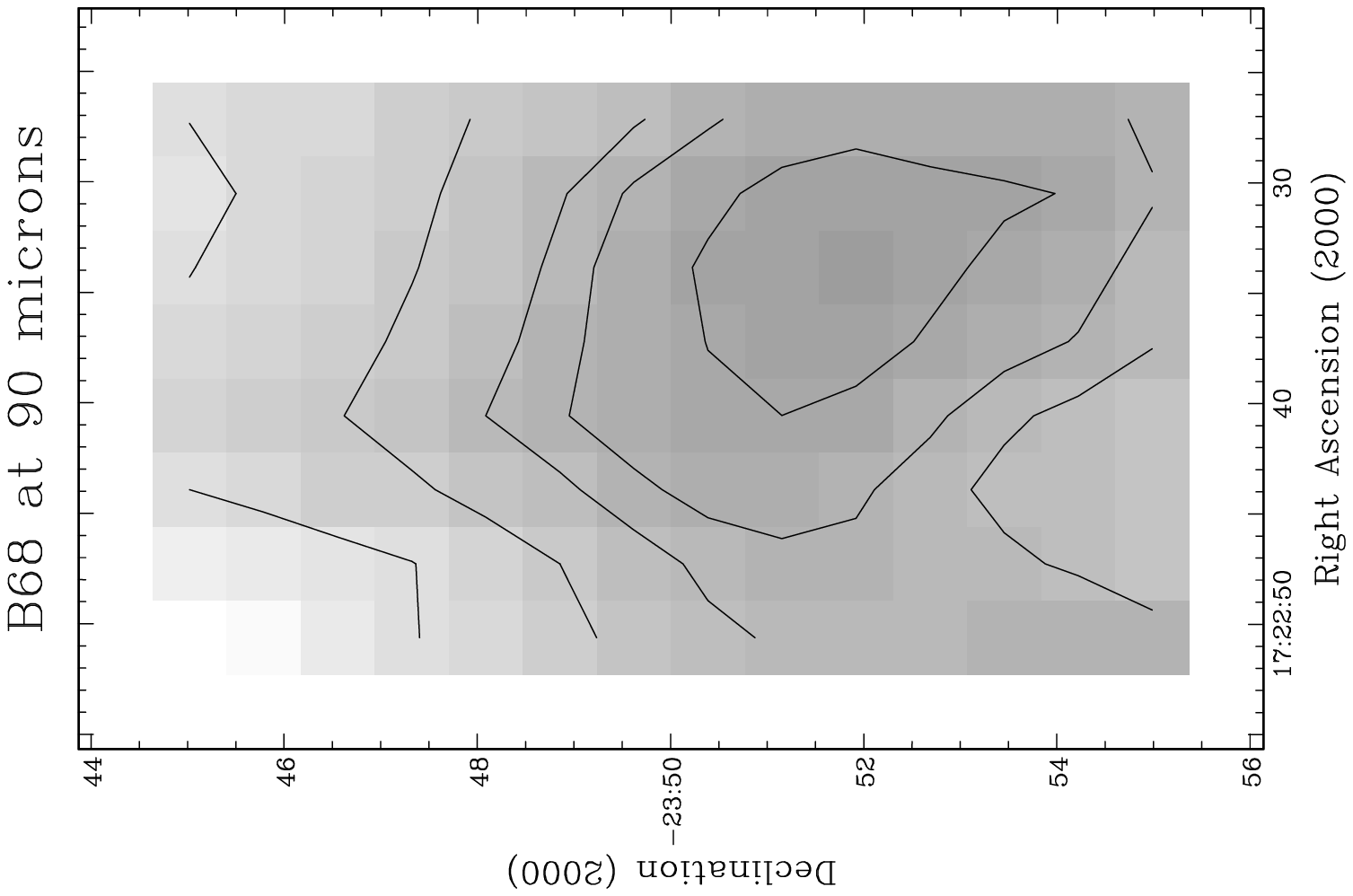}
\end{picture}
\caption{Images of L63 (left) and B68 (right). Details as in Figure 1.}
\end{figure*}

\begin{figure*}
\setlength{\unitlength}{1mm}
\begin{picture}(230,230)
\includegraphics{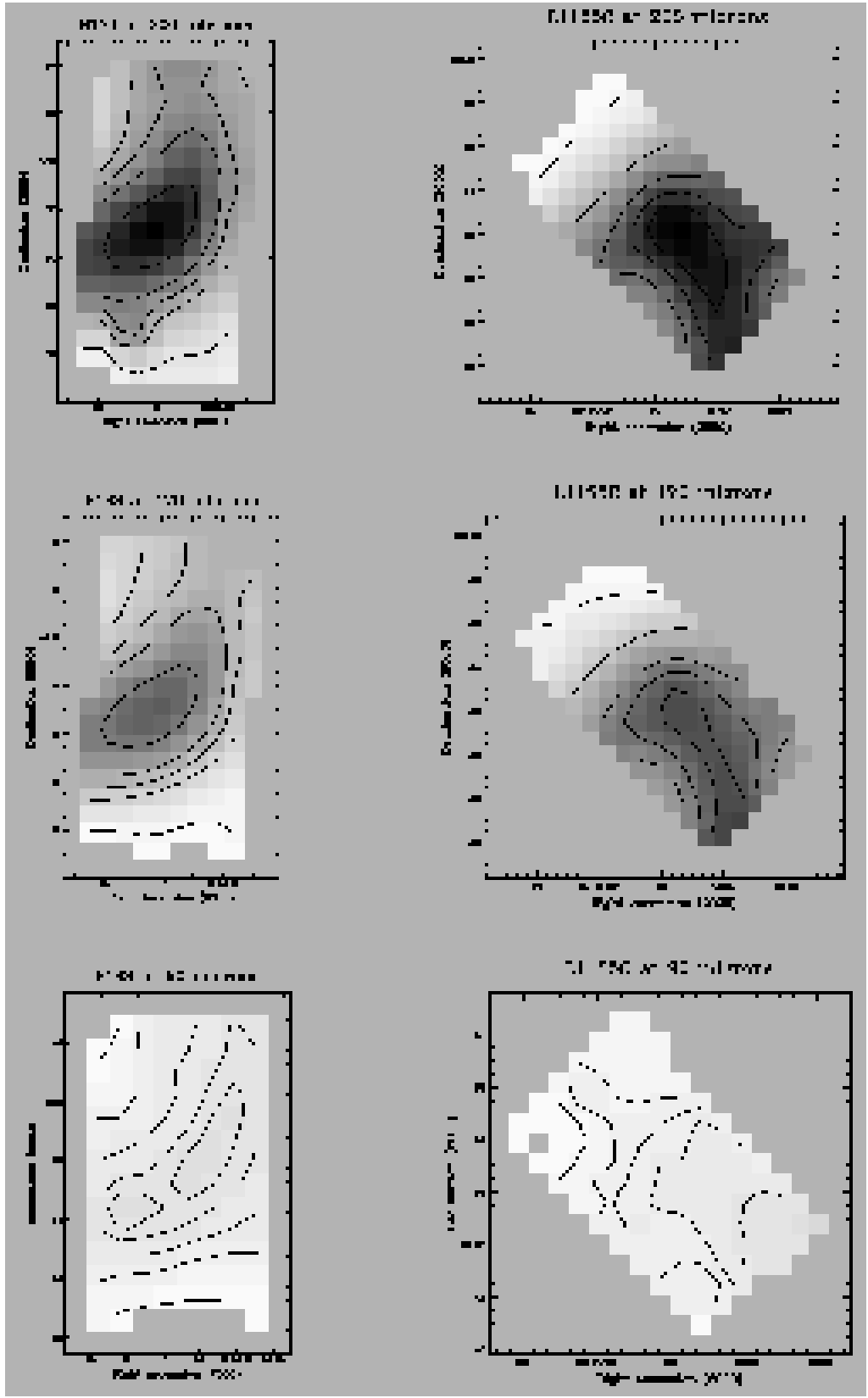}
\end{picture}
\caption{Images of B133 (left) and L1155C (right). Details as in Figure 1.}
\end{figure*}

\begin{figure}
\setlength{\unitlength}{1mm}
\begin{picture}(80,90)
\includegraphics{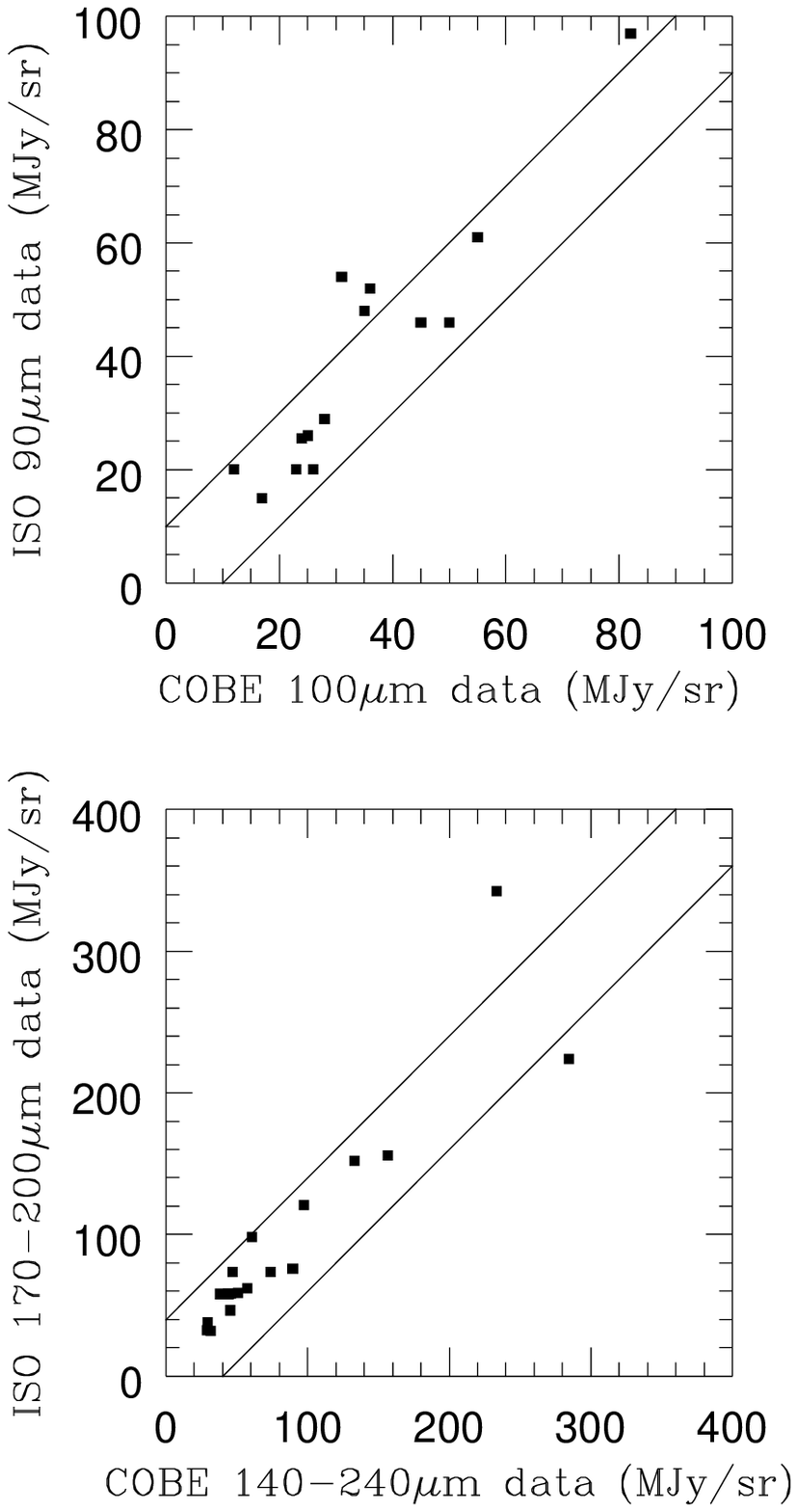}
\end{picture}
\label{fig9}
\caption{Comparison of ISO calibration with COBE data. The upper panel shows
our estimated 90-$\mu$m background plotted against the DIRBE 100-$\mu$m
background for each mapped region.
The lower panel shows the mean of our 170- and 200-$\mu$m
backgrounds plotted against the mean of the DIRBE 140- and 240-$\mu$m
backgrounds in each region.}
\end{figure}

It can be seen that the 
cores show a variety of shapes and appearances. However, the first 
observation
which strikes one when studying these images is that the cores are very 
much less clearly defined
at 90~$\mu$m than at the other two wavelengths. In fact, most 
of the cores are not significantly
detected at 90~$\mu$m, despite appearing
strongly in the images at the other two wavebands.
L1689A is detected at 90~$\mu$m. We have classified
L1582A, L1709C, B68 and B133 as marginal detections at 90~$\mu$m,
because they either have signal-to-noise ratios of $\leq$5$\sigma$,
or the peak at 90~$\mu$m does not coincide in position with the peak 
at the longer wavelengths. Consequently,
in each of these cases, it is not clear that the
actual core itself has been detected. 
We may either be detecting the extended cloud, or a different, warmer source
in the same cloud. As a result,
we do not include any of these sources in our subsequent analysis.

Table 1 lists some of the principle measurements of our target sources.
Column 1 lists the source name, wherein we follow the naming
convention of Benson \& Myers (1989). Columns 2 \& 3 list the source
positions as measured from the images in Figures 1--8. If there was a 
discrepancy in source peak position from one wavelength to another, then
the 200-$\mu$m peak position was taken. The final six columns list 
the peak flux density of each source at each wavelength (in MJy/sr) and 
the total integrated flux density measured in the whole of each 
mapped area. In the case of the 90-$\mu$m data 
the brightest point does not coincide with the core in most sources, 
so each one is listed as an upper limit. The exception is L1689A,
which was the only core to be clearly detected at 90$\mu$m.

The cores exhibit a variety of structures, and in some cases extend beyond
the boundaries of the mapped areas. This is because before ISO no-one had
imaged these objects at any infra-red wavelength -- none were detected by
IRAS.
Consequently, in setting up these observations we had little guidance as
to how big to make the maps, and exactly where to position them. We used
as our guide the previous CO and ammonia maps (Benson \& Myers 1989 and 
references therein). Nonetheless, in most cases we have successfully
imaged the cores and extended the maps far enough to estimate the
background level at each waveband in each field, as we now describe.

\begin{figure}
\setlength{\unitlength}{1mm}
\begin{picture}(80,60)
\includegraphics{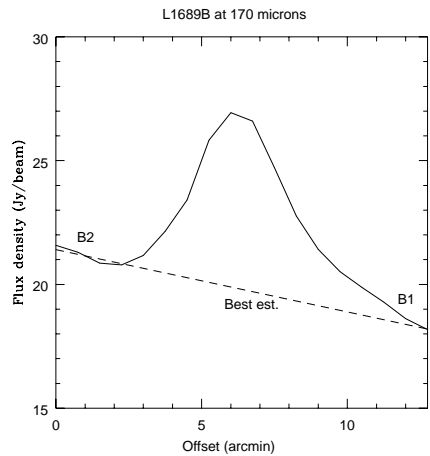}
\end{picture}
\label{fig10}
\caption{North-south flux density profile through L1689B at 170$\mu$m,
at R.A. (2000) of 16$^{\rm h}$ 34$^{\rm m}$ 48.8$^{\rm s}$, and from
Dec. (2000) of $-$24$^\circ$ 31$^\prime$ 50$^{\prime\prime}$ to
$-$24$^\circ$ 44$^\prime$ 40$^{\prime\prime}$.
Notice how the core sits on a sloping background from the B2
level in the north to the B1 level in the south. 
The 1$\sigma$ standard deviation in the B1 and B2 regions 
after removing this sloping background is 0.56Jy/beam.}
\end{figure}

\subsection{Background subtraction}

The infrared background arises from a number of causes, including Zodiacal
emission, and large-scale Galactic plane emission. IRAS detected these
extended background emissions (Beichman et al 1988; Jessop \& Ward-Thompson
2000), as did the Diffuse Infrared Background
Experiment (DIRBE) on the COBE satellite. IRAS did not cover wavelengths
beyond 100~$\mu$m, but DIRBE extended out to 240~$\mu$m. Consequently, we
chose to compare our background measurements with DIRBE. 
We estimated the background from the lowest emission region of each map
(the same region at each wavelength)
and called this background estimate B1.
The values of the backgrounds that we measured are listed in Table 2.
A comparison between our B1 background estimates and the
COBE data is plotted in Figure 9.

Figure 9(a) shows the background (B1) determined from our ISO 90-$\mu$m data
plotted against the DIRBE measurement at 100~$\mu$m which was closest on
the sky to our source. Figure 9(b) plots the
mean of our 170- and 200-$\mu$m background measurements against the mean
of the DIRBE 140- and 240-$\mu$m background determinations at the closest
position on the sky. 

In each case
we are not comparing exactly like with like. The exact wavelength and
filter bandwidths do not match. Likewise the resolutions of the two
telescopes do not match -- DIRBE had much lower resolution.
In addition our background measurements may be
slight over-estimates due to the finite size of our images. 

Consequently
we would not expect an exact correlation between the two data-sets,
and we would expect a certain degree of scatter.
Nonetheless, the broad agreement at all wavelengths is very encouraging.
The spread indicated by the two solid lines on each plot corresponds to
roughly
$\pm$30 per cent for mean background levels. This tends to support our
assertion above that the absolute calibration uncertainty is no worse than
$\pm$30 per cent.

Table 3 lists the extended source flux densities at each wavelength.
In an attempt to separate the compact emission due to the pre-stellar 
cores from the extended background emission at 170 \& 200$\mu$m,
we subtracted the background level (B1) from every field as
estimated from the lowest emission region in each image.
We then measured the FWHM extent of each source at 200$\mu$m along
the maximum and minimum directions. 
These were occasionally tricky to estimate if the source extended beyond 
the edge of the mapped region, and some judgement was required in these 
cases. We measured the flux density at each wavelength for each object 
within this FWHM in the resulting images. These flux densities are listed 
in Table 3 under the columns headed B1. 

We note that in every case the pre-stellar cores also
appear to sit on top of more extended emission that is probably
related to the larger scale cloud in which the core is embedded. 
This can be seen if we take a 1-D north-south cut through the core
L1689B at 170$\mu$m. This is shown in Figure 10. It can be seen from
this plot that the background in the north is considerably higher than
that in the south. We estimated this higher `background' region due to
extended cloud emission and called this background B2. The two regions
B1 \& B2 are shown in Figure 6(a) to illustrate this process.

We attempted to estimate the value of the
B2 background emission by measuring the mean
flux density in the region of each image where there was extended
emission, but away from the core itself. The measured values of the B2
backgrounds are listed in Table 2. We then subtracted these
values from each image and re-measured the flux density within the
FWHM. These flux densities are listed in Table 3 under
the columns headed B2.

Finally, we made our best estimate of the background level in each region,
by fitting a sloping background to each region and subtracting this from
the core emission. The example of this sloping background in the case of 
L1689B is illustrated in Figure 10. We measured the FWHM of each source
on the resulting images, and list these FWHM in Table 3. We measured
the flux densities within these FWHM for each
core and list these in Table 3 under the columns headed
`Best est'. 

Thus Table 3 lists the full variation in flux densities
at 170 \& 200~$\mu$m that
could be associated with our cores, given the limited sizes of the mapped
areas in each case. The errors we list with each measurement are the 
statistical errors associated with each individual background subtraction.
At 90$\mu$m we did not detect most of the
200-$\mu$m sources, so we simply list our
best estimate of the upper limit to the flux density in each region,
based on the 3$\sigma$ variation in the local background.

\subsection{Colour temperatures}

We measured the colour temperature variation across each
core by first subtracting the background level 
that we had measured in each image according to our best estimate,
as described above. We then ratio-ed the 
background-subtracted images at 170 and 200~$\mu$m. We converted this to
a colour temperature using the assumption of optically thin 
grey-body emission (see section 3.4 below), namely:

\[ (F_{\nu_1}/F_{\nu_2}) = 
\frac{\nu_{1}^{3+\beta}(e^{[h\nu_{2}/kT]} - 1)}
{\nu_{2}^{3+\beta}(e^{[h\nu_{1}/kT]} - 1)}, \]

\begin{figure*}
\setlength{\unitlength}{1mm}
\begin{picture}(230,230)
\includegraphics{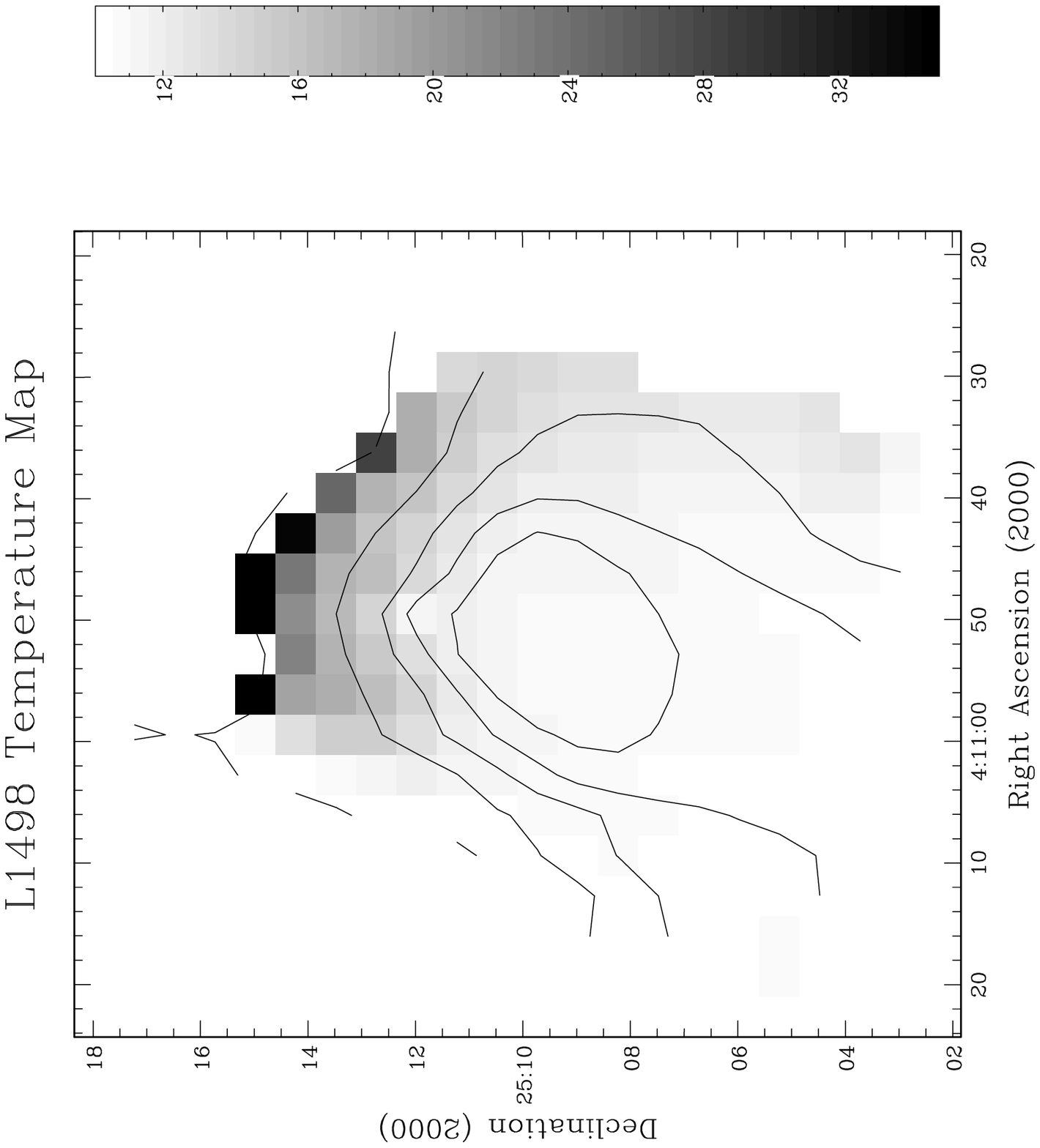}
\includegraphics{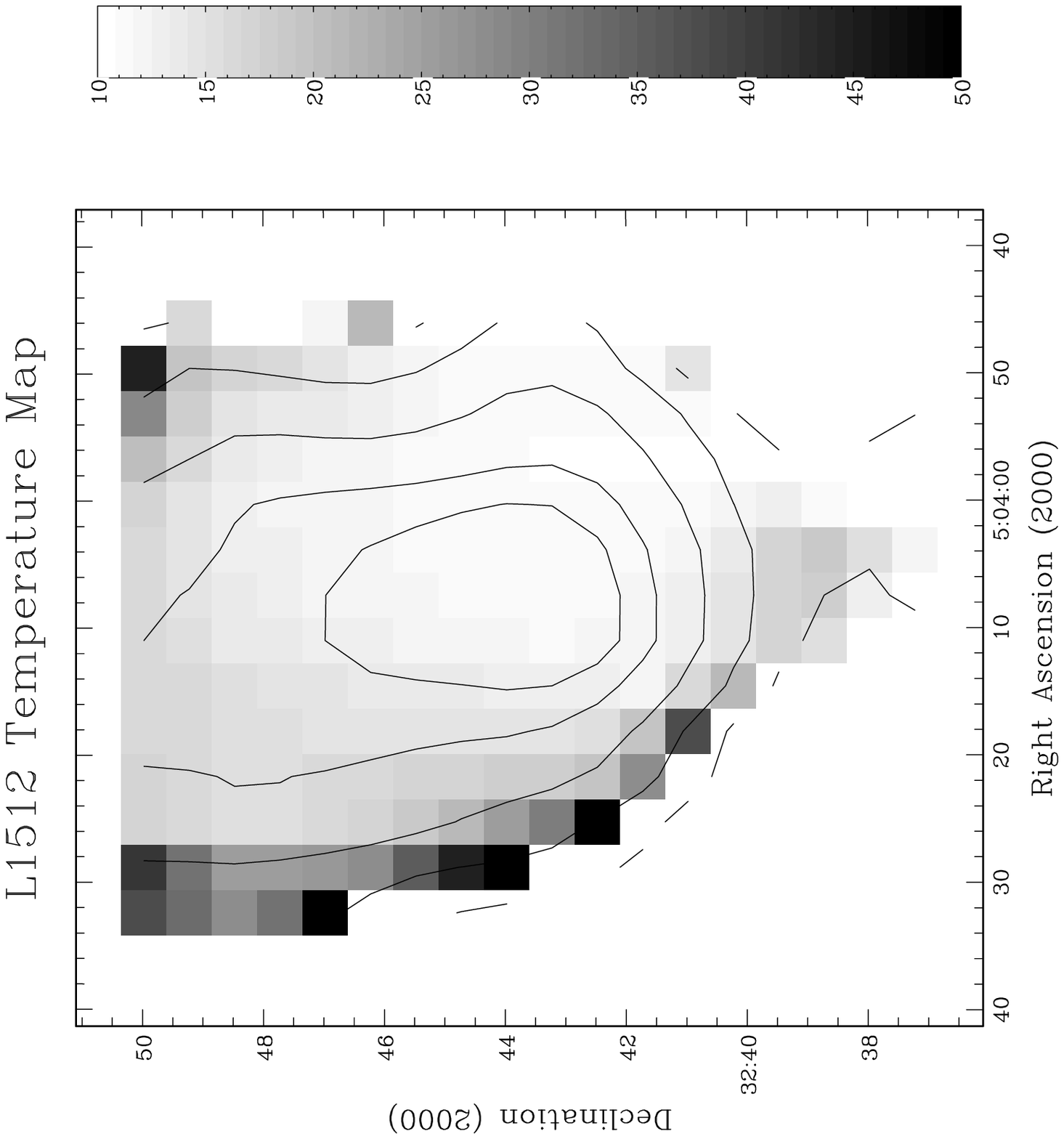}
\includegraphics{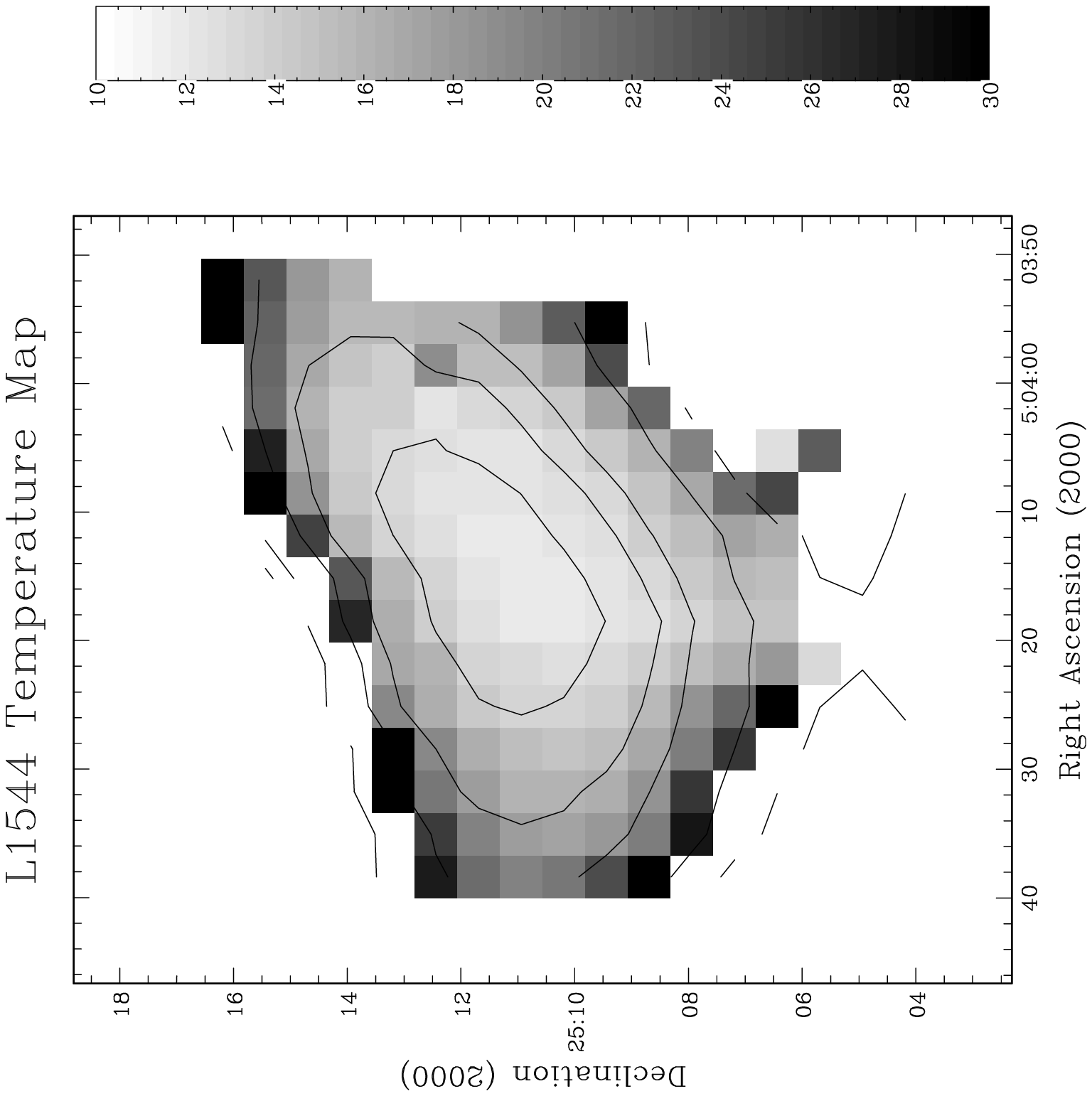}
\includegraphics{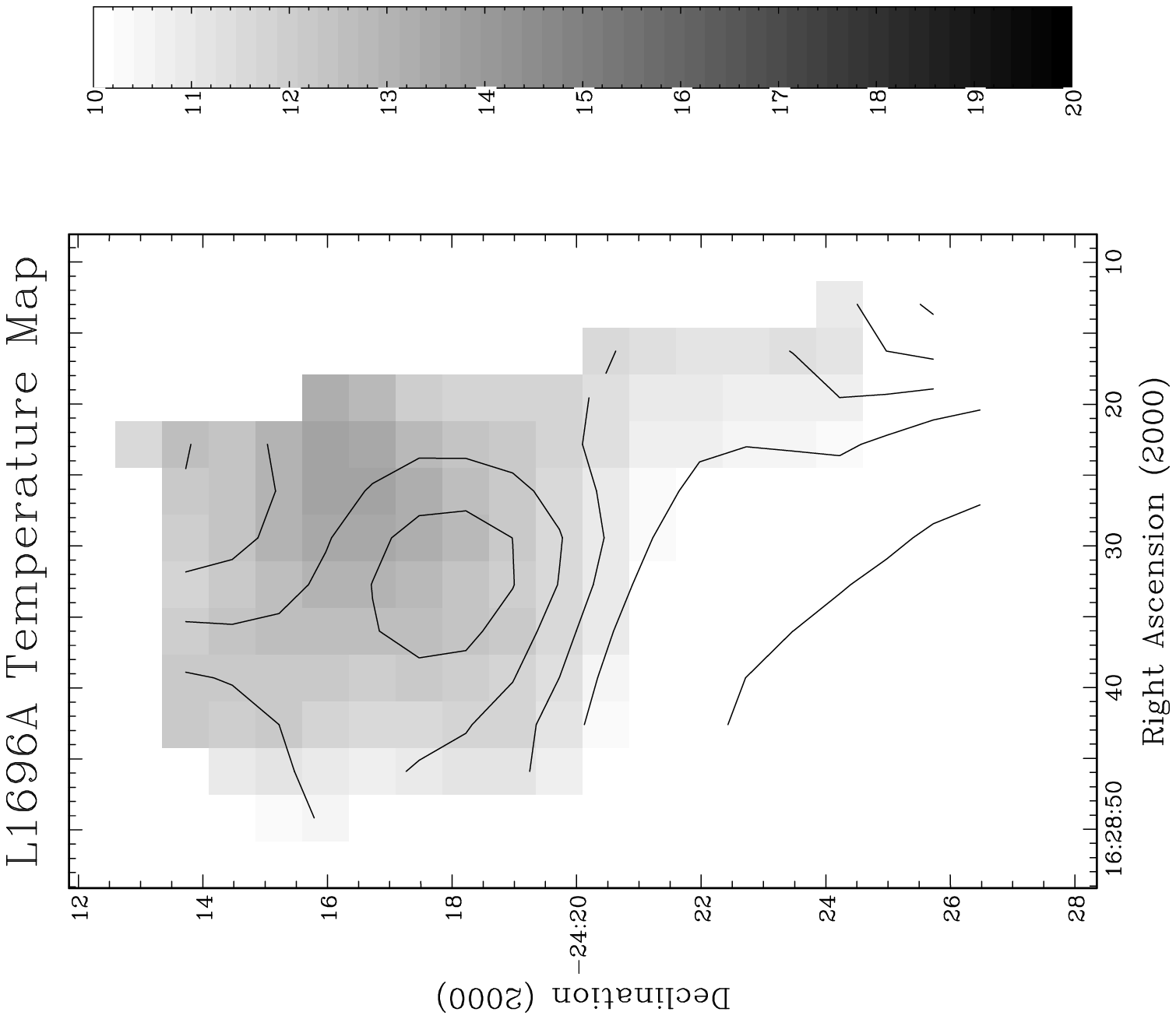}
\includegraphics{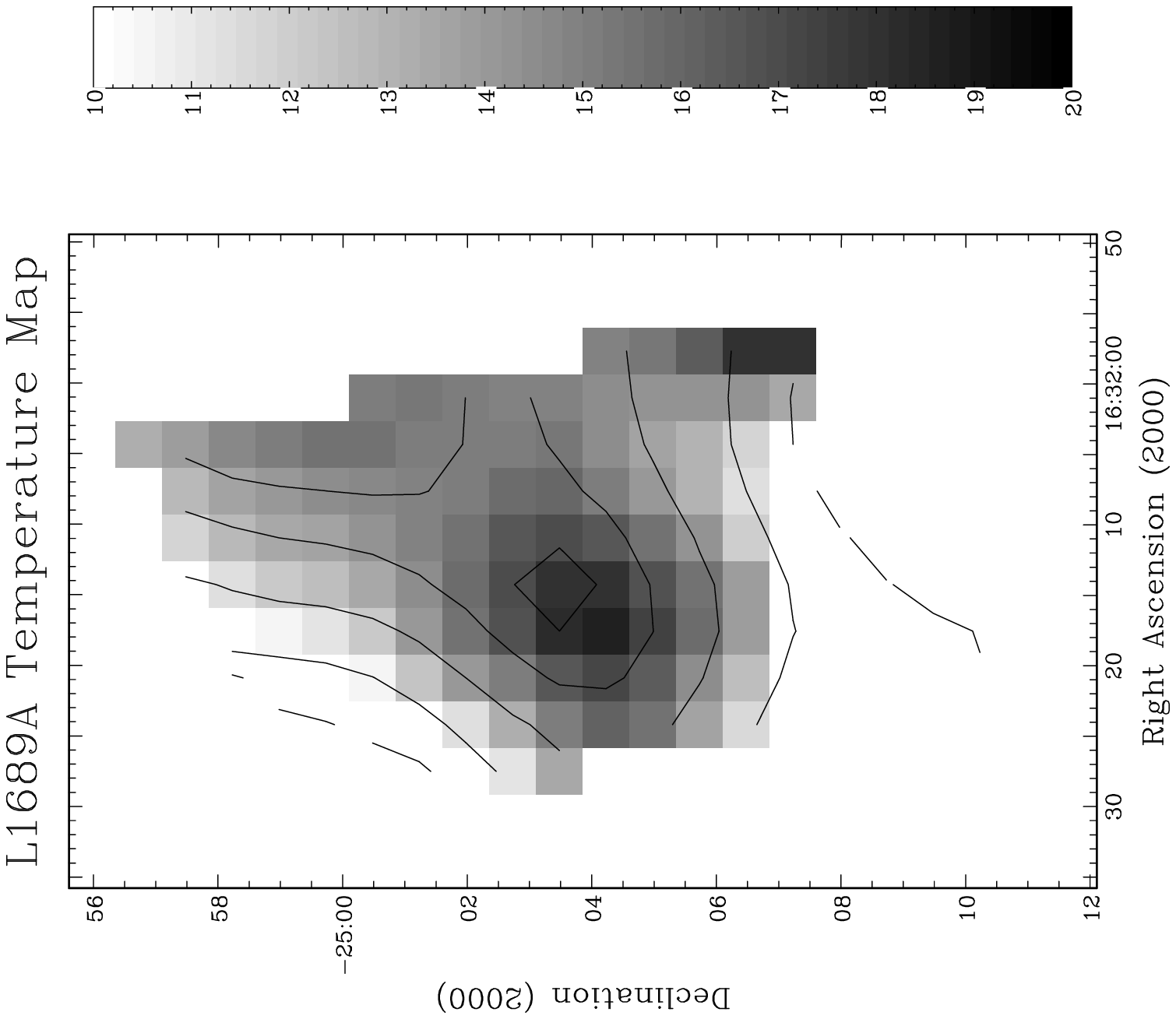}
\includegraphics{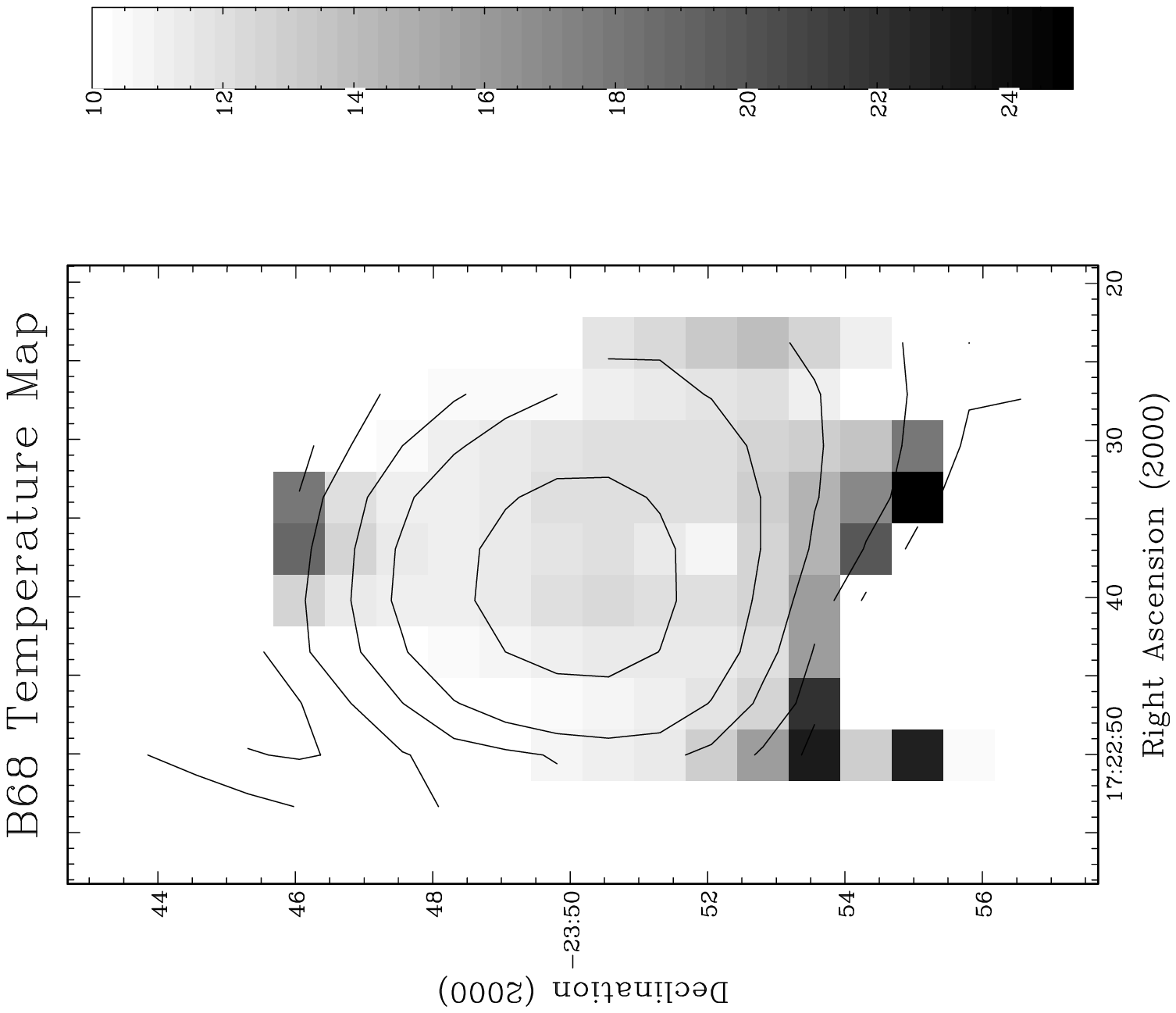}
\end{picture}
\caption{ISOPHOT 170/200-$\mu$m colour temperature grey-scale images
of L1498, L1512, L1544, L1696A, L1689A \& B68, superposed on contours of 
170-$\mu$m flux density. The grey-scale bars in each figure show the values 
of the colour temperatures in K.}
\end{figure*}

\begin{figure*}
\setlength{\unitlength}{1mm}
\begin{picture}(140,140)
\includegraphics{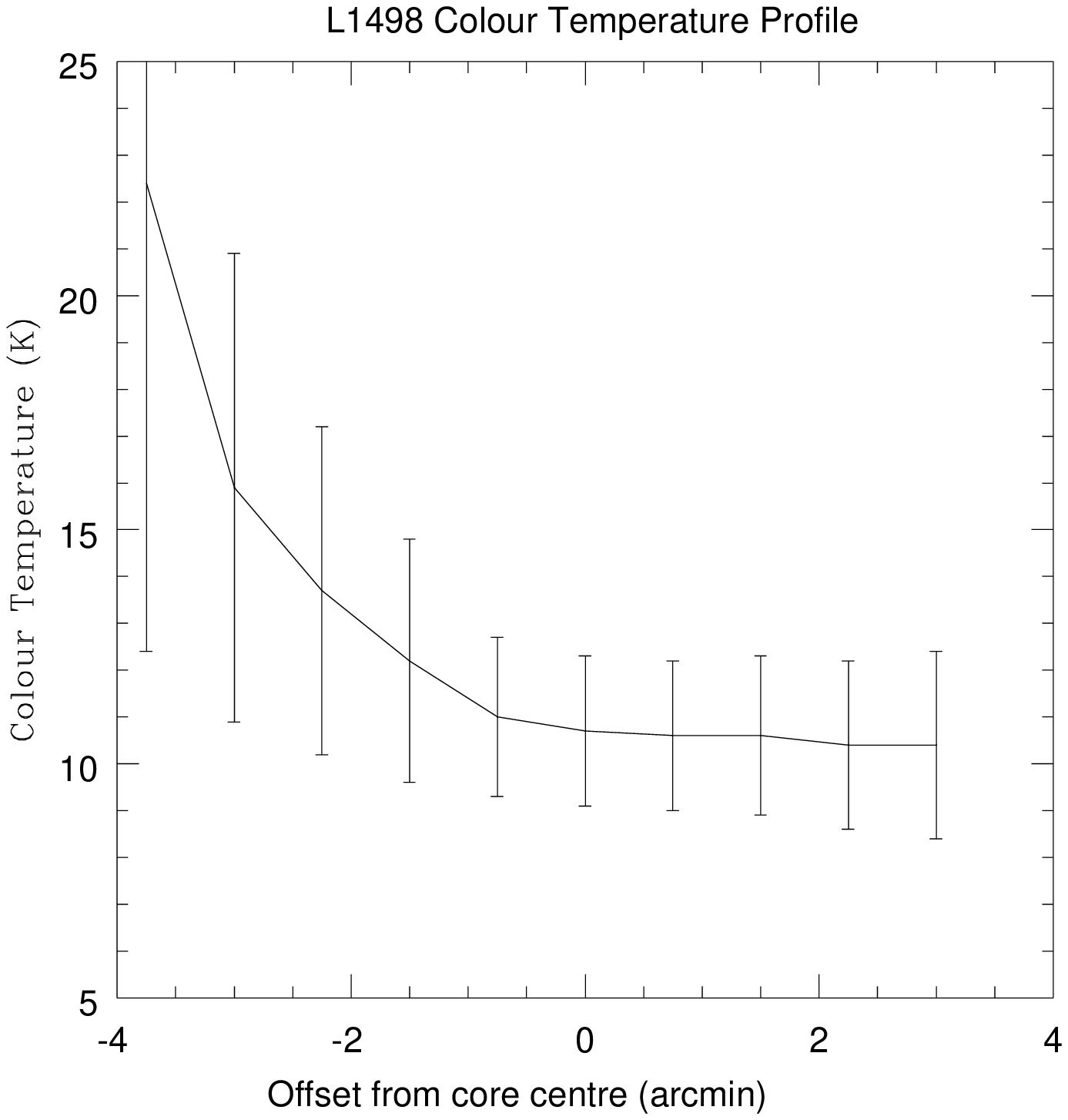}
\includegraphics{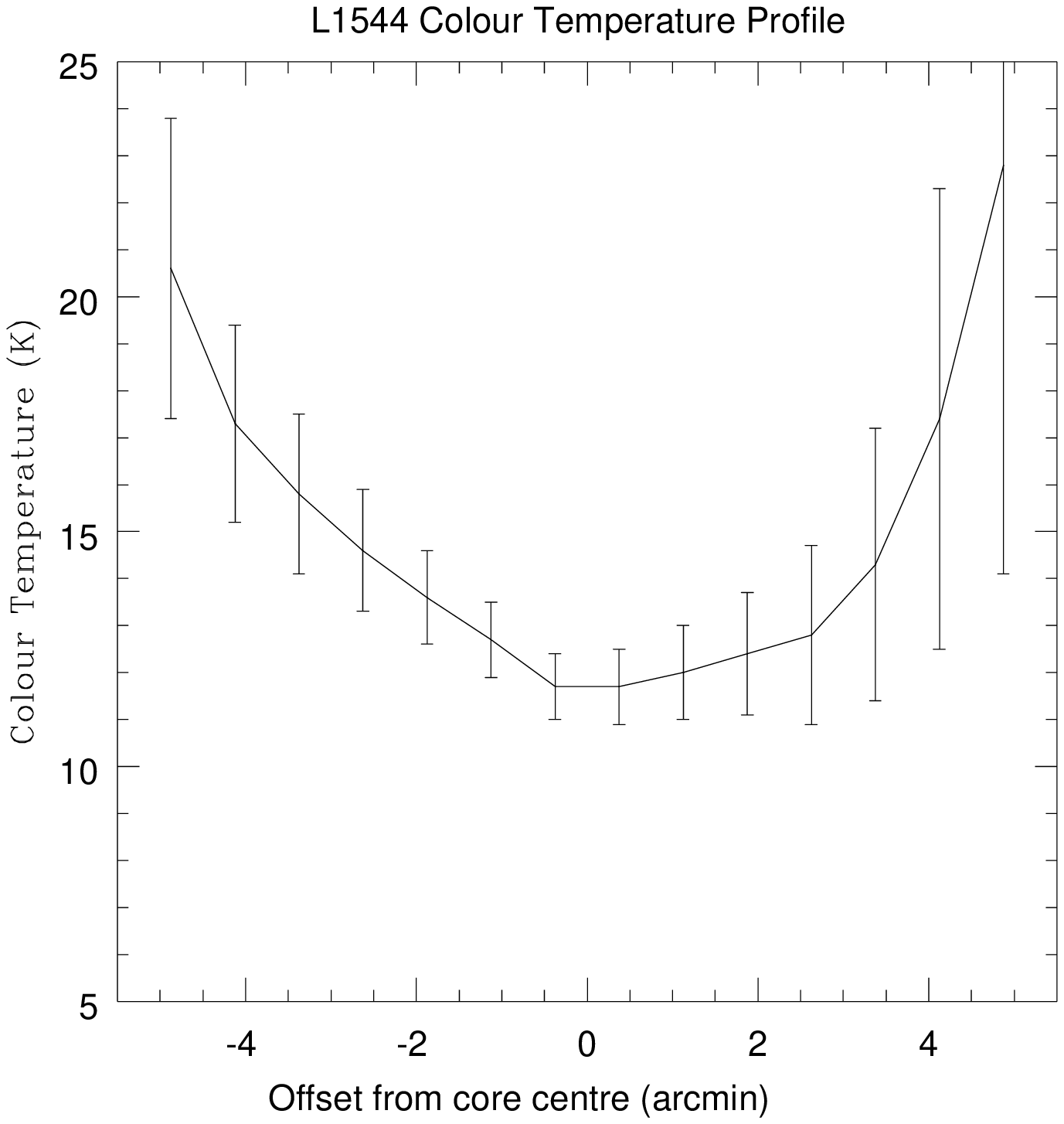}
\includegraphics{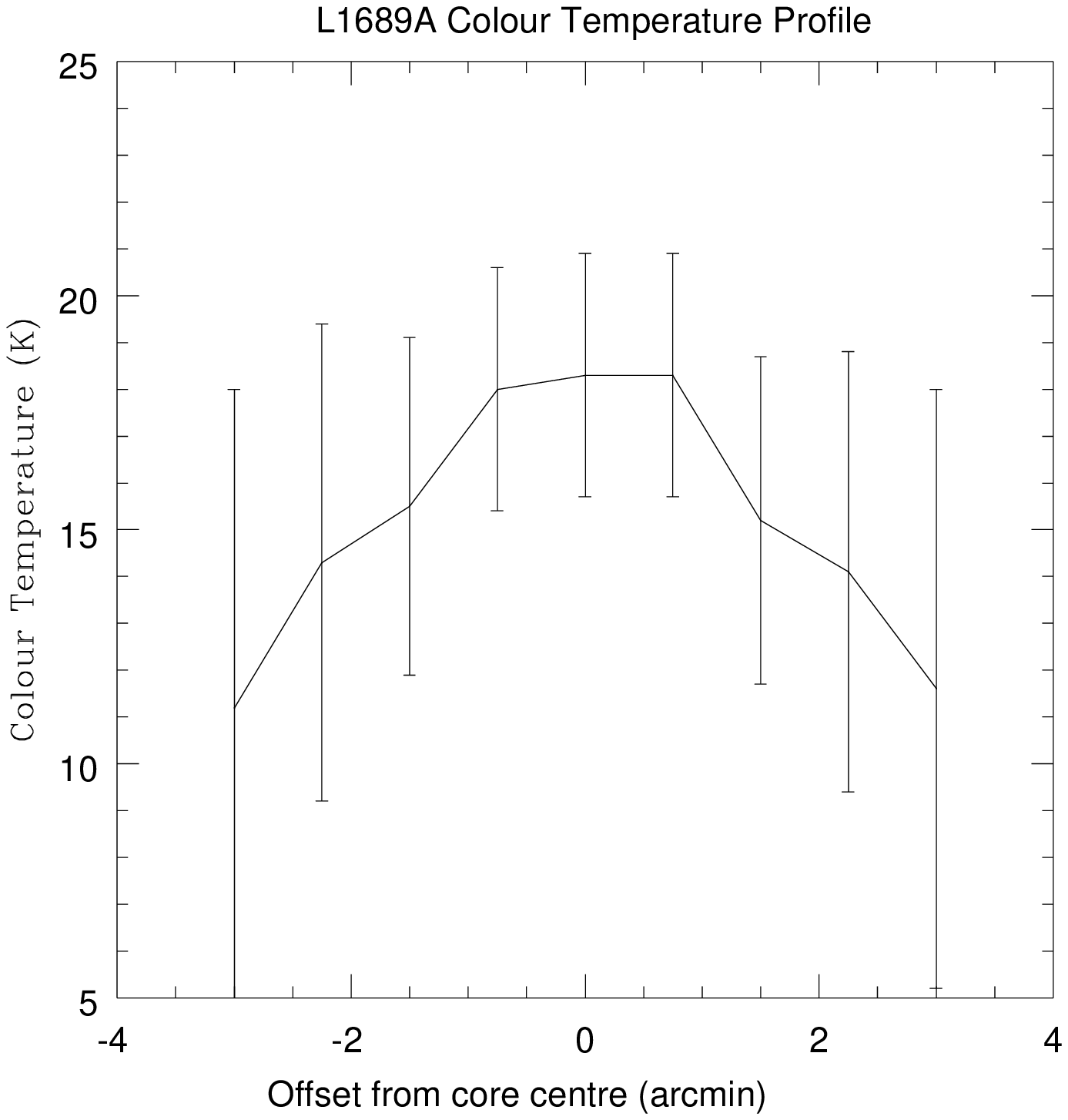}
\includegraphics{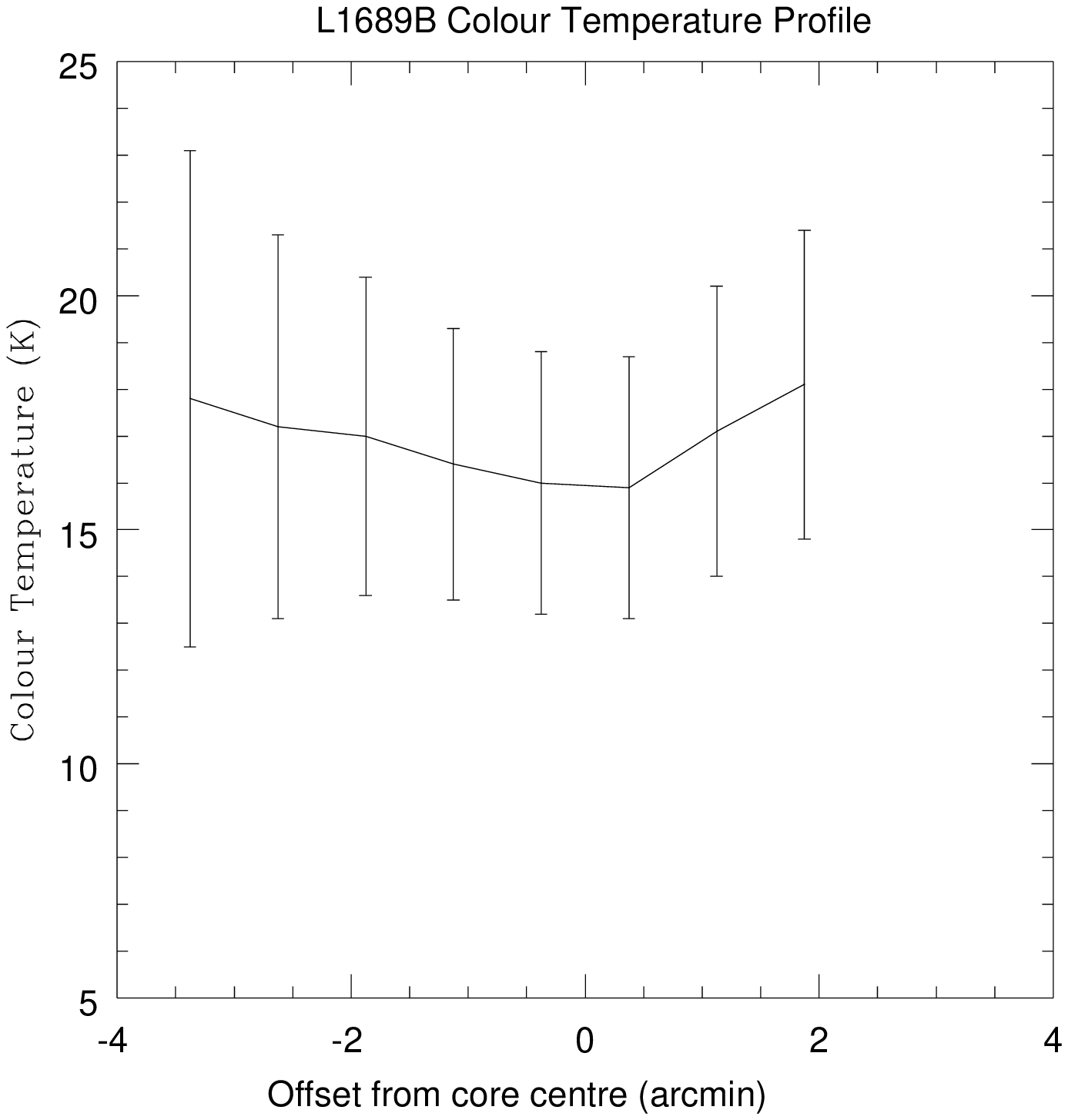}
\end{picture}
\label{fig12}
\caption{1-D colour temperature profiles through 
L1498 (NW$\rightarrow$SE), L1544 (E$\rightarrow$W), L1689A 
(SW$\rightarrow$NE) and L1689B (E$\rightarrow$W).}
\end{figure*}

\noindent
where the frequencies $\nu_{1}$ and $\nu_{2}$ correspond to
wavelengths of 200 and 170~$\mu$m respectively, and $F_{\nu_1}$ and
$F_{\nu_2}$ are the flux densities at each of these frequencies.
T is the dust temperature, $\beta$ is the dust emissivity index,
and $h$ and $k$ are the Planck and Boltzmann constants respectively.
This is one of the simplest sets of assumptions that can be made, as
it assumes that all of the dust in a 
given pixel is at a single temperature, T, and that it is
exactly the same dust which is emitting at both wavelengths.
We used $\beta$=2 to fit the data, as we
found this to be the typical value for these cores (see section 3.4 below).
Using these assumptions we constructed a series of colour temperature
maps.

Examination of these maps showed us that for 8 of the 18 cores in our
target sample there is no structure in the colour
temperature maps that could be
associated with the cores. Either there is just a uniform temperature 
across the field, or else a uniform temperature gradient
that is clearly associated with the larger scale cloud, rather
than the core itself. These 8 sources are: L1517C, L183, L1696A, L1709C,
L63, B68, B133 \& L1155C. The colour temperature map of L1689B appears
to have a saddle-like geometry, with an increasing gradient from the centre
in one direction, and a decreasing gradient orthogonal to this. We
believe this may arise due to not having mapped the whole core or
a lack of signal-to-noise off source. In the regions of high signal-to-noise
ratio, it is slightly cooler at the centre than the edge.

All of the remaining 9 cores show a temperature gradient approximately
centred on the core. 8 sources show a gradient which is cooler at
the centre than the edge.
These are L1498, L1517A\&B, L1512, L1544, L1582A, L1709A \& L204B.
Only one core appears warmer at the centre than the edge,
L1689A. We note that cores which are cooler at
their centres than their edges are qualitatively
consistent with external heating by the local inter-stellar
radiation field (ISRF).

\begin{table}
\begin{center}
\caption{Extended 
source flux densities measured in a 150-arcsec diameter aperture 
centred on the 200-$\mu$m peak position, as measured from the 
`Best est.' background-subtracted maps. These are used in combination with
mm/submm data to generate spectral energy distributions.
The upper limits and error-bars are based on background fluctuations.
Calibration errors are $\pm$30\%.}
\begin{tabular}{lccc}
\hline
Source & $S_{90}$ (Jy) & $S_{170}$ (Jy) & $S_{200}$ (Jy) \\ \hline
L1498  & $\leq$1.02  & 10.4$\pm$0.4 & 15.2$\pm$0.3  \\
L1517C &      --     & 5.79$\pm$0.4 & 11.4$\pm$2.0  \\
L1517A & $\leq$1.23  & 9.64$\pm$0.4 & 14.1$\pm$2.0  \\
L1517B & $\leq$0.78  & 5.90$\pm$0.4 & 8.39$\pm$2.0  \\
L1512  & $\leq$1.40  & 11.8$\pm$0.2 & 16.2$\pm$0.3  \\
L1544  & $\leq$0.57  & 14.9$\pm$0.2 & 18.6$\pm$0.3  \\
L1582A & $\leq$10.6  & 53.7$\pm$1.0 & 59.8$\pm$1.0  \\
L183   & $\leq$0.28  & 10.4$\pm$0.1 & 12.4$\pm$0.3  \\
L1696A & $\leq$2.07  & 38.7$\pm$0.8 & 47.5$\pm$1.0  \\
L1709A & $\leq$1.60  & 10.3$\pm$3.0 & 17.5$\pm$0.7  \\
L1689A & 34.5$\pm$1  & 125$\pm$3.0  & 117$\pm$3.0  \\
L1709C & $\leq$5.72  & 19.9$\pm$0.6 & 19.6$\pm$0.8  \\
L1689B & $\leq$0.87  & 29.1$\pm$0.8 & 29.0$\pm$0.9  \\
L204B  & $\leq$1.05  & 11.1$\pm$0.5 & 24.9$\pm$0.9  \\
L63    & $\leq$2.69  & 19.2$\pm$0.5 & 28.5$\pm$0.4  \\
B68    & $\leq$2.79  & 12.0$\pm$0.2 & 15.8$\pm$0.4  \\
B133   & $\leq$4.46  & 22.6$\pm$0.5 & 34.6$\pm$0.7  \\
L1155C & $\leq$0.69  & 8.12$\pm$0.1 & 7.42$\pm$1.0  \\ \hline
\end{tabular}
\end{center}
\end{table}

Figure 11 shows colour temperature maps of six cores 
as grey-scale images with contours from the 170-$\mu$m background-subtracted
(`Best est.') 
flux density images superposed. The values of the colour temperatures
were not calculated above 50K, as the 170- to 200-$\mu$m flux density
ratio is not sensitive to temperatures in this range. Typically values
of this sort were only found at the edges of the images, where the
signal-to-noise ratio was low, and hence such values were cut out of the
images in Figure 11.

We show one example of a uniform temperature core, B68,
although we note that even this source has tentative evidence for being
warmer at its edge. We show
one example of a large-scale temperature gradient, L1696A.
We note that this source has a temperature peak to the northwest of
the core. This may be due to asymmetric heating by nearby young stars
causing one side of the core to be warmer than the other.
We illustrate 3 cores that are cooler at the centre 
than the edge, L1498, L1512 \& L1544 (we note that in the
case of L1498 we only see the temperature gradient on one side
of the core due to only having partial maps of the region).
We also show the core that appears warmer at the centre than the
edge, L1689A. Thus these six sources illustrate the variation in
the colour temperature maps that we see.

The fact that none of the cores except L1689A
shows a strong central temperature peak 
indicates that none (except possibly L1689A)
has a central protostar yet, as was suspected, and
hence they are confirmed as pre-stellar in nature. The cores which are
clearly cooler at their centres than their edges show the behaviour
one would qualitatively
expect for externally heated cores. The remainder are consistent
with being in isothermal equilibrium. We also note that
the only core with a strong temperature gradient where the centre is warmer
than the edge, L1689A, is also the only core with a clear 90-$\mu$m 
detection. Therefore L1689A is clearly different from the rest of our sample.
It may be more evolved than the other cores and contain an embedded protostar.
We think this is unlikely since its submillimetre flux densities are lower
than some of the other cores, such as L1689B (Paper I).
Alternatively, it may simply be warmer due being more diffuse, and hence
more prone to heating throughout the core by the surrounding environment.

\begin{figure*}
\setlength{\unitlength}{1mm}
\begin{picture}(200,200)
\includegraphics{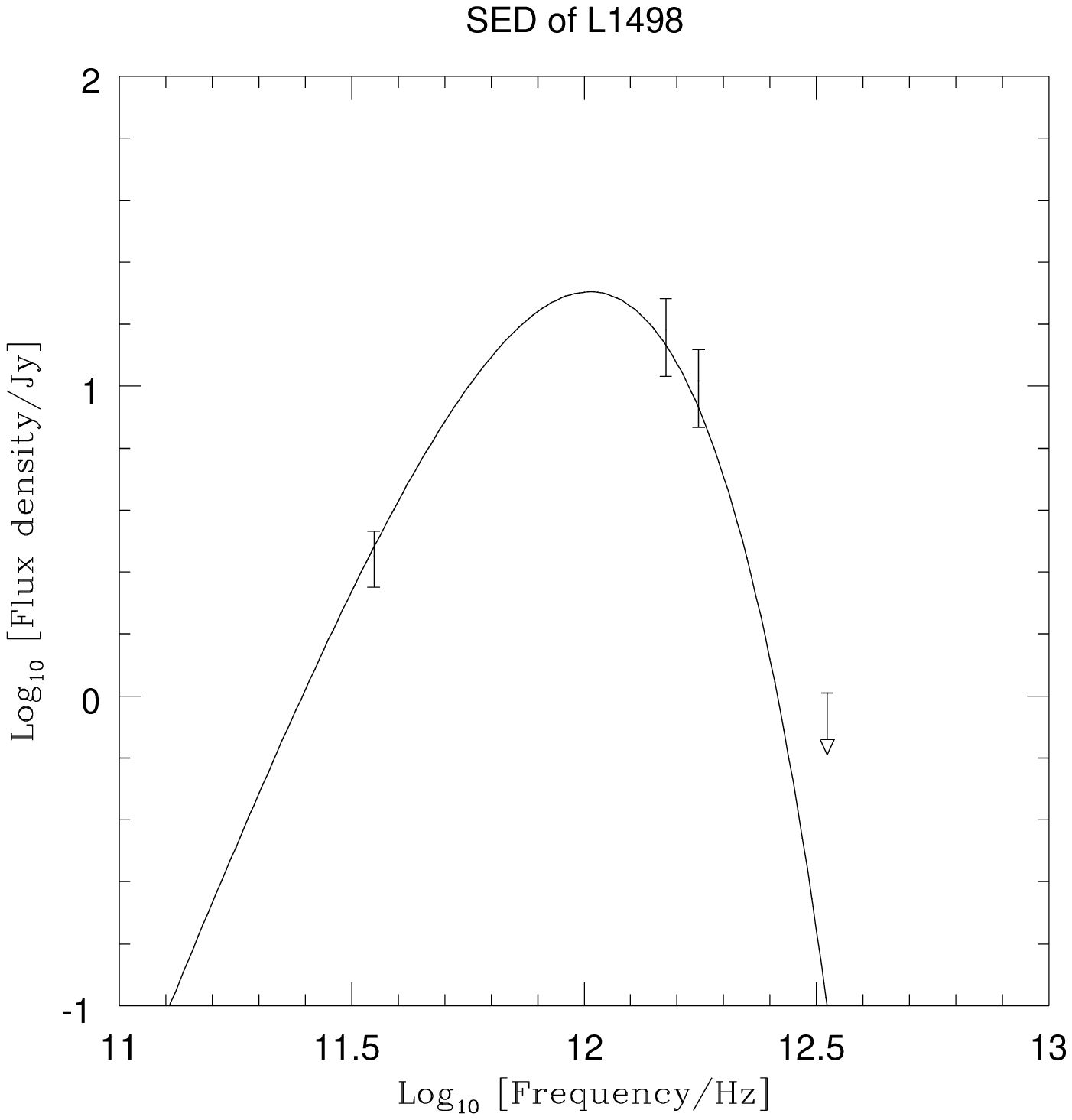}
\includegraphics{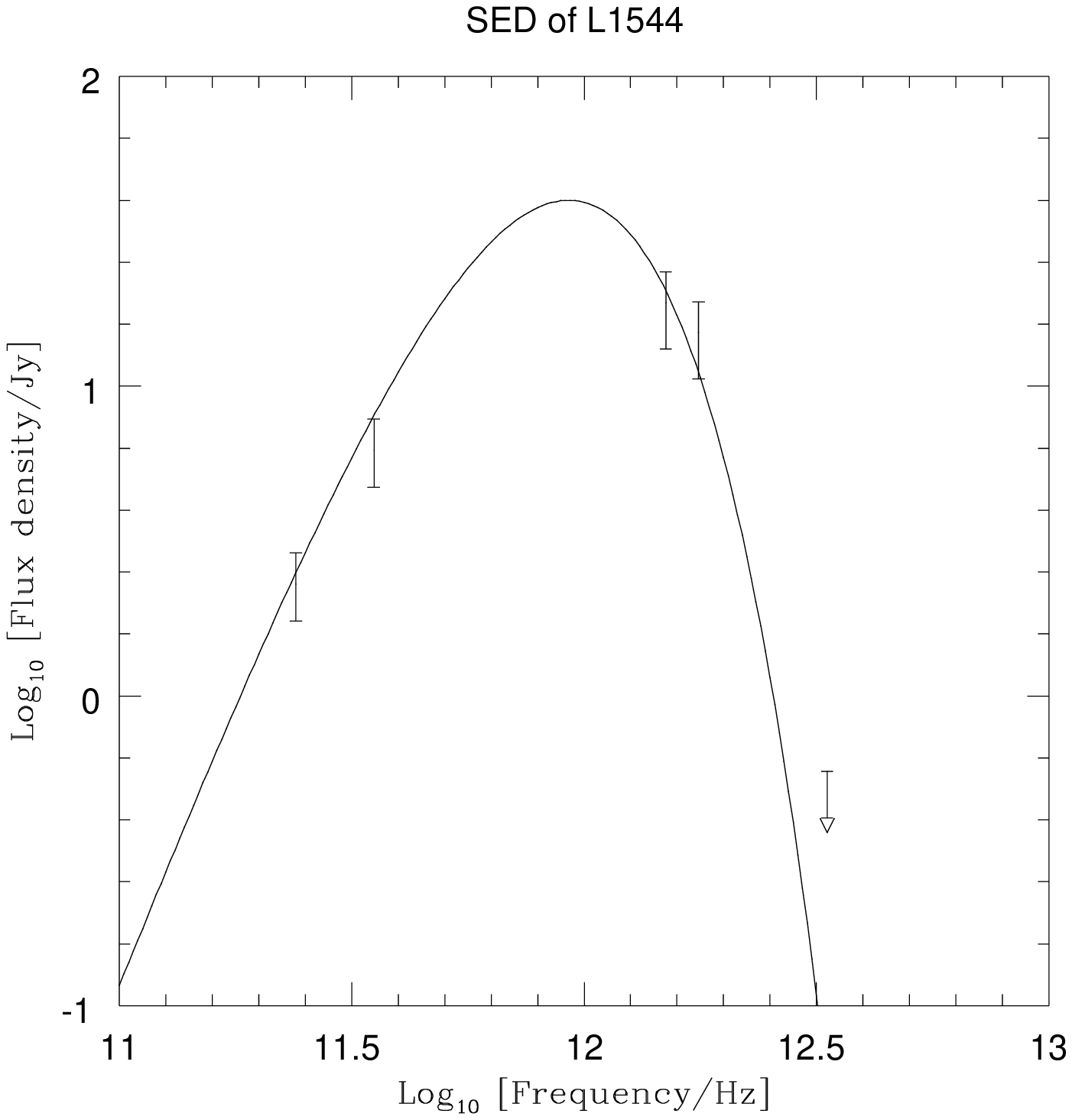}
\includegraphics{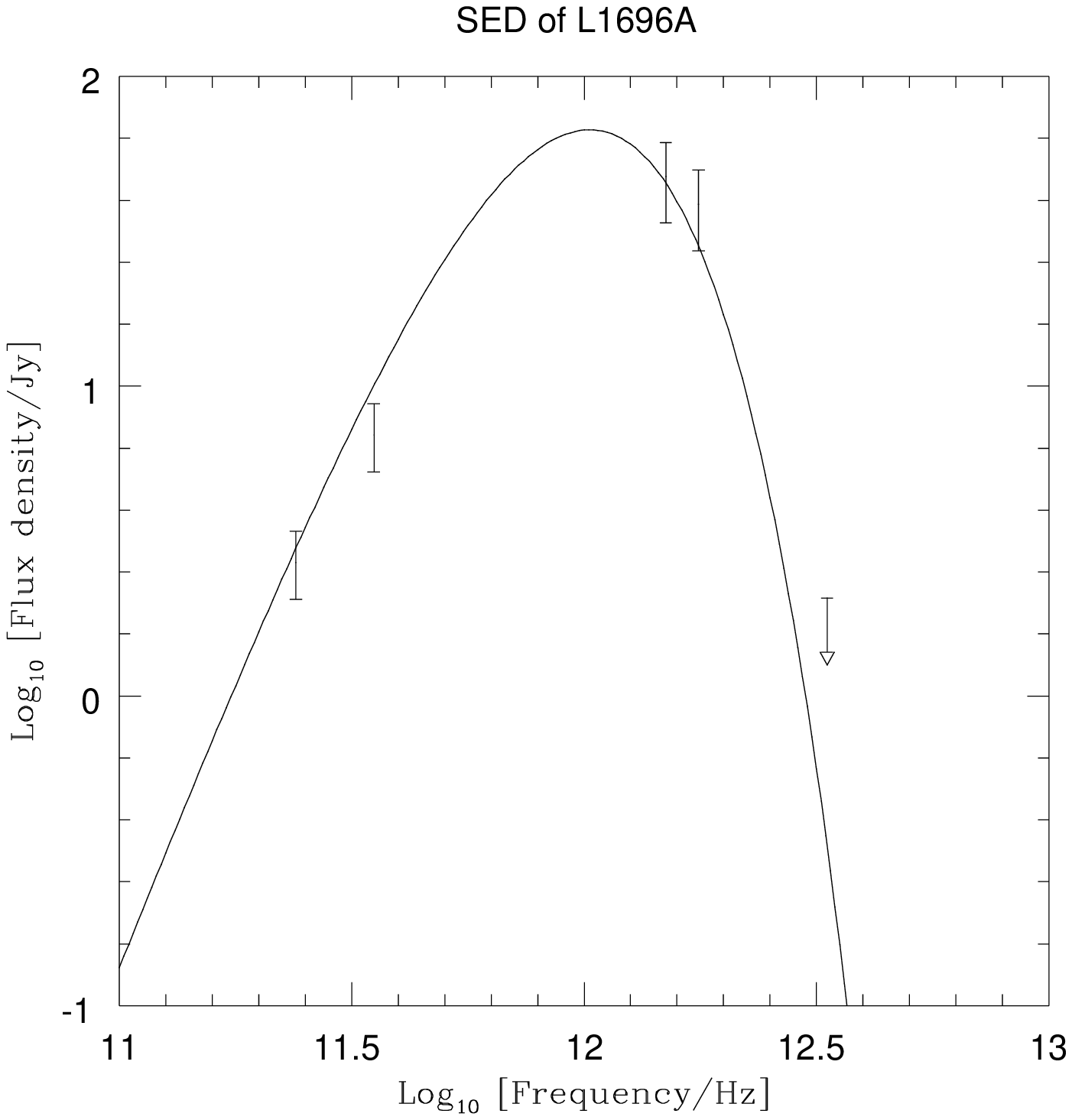}
\includegraphics{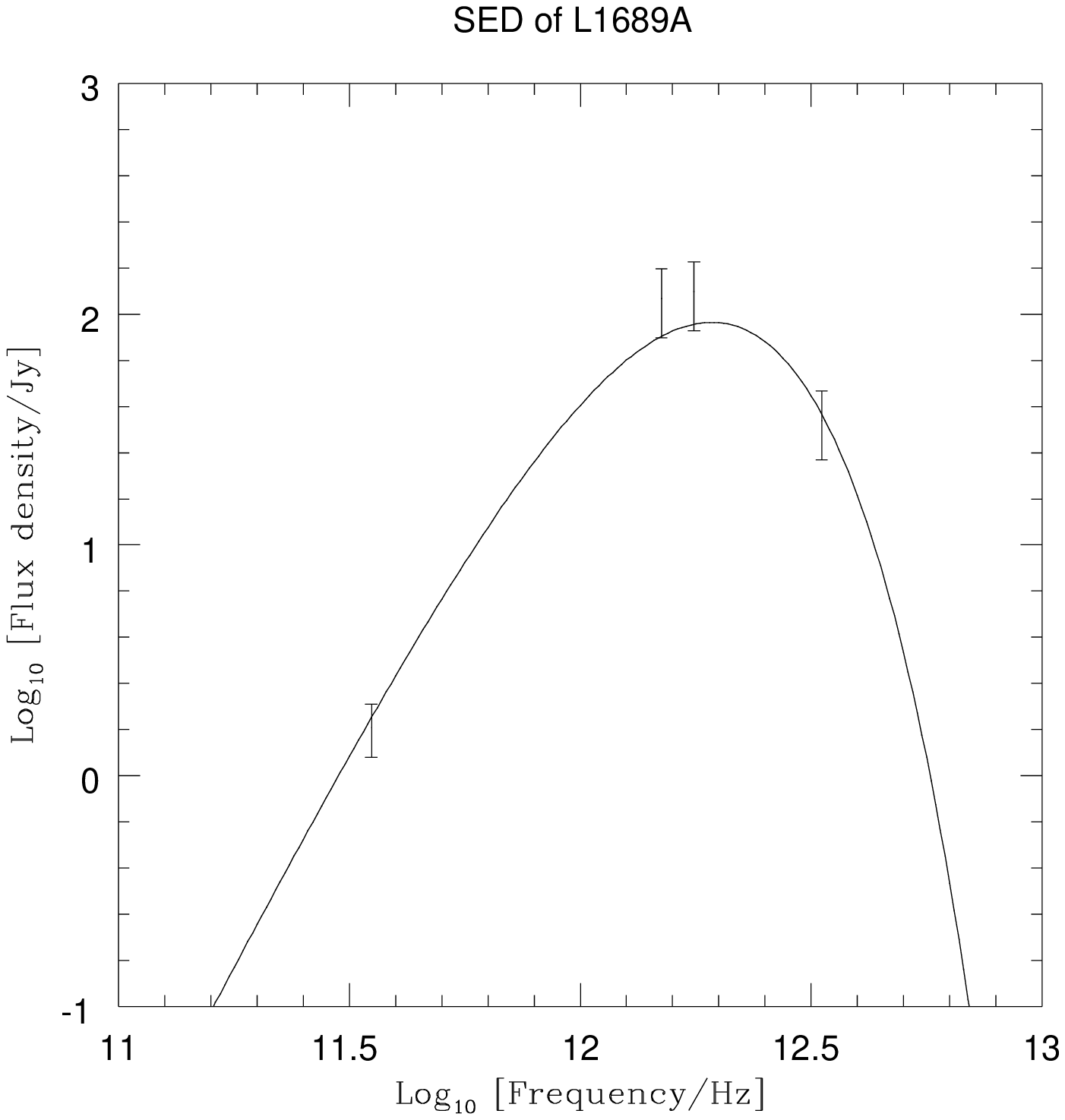}
\includegraphics{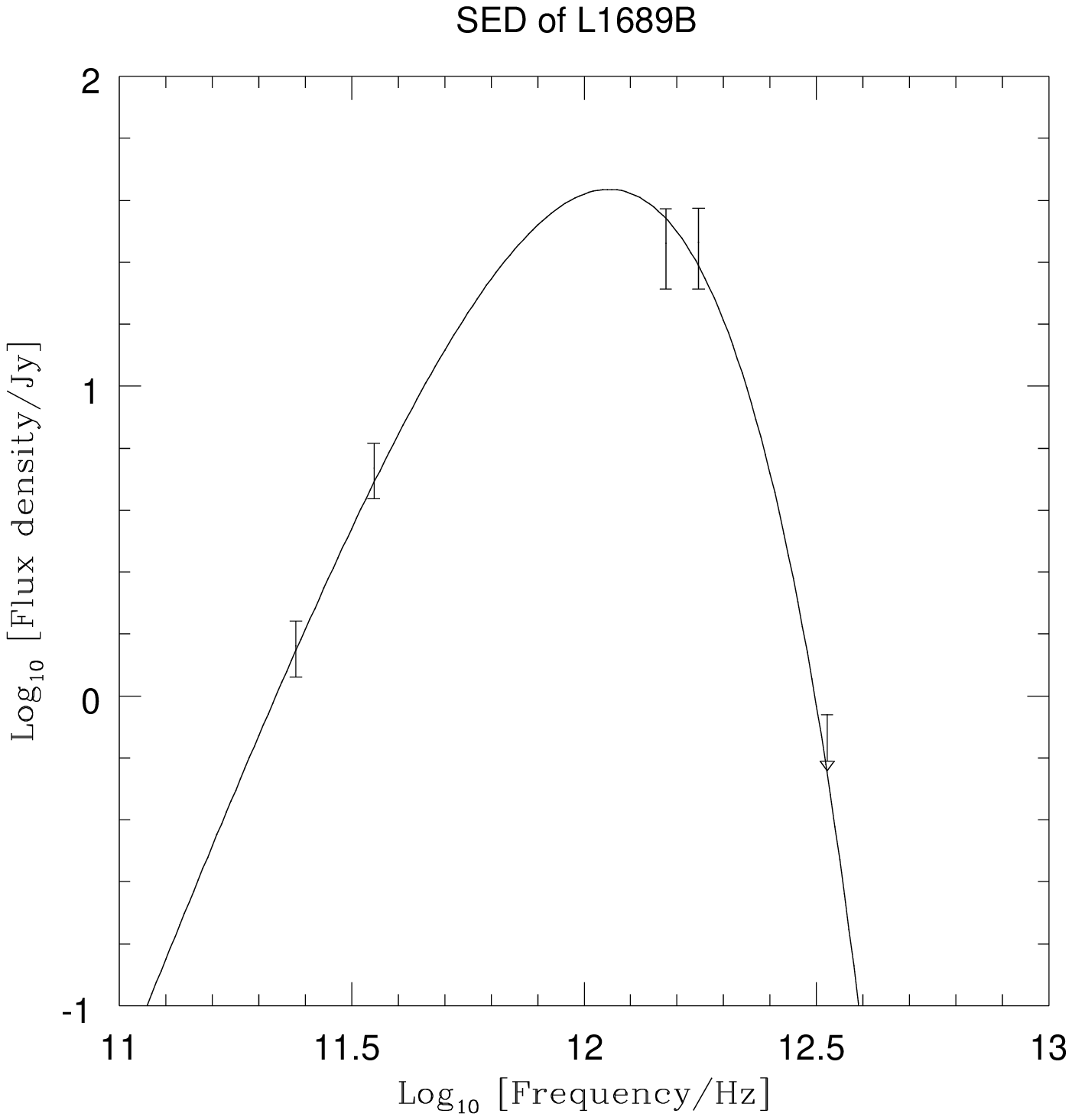}
\includegraphics{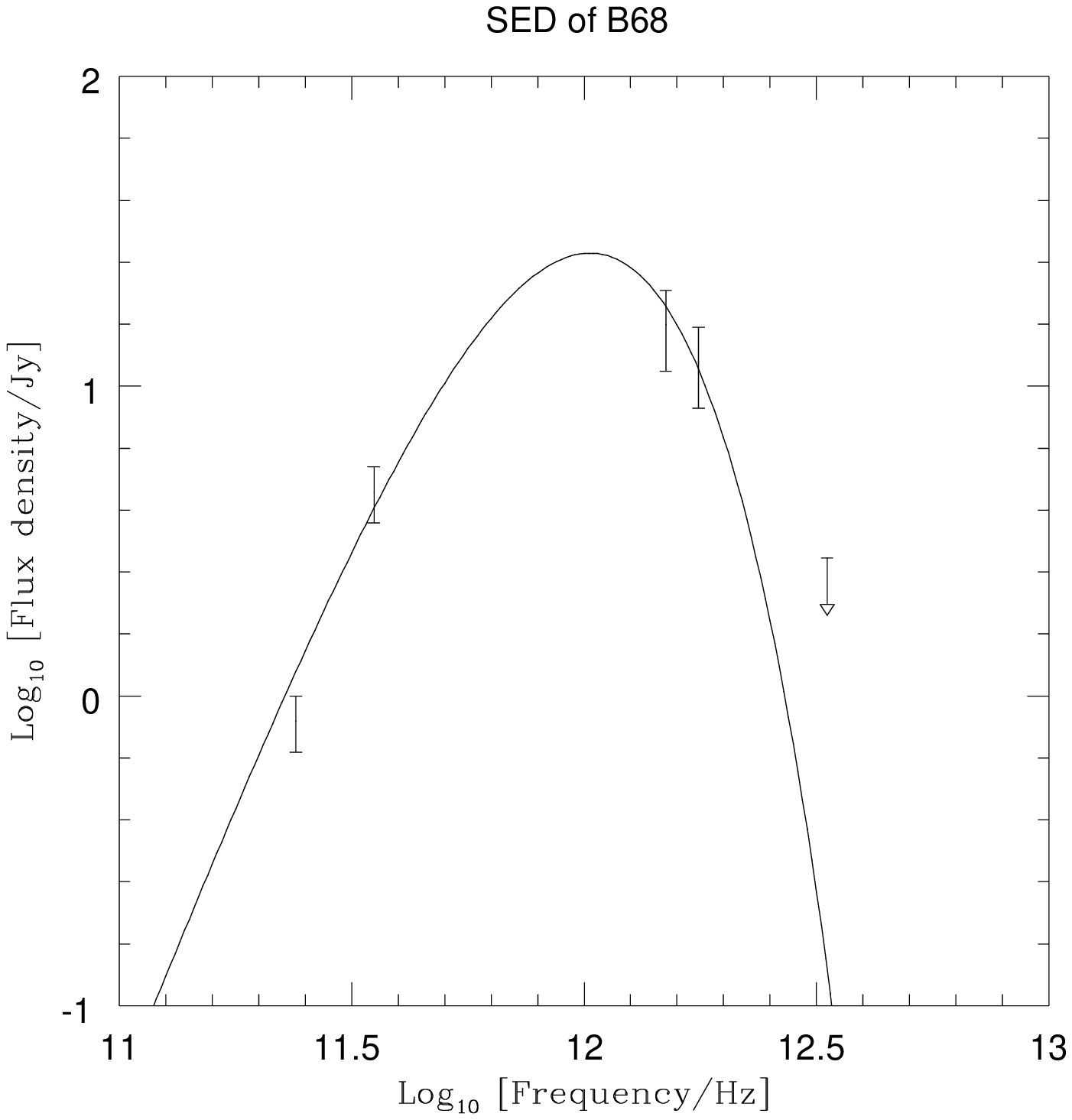}
\end{picture}
\caption{Spectral energy distributions of the cores L1498,
L1544, L1696A, L1689A\&B and B68. See text for discussion.}
\end{figure*}

Figure 12 shows cuts through the colour temperature maps, with error-bars,
of three cores that appear cooler at their centres,
L1498, L1544 and L1689B, and the core that is warmer at the centre, L1689A.
The difference in appearance between the colour temperature profiles of 
L1544 and L1689A is most marked. The temperature gradients we observe are
larger than our calculated error-bars for all except L1689B, which
shows the least temperature variation of the four. In all cases there
appears to be a central region of roughly uniform temperature.
In the case of L1498 
the core appears to lie at the end of a ridge extending to the southeast,
and the temperature appears uniform along this rige, rising at the
northwestern side of the core. Of the 9 cores (including L1689B) which 
appear cooler at their centres, 4 have gradients larger than the error-bars,
L1498, L1544, L1512 and L1582A. The core that is warmer at its centre,
L1689A, also has a gradient larger than the error-bars.
All of the remaining cores are consistent with being isothermal to within
a 1$\sigma$ error of typically 2--4~K.

\subsection{Spectral energy distributions and luminosities}

We can combine the ISOPHOT data presented in this paper with data taken at 
other wavelengths to calculate the spectral energy distribution (SED) of
some of these pre-stellar cores. We also have millimetre and
submillimetre data (Papers I--III, 
Kirk et al. in prep.) of some of the cores. These data only cover the
central region of each core, in which even those cores with
temperature gradients at their edges appear isothermal. Thus we used the
isothermal approximation in modelling the SEDs of the cores, and
concentrated on those cores which have been best studied.

We measured the flux density on the
best estimate background-subtracted images in 
a circular aperture of diameter 150 arcsec
at each wavelength for each of our sources, to match the largest
aperture for which we have mm/submm flux densities available,
and list the measured flux densities in Table 4.
We compared these with the
flux densities measured at millimetre (Papers II \& III) and submillimetre 
wavelengths (Paper I; Kirk et al. in prep.) measured in the same size 
apertures.
We plot these data on a logarithmic scale, as a function of frequency, 
in Figure 13, for the sources L1498, L1544, L1696A, L1689A\&B and B68.

Also plotted on Figure 13, are a series of modified black-body, or grey-body,
curves of the form:

\[ 
F_{\nu} = B_{\nu,T} (1 - e^{-\tau_\nu}) \Omega
\]

\noindent
where $F_{\nu}$ and
$B_{\nu,T}$ are the flux density and
black-body emission respectively, at frequency $\nu$.
T is the dust temperature,
$\Omega$ is the solid angle of the source
and $\tau_\nu$ is the optical depth at 
frequency $\nu$, which is usually given in the form 
$\tau_\nu\propto\nu^\beta$. The solid curves shown in Figure 13 all have
$\beta$=2 and $\tau_{200\mu m}$=0.06,
and have dust temperatures of T=10K for L1498,
T=9K for L1544, T=10K for L1696A, T=19K for L1689A, T=11K for L1689B, and
T=10K for B68. We note that $\tau_{200\mu m}$=0.06, making
typical assumptions about dust mass opacity (e.g. Hildebrand 1983),
corresponds to a central column density of
N$^{Central}_{H_2}$=7$\times$10$^{22}$cm$^{-2}$. This is
consistent with the values we estimate in these cores (see Table 5).

We estimate that
the error-bars on the temperatures are typically $\pm$3~K.
These were calculated by measuring the flux densities in a 150-arcsec
aperture for each source at each wavelength on the images derived from
the B1 and B2 background subtractions, since this represents the dominant
source of error. Then grey-body curves were fitted through these flux
densities and their
temperatures were derived. In this way the maximum possible 
shift in temperature was calculated, and found to be around $\pm$3~K.
We note that the hottest
core, L1689A, is the only core where we see a temperature gradient
that decreases towards the edge. It is therefore obvious that L1689A is
different from the other cores in more than one way, and its status is unclear.

\begin{table*}
\begin{center}
\caption{Source properties of some of the best studied cores. The SED
temperature comes from fits such as those
in Figure 13. The temperature gradient
was derived from studies such as those shown in
Figures 11 and 12, and is marked `in' if the centre is
warmer than the edge, `out' if the centre is cooler, and `0' if there is
no gradient. The output luminosity $L_{out}^{SED}$ is measured by integrating
under grey-body fits such as those shown in
Figure 13. The input luminosity $L_{in}^{ISRF}$ is
derived from the estimated local radiation field incident upon the 
core (see text). The central column
density N$^{Central}_{H_2}$ is taken from
Paper III, Bacmann et al. (2000) \& Kirk et al. (in prep). 
The column headed ISOCAM absorption indicates whether the cores were
seen in absorption by ISOCAM (Bacmann et al. 2000).
The typical error-bars on the temperature 
estimates are $\pm$3~K (see text for details).}
\begin{tabular}{lccccccc}
\hline
Core & SED & Colour Temp. & $L_{out}^{SED}$ (L$_\odot$) & 
$L_{in}^{ISRF}$ (L$_\odot$) & Assumed & N$^{Central}_{H_2}$ & ISOCAM \\ 
Name & Temp. (K) & Gradient & (150-arcsec) &  (150-arcsec) & Distance (pc) 
& (cm$^{-2}$) & Absorption \\ \hline
L1498  & 10 & out & 0.1 & 0.06 & 140 & 5$\times$10$^{22} $ & $-$ \\ 
L1517B & 10 & out & 0.08 & 0.05 & 140 & 4$\times$10$^{22}$ & Y \\ 
L1544  & 9 & out & 0.2 & 0.1 & 140 & 6$\times$10$^{22}$ & Y \\ 
L1582A & 15 & out & 4.0  & 1.5 & 400 & 3$\times$10$^{22}$ & Y \\
L183   & 10 & 0  & 0.15 & 0.05 & 150 & 7$\times$10$^{22}$ & $-$ \\
L1696A & 10 & 0  & 0.3 & 0.6 & 140 & 7$\times$10$^{22}$ & Y \\ 
L1689A & 19 & in & 0.8 & 0.5 & 140 & 2$\times$10$^{22}$ & N \\ 
L1689B & 11 & out & 0.3 & 0.5 & 140 & 6$\times$10$^{22}$ & Y \\ 
L63    & 11 & 0 & 0.2 & 0.06 & 160 & 5$\times$10$^{22}$ & $-$ \\ 
B68    & 10 & 0 & 0.3 & 0.2 & 200 & 3$\times$10$^{22}$ & N \\ 
B133   & 13 & 0 & 2.3 & 0.6 & 400 & 4$\times$10$^{22}$ & N \\ 
\hline
\end{tabular}
\end{center}
\end{table*}

The total far-infrared and mm/submm
luminosity radiated by each core, $L_{out}^{SED}$, 
can be calculated by integrating under 
the solid curves of Figure 13. We list these luminosities 
in Table 5. We also list the central column density (e.g. Paper III;
Bacmann et al. 2000) and assumed distance of each of the cores, as well
as the direction of the temperature gradient derived above, and details
of whether the core was seen in absorption by ISOCAM (Bacmann et al. 2000).
The incident luminosity on each core is listed in column 5 of Table 5,
and we now address the manner in which this was calculated.

\section{Energy budget of the cores}

If the cores in our sample are truly starless, the total luminosity
emitted by each core is not expected to exceed the amount of radiation
incident upon the core. The present ISO results allow us to check 
this quantitatively for the first time in these cores.
We can indeed compare the measured output SED luminosities ($L_{out}^{SED}$)
with estimates of the input luminosities ($L_{in}^{ISRF}$) provided by 
the local inter-stellar radiation field (ISRF). 
The local ISRF directly incident on each core is not precisely known but 
can be roughly estimated from the mid-IR and far-IR 
backgrounds observed towards each core by ISOCAM (Bacmann et 
al. 2000) and ISOPHOT (this paper). 

The radiation field within a molecular cloud is believed to consist of 
two main components (Mathis, Mezger, \& Panagia 1983): 
one component that we shall refer to as the
`near-IR' component, since it peaks at $\sim $~1--5~$\mu $m,
(but note that this also includes the optical and UV radiation);
and another component that we shall refer to as the
`far-IR' component,
since it peaks at $\sim $~100--200~$\mu $m. 

For simplicity, 
we assume that the spectral shapes of these components are the same as 
for the average ISRF in the solar vicinity 
(see figure~1 of Mathis et al. 1983). However, 
we can allow for the fact that the strengths of the two ISRF components 
vary from core to core and sometimes differ substantially from the strengths 
estimated by Mathis et al. for the average solar ISRF (outside molecular 
clouds), which are: $I_{ISRF}^{Mathis} (7\, \mu {\rm m}) \sim 0.1 $~MJy/sr 
and $I_{ISRF}^{Mathis} (200\, \mu {\rm m}) \sim 6 $~MJy/sr. 

The mid-IR
Zodiacal emission was taken from Bacmann et al. (2000). The 200-$\mu$m
Zodiacal emission was estimated to be typically only $\sim$1.5MJy/sr
from a model based on COBE data (c.f. Kelsall et al. 1998), which is
much smaller than the values listed in Table 2.
In fact we see that
the near- to mid-IR and far-IR backgrounds 
(after subtraction of the Zodiacal foreground emission) measured by ISO 
towards the cores are indeed significantly larger than the Mathis et al. 
values in all cases. 

This is not surprising since the mid-IR background 
emission is thought to arise from very small (PAH-like) grains excited 
by the local far-UV radiation field in the outer ($A_V \simlt 1$) envelopes of 
molecular clouds (e.g. Bernard et al. 1993), and the far-IR radiation 
field deep inside a cloud is known to be much stronger than the Galactic 
far-IR radiation field (Mathis et al. 1983). Furthermore, in 
the $\rho$~Oph complex, the external far--UV field is estimated to 
be a factor of $\sim $~10--100 times 
brighter than the average far--UV field in the solar 
neighbourhood (Liseau et al. 1999). 

To obtain a realistic estimate of the 
local ISRF around each core, we have scaled the strengths of the 
near-IR and far-IR components of the Mathis et al. radiation field according 
to the observed values of the mid-IR and far-IR backgrounds (c.f. table~2 of 
Bacmann et al. 2000 and Table~2 of this paper). We also account  
for the fact that a substantial fraction of the near-IR 
(plus UV and optical) component is absorbed
by the parent molecular cloud before reaching the surface of the core, 
which is at a typical depth of $A_V^{depth} \sim $~2--10. 

We thus assume that the effective radiation field incident on the core is:

\[ I_{ISRF}^{eff} (\lambda) = 
[I_{back}^{ISOCAM} (7\, \mu {\rm m})/I_{ISRF}^{Mathis} (7\, \mu {\rm m})] \]

\[ \hspace*{3cm} \times I_{ISRF}^{Mathis} (\lambda) \times 
{\rm exp}(-\tau^{depth}_{\lambda}) \]

\noindent
from 0.09~$\mu $m to 20~$\mu $m, and:

\[ I_{ISRF}^{eff} (\lambda) = 
[I_{back}^{ISOCAM} (200\, \mu m)/I_{ISRF}^{Mathis} (200\, \mu m)] \]

\[ \hspace*{3cm} \times I_{ISRF}^{Mathis} (\lambda) \] 

\noindent
between 20~$\mu $m and 1000~$\mu $m. 
Adopting a spherical core geometry, we compute the effective ISRF flux 
density absorbed by the core as:

\[ F_{ISRF}^{abs} (\lambda) = 2\pi\, I_{ISRF}^{eff} (\lambda) 
\int_{0}^{1}\, [1-{\rm exp}(-\tau_{core}(\mu,\lambda))]\mu\, d\mu , \]

\noindent
where $ \mu = cos \theta $ is the cosine of the angle that a ray makes with 
the normal to the surface of the core, and $\tau_{core}(\mu,\lambda)$ is 
the core optical depth along that ray (see Lehtinen et al. 1998). 
The integral over angles is calculated numerically using the typical 
radial column density profile measured by Bacmann et al. (2000).

For example, in the case of L1689B,
figure 5(e) of Bacmann et al. (2000) 
indicates a column density $N_{H_2}\sim 2\times 10^{22}$~cm$^{-2}$
(i.e. A$_V\sim$20) at an angular radius of 75~arcsec
from the core centre. Assuming that the 
core is embedded roughly
mid-way through the cloud, this leads to a depth A$_V\sim$10
from the cloud surface to the core surface at 75~arcsec radius.
$\tau^{depth}_{\lambda}$ is then
derived from this A$_V$ using the extinction curve of Mathis et al. 
(1983 -- c.f. their appendix C).

The total input luminosity provided to the core by the effective 
incident radiation field is then:

\[ L_{in}^{ISRF} = 4\pi\ R_{core}^2 \, 
\int_{0.09\mu m}^{1000\mu m}\, F_{ISRF}^{abs} (\lambda) d\lambda, \]

\noindent
where $R_{core} $ is taken to be the radius corresponding to a 150$\arcsec$
diameter aperture.

The values of $L_{in}^{ISRF}$ estimated in this way are listed in Table 5. 
However, note that the spectrum of the ISRF within molecular clouds
that we have used in the above calculations is not 
accurately known (Mathis et al. 1983), as stated above,
and dust grain absorption
properties are extremely uncertain. 

Similarly, the radial density
distributions of the cores is not well known in all cases.
For three cores (L1544, L1696A, L1689B), $L_{in}^{ISRF}$ was estimated 
using the radial density structure measured by Bacmann et al. (2000). 
For the remaining cores the exact radial profile was not so clear,
and a typical structure matching the measured central column density 
in Table~5 was assumed. The values of $L_{in}^{ISRF}$ are more uncertain in 
these cases. 

The density profiles are crucially important in determining the exact
fraction of the ISRF that is absorbed by the cores. For example, in the
case of L1689B in which we believe we know the density profile quite well,
we calculate that only $\sim$10\% of the incident flux is absorbed. In
this case we see excellent agreement between our calculated value of
$L_{in}^{ISRF}$ and our measured value of $L_{out}^{SED}$. For some of the
other cores, given all of the uncertainties,
we estimate that the luminosity values
in Table~5 may be uncertain by at least a factor of $\sim 2-4$. 

A detailed submillimetre survey of the cores would help to remove some
of the uncertainties associated with the density structure. Nevertheless,
it is encouraging to see that $L_{in}^{ISRF}$ is found to be 
similar to $L_{out}^{SED}$ in most cases. 

We therefore conclude that the cores are generally
consistent with heating
by the local ISRF, and require no additional central protostellar 
heating sources. 
Note that this is true even for those pre-stellar cores, such as L1544,
which already have evidence for infall motions
(Tafalla et al. 1998). This might be expected, according to some models, since
the compressional luminosity due to the work done by the gravitational forces
is indeed estimated to be negligible ($\sim 10^{-3} \, L_\odot$) as long as
the collapse remains roughly isothermal (e.g. Henriksen 1994).
However, it is not consistent with models that require a significant
protostar to have formed at the centre of collapsing cores such as L1544.
Thus the cores are all consistent with being pre-stellar in nature.

\section{Conclusions}

We have presented data taken with ISO, using the long wavelength 
photo-polarimeter
ISOPHOT, of a series of starless cores. We have measured their flux
densities and colour temperatures at 90--200$\mu$m. 
The higher resolution and longer wavelength coverage of ISO compared
to IRAS have allowed us to make colour temperature maps of 
the cores. We have found that most of the cores are
either isothermal, or have a temperature gradient that is cooler at the
centre than the edge. 

We have fitted spectral energy distributions to the cores that allow
us to ascribe dust temperatures to the densest regions of the cores.
Subsequent work will be able to use these temperatures to calculate
other properties of the cores, such as their masses.
We have also used the spectral energy distributions to calculate the
luminosities of the cores.

We have compared our ISOPHOT images of emission from cool dust in the
cores with our previously published ISOCAM images of absorption by the
same dust. We found that
by comparing the output luminosity of each core with an
estimate of its energy input from the local ISRF,
we can conclude that the cores are generally
consistent with being pre-stellar in
nature and externally heated, with no need for a central heating 
protostellar source.

\section*{Acknowledgments}

The authors wish to thank the ISOPHOT team, and in particular Martin Haas,
at the Max Planck Institut fur Astronomie (MPIA), Heidelberg, for assistance 
with calibration of the data. 

We would also like to thank the UK ISO support 
team at the CCLRC (formerly RAL) for providing much necessary
information and support. In particular we thank Helen Walker for the
answers to many questions, and Phil Richards for software assistance,
especially with regard to the calibration errors and the
colour temperature analysis. 

We are also grateful to Charles Ryter for an enlightening discussion on the
ISRF and to Aurore Bacmann for estimates of the mid-IR background. We should
like to thank the referee for a number of helpful suggestions.

The Infrared Space Observatory is an ESA project with instruments funded
by ESA member states, particularly the PI countries, France, Germany,
Netherlands and the United Kingdom, and with the participation of ISAS
(Japan) and NASA (USA). 

ISOPHOT was built by the ISOPHOT Consortium, and
the data presented in this paper were reduced using PIA, which is a joint 
development by the ESA Astrophysics Division and the ISOPHOT
Consortium with the collaboration of the Infrared Processing and Analysis 
Center (IPAC). 

Contributing ISOPHOT Consortium institutes are the Dublin IAS, 
the UK CCLRC (formerly RAL),
AI Potsdam, MPI Kernphysik, and MPIA Heidelberg.

\end{document}